\documentclass[a4paper,11pt]{article}

\usepackage{jheppub_mod} 
\usepackage[T1]{fontenc} 

\usepackage{amsmath,booktabs}
\usepackage{graphicx}
\usepackage{xspace}
\usepackage{color}
\pagecolor{white}
\usepackage[dvipsnames]{xcolor}
\usepackage{url}
\usepackage[utf8]{inputenc}
\usepackage{epstopdf}
\usepackage{hyperref}
\usepackage{multirow}
\usepackage{gensymb}

\usepackage{physics}
\usepackage{subcaption}
\usepackage[shortlabels]{enumitem}
\usepackage{siunitx}

\usepackage{ifmtarg}

\newcommand{\Order}{\mathcal{O}}

\newcommand{\mpi}{M_\pi}

\newcommand{\Br}{\text{Br}}
\newcommand{\iu}{\textit{i}}
\newcommand{\eu}{\textit{e}}

\renewcommand{\Re}{\text{Re}\,}
\renewcommand{\Im}{\text{Im}\,}
\newcommand{\disc}{\text{disc}\,}
\newcommand{\disca}{\text{disc}_\text{an}\,}

\providecommand{\MeV}{\,\text{MeV}}
\providecommand{\GeV}{\,\text{GeV}}

\allowdisplaybreaks[1]

\title{Anomalous thresholds in $\boldsymbol{B\to (P,V)\,\gamma^*}$ form factors}

\author[a]{Simon Mutke,}
\author[b]{Martin Hoferichter,}
\author[a]{and Bastian Kubis}

\affiliation[a]{Helmholtz-Institut f\"ur Strahlen- und Kernphysik (Theorie) and
Bethe Center for Theoretical Physics, Universit\"at Bonn, 53115 Bonn, Germany}

\affiliation[b]{Albert Einstein Center for Fundamental Physics, Institute for Theoretical Physics, University of Bern, Sidlerstrasse 5, 3012 Bern, Switzerland}

\emailAdd{mutke@hiskp.uni-bonn.de}
\emailAdd{hoferichter@itp.unibe.ch}
\emailAdd{kubis@hiskp.uni-bonn.de}

\abstract{We study the effects of anomalous thresholds on the non-local form factors describing the hadronization of the light-quark contribution to $B\to(P,V)\,\gamma^*$ transitions. Starting from a comprehensive discussion of anomalous thresholds in the triangle loop function for different mass configurations, we detail how the dispersion relation for $\pi\pi$ intermediate states is affected by contour deformations mandated by the anomalous branch points. Phenomenological estimates of the size of the anomalous contributions to the form factors are provided with couplings determined from measured branching fractions and Dalitz plot distributions. Our key finding is that anomalous effects are suppressed on the $\rho(770)$ resonance, while off-peak the effects can become as large as $\Order(10\%)$ of the full (light-quark-loop induced)  non-local form factors. We comment on future generalizations towards higher intermediate states and the charm loop, outlining how the dispersive framework established in this work could help improve the non-local form factors needed as input for a robust interpretation of $B\to (P,V)\,\ell^+\ell^-$ decays.}

\begin{document}
\maketitle

\addtocontents{toc}{\protect\setcounter{tocdepth}{2}}
	
\section{Introduction}
\label{sec:intro}

Flavor-changing neutral-current decays of $B$-mesons constitute an appealing set of observables to search for physics beyond the Standard Model (BSM), given that in the SM only contributions suppressed by loop factors and CKM matrix elements are possible. Especially manifestations of the $b\to s\ell\ell$ transition have attracted significant attention as rare processes  in which potential BSM contributions might be discovered, in view of the appearance of anomalies suggesting the violation of lepton flavor universality~\cite{Bobeth:2007dw,LHCb:2017avl,LHCb:2019hip,Alguero:2019ptt,Aebischer:2019mlg,Ciuchini:2019usw,Arbey:2019duh,LHCb:2021trn,Crivellin:2021sff,Crivellin:2023zui}. While these hints have disappeared in the meantime~\cite{LHCb:2022qnv}, angular observables and decay rates for $b\to s\mu\mu$ still show sizable deviations from their SM predictions~\cite{Descotes-Genon:2012isb,Descotes-Genon:2013vna,LHCb:2014cxe,LHCb:2020lmf,LHCb:2021zwz,Parrott:2022zte,Gubernari:2022hxn}, see Ref.~\cite{CMS:2024tbn} for the most recent $P_5'$ measurement by CMS, preserving the preference for BSM contributions in global fits at a high level of significance~\cite{Buras:2022qip,Ciuchini:2022wbq,Alguero:2023jeh}.

A key point in assessing the significance of these tensions concerns control over the hadronic matrix elements, usually separated into local and non-local form factors. While $B\to (P,V)\,\ell^+\ell^-$ form factors have been studied in detail using the operator product expansion, light-cone sum rules, and quark-hadron duality, non-local effects therein, such as originating from charm loops, have received particular attention in recent years, given that they could potentially mimic BSM contributions~\cite{Beneke:2001at,Beneke:2004dp,Khodjamirian:2010vf,Khodjamirian:2012rm,Asatrian:2019kbk}. In order to address this concern, a dispersive approach for the corresponding $B\to (P,V)\,\gamma^*$ form factors, with pseudoscalars $P=K,\pi,\ldots$ and vectors $V=K^*,\phi,\rho,\omega,\ldots$ has been developed~\cite{Bobeth:2017vxj,Gubernari:2020eft,Gubernari:2022hxn,Gubernari:2023puw}, to incorporate the constraints from analyticity and unitarity, including experimental information on the residues of the $J/\psi$ and $\psi(2S)$ poles.\footnote{See Ref.~\cite{Isidori:2024lng} for a recent estimate of charm rescattering effects in $B^0\to K^0\ell^+\ell^-$ using heavy-hadron chiral perturbation theory (ChPT).} 
While the use of dispersive techniques has become more common, see, e.g., Refs.~\cite{Cornella:2020aoq,Marshall:2023aje,Hanhart:2023fud,Bordone:2024hui}, a common objection concerns the possible distortion of the analytic structure by anomalous thresholds~\cite{Ciuchini:2022wbq,Ladisa:2022vmh}. However, up to now no actual calculations have been performed to try and estimate the size of the impact on $B\to(P,V)\,\gamma^*$ form factors. The aim of this work is to make the first step in this direction. 

To this end, we first present a comprehensive analysis of anomalous thresholds in the triangle loop function $C_0$, including all the different analytic continuations required for the various possible mass configurations. While, in principle, it is well understood how to account for the anomalous cuts in its dispersive representation~\cite{Lucha:2006vc,Hoferichter:2013ama,Colangelo:2015ama,Hoferichter:2019nlq}, we are not aware of a systematic presentation of all the different cases, a number of which become relevant for the $B\to(P,V)\,\gamma^*$ application due to a host of possible intermediate states and external particles. These results are presented in Sec.~\ref{sec:anom_triangle}. 

In Sec.~\ref{sec:light}, we turn to the dispersive framework for the conceptually simplest case, that is, the hadronization of the light-quark loop in terms of $\pi\pi$ intermediate states. In particular, we perform the decomposition of the non-local matrix elements into Bardeen--Tung--Tarrach (BTT)~\cite{Bardeen:1968ebo,Tarrach:1975tu} structures, whose scalar functions are suitable for a dispersive representation, and derive the unitarity relations as well as their solutions using Muskhelishvili--Omn\`es (MO) techniques~\cite{Muskhelisvili:1953,Omnes:1958hv}, including the additional contributions required by anomalous cuts. We present a first phenomenological analysis in Sec.~\ref{sec:pheno}, using measured decay rates and Dalitz plot distributions to determine the free parameters in the solution of the dispersion relation, as an estimate of how large the anomalous contributions could potentially become in this case. 
In Sec.~\ref{sec:generalizations}, we outline future generalizations regarding higher intermediate states, isoscalar, and $B_s$ decays, as well as the charm loop, before closing with a summary and outlook in Sec.~\ref{sec:summary}.

\section[Anomalous thresholds in the triangle loop function]{Anomalous thresholds in the triangle loop function\footnote{For an extended version of this section, see Ref.~\cite{Mutke:2024}. The presentation is based upon Refs.~\cite{Eden:1966dnq,Itzykson:1980rh,Gribov:2009cfk,Guo:2019twa}.}}
\label{sec:anom_triangle}

\subsection{Landau equations and classification of singularities}
\label{sec:landau_equations}

The singularities of a given Feynman diagram can be determined by means of the 
Landau equations~\cite{Landau:1959fi}
\begin{equation}\label{eq:landau_equations}
	\sum_{j} \pm \alpha_j k_j = 0,\quad \text{for each loop, and either}~\alpha_i=0~\text{or}~k_i^2=m_i^2~\text{for each~}i,
\end{equation}
where $k_j$ denotes the momentum associated with propagator $j$. Leading Landau singularities are the ones for which all internal particles of the graph are on-shell and thus all Feynman parameters $\alpha_i\neq 0$, while subleading singularities have at least one 
$\alpha_i=0$.

\begin{figure}[tb]
	\centering
	\begin{subfigure}[b]{0.14\textwidth}
		\centering
		\includegraphics[width=\textwidth]{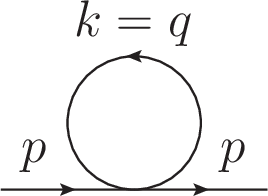}
		\caption{}
		\label{fig:bubble_diagram_single}
	\end{subfigure}
	\hspace{25pt}
	\begin{subfigure}[b]{0.19\textwidth}
		\centering
		\includegraphics[width=\textwidth]{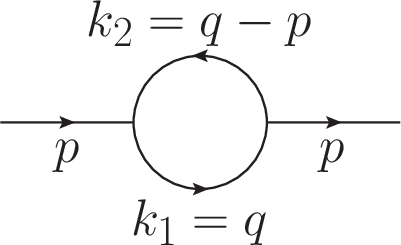}
		\caption{}
		\label{fig:bubble_diagram_two}
	\end{subfigure}
	\hspace{25pt}
	\begin{subfigure}[b]{0.28\textwidth}
		\centering
		\includegraphics[width=\textwidth]{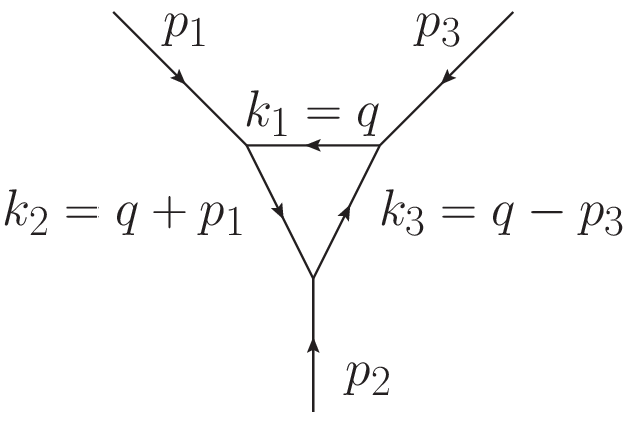}
		\caption{}
		\label{fig:triangle_diagram}
	\end{subfigure}
	\caption{Scalar diagrams corresponding to (a) Eq.~\eqref{eq:one_point_function}, (b) Eq.~\eqref{eq:two_point_function}, and (c) Eq.~\eqref{eq:triangle_function}.}
	\label{fig:bubble_diagrams}
\end{figure}
Before turning to the triangle loop function, we
establish the nature of the singularities of the simpler cases in this language, see Fig.~\ref{fig:bubble_diagrams}. Using the conventions from Ref.~\cite{Denner:1991kt} and setting $t\equiv p^2$, the leading Landau singularities of the bubble diagram
\begin{equation}
\label{eq:two_point_function}
	B_0(t) \equiv B_0(p^2,m_1^2,m_2^2) \equiv \frac{1}{\iu\pi^2} \int \dd[4]{q} \frac{1}{(q^2-m_1^2+\iu\epsilon)((q-p)^2-m_2^2+\iu\epsilon)}
\end{equation}
 take the form
\begin{equation}
	\begin{Bmatrix}
		0=\alpha_1 k_1 + \alpha_2 k_2\\
		0 = k_1^2 - m_1^2 = k_2^2 - m_2^2
	\end{Bmatrix}
	\quad
	\Rightarrow
	\quad
	\begin{Bmatrix}
		2 m_1^2 \alpha_1 + (m_1^2+m_2^2-t) \alpha_2 = 0\\
		(m_1^2+m_2^2-t) \alpha_1 + 2 m_2^2 \alpha_2 = 0
	\end{Bmatrix},
\end{equation}
and thus
\begin{equation}
 \lambda(t,m_1^2,m_2^2)\equiv (m_1^2+m_2^2-t)^2 - 4m_1^2m_2^2=0,
\end{equation}
with solutions at $t=(m_1\pm m_2)^2$. 
Using $\alpha_1 + \alpha_2 = 1$, the corresponding parameters are
\begin{equation}
	\begin{cases}
		\alpha_1 = \frac{m_2}{m_1+m_2},\hspace{5pt}\alpha_2 = \frac{m_1}{m_1+m_2}, & \text{for}~t = (m_1+m_2)^2, \\
		\alpha_1 = \frac{m_2}{m_2-m_1},\hspace{5pt}\alpha_2 = \frac{m_1}{m_1-m_2}, & \text{for}~t = (m_1-m_2)^2.
	\end{cases}
\end{equation}
Accordingly, at the threshold $t_\text{thr} \equiv (m_1+m_2)^2$ both $\alpha_1$ and $\alpha_2$ have positive real solutions and, thus, lie inside the integration range. The two-particle threshold 
 $t_\text{thr}$ is a square-root branch point that appears on every Riemann sheet of $B_0(t)$ and gives rise to the right-hand unitarity cut. In contrast, for the pseudothreshold at $t=(m_1-m_2)^2$ exactly one out of $\alpha_1$ and $\alpha_2$ must be negative, which means that the corresponding singularity cannot appear on the principal Riemann sheet. However, on every other sheet that can be reached after crossing the unitarity cut it appears as another square-root branch point with a corresponding left-hand cut. 

 The subleading Landau singularities stem from setting either $\alpha_1$ or $\alpha_2$ to zero and, thus, correspond to the leading singularities of the single-particle one-loop diagram in Fig.~\ref{fig:bubble_diagram_single} represented by 
\begin{equation}~\label{eq:one_point_function}
	A_0(m^2) \equiv \frac{1}{\iu\pi^2} \int \dd[4]{q} \frac{1}{q^2-m^2+\iu\epsilon},
\end{equation}
which is independent of $t$. They appear if either $m_1^2=0$ or $m_2^2=0$, and behave as $m_1^2 \log\,(m_1^2)$ or $m_2^2 \log\,(m_2^2)$, respectively, for either $m_i^2$ close to zero.

Anomalous thresholds first arise as the leading Landau singularity for the 
triangle diagram in Fig.~\ref{fig:triangle_diagram}, corresponding to the triangle function
\begin{align}\label{eq:triangle_function}
	C_0(t) &\equiv C_0(p_1^2,p_2^2,p_3^2,m_1^2,m_2^2,m_3^2) \equiv \frac{1}{\iu\pi^2} \int \dd[4]{q} \frac{1}{D},\notag\\
	D &= (q^2-m_1^2+\iu\epsilon)((q+p_1)^2-m_2^2+\iu\epsilon)((q-p_3)^2-m_3^2+\iu\epsilon),
\end{align}
where $t \equiv p_2^2$. The subleading Landau singularities can be read off from the previous discussion, i.e., 
 the two-particle threshold $t_\text{thr} = (m_2+m_3)^2$ gives rise to the unitarity cut, 
 and there are also singularities for $p_1^2 = (m_1+m_2)^2$ or $p_3^2 = (m_1+m_3)^2$, which need to be taken into account 
 when performing an analytic continuation in either of these masses. The pseudothresholds at $t = (m_2-m_3)^2$, $p_1^2 = (m_1-m_2)^2$, and $p_3^2 = (m_1-m_3)^2$ again do not appear on the principal Riemann sheet. Lastly, there are the three singularities $m_1^2=0$, $m_2^2=0$, and $m_3^2=0$ inherited from $A_0(t)$. 
The singularities at $t_\text{thr} = (m_2+m_3)^2$, $p_1^2 = (m_1+m_2)^2$, and $p_3^2 = (m_1+m_3)^2$ are again square-root branch points, leading to right-hand cuts in the respective variable. The correct branch for $p_1^2 \geq (m_1+m_2)^2$ or $p_3^2 \geq (m_1+m_3)^2$ is obtained by introducing an infinitesimal imaginary part $p_1^2 \to p_1^2 + \iu\delta$ or $p_3^2 \to p_3^2 + \iu\delta$, respectively~\cite{Gribov:1962fu,Bronzan:1963mby}.

To find the leading Landau singularities of $C_0(t)$, one needs to solve
\begin{equation}
	\begin{Bmatrix}
		\alpha_1 k_1 + \alpha_2 k_2 + \alpha_3 k_3 = 0 \\
		0 = k_1^2 - m_1^2 = k_2^2 - m_2^2 = k_3^2 - m_3^2
	\end{Bmatrix},
\end{equation}
which, defining $y_{ij} \equiv (k_i \cdot k_j) / (m_i m_j)$, requires 
\begin{equation} 
\label{eq:triangle_y_condition}
	0 = 1 + 2 y_{12} y_{13} y_{23} - y_{12}^2 - y_{13}^2 - y_{23}^2.
\end{equation}
Solving this condition for $y_{23}$, one finds that the triangle singularities can appear at
\begin{align}\label{eq:triangle_singularities}
	t_\pm&=p_1^2 \frac{m_1^2 + m_3^2}{2m_1^2} + p_3^2 \frac{m_1^2+m_2^2}{2m_1^2} - \frac{p_1^2 p_3^2}{2m_1^2} - \frac{(m_1^2-m_2^2)(m_1^2-m_3^2)}{2m_1^2} \notag\\
	&\pm \frac{1}{2m_1^2} \sqrt{\lambda(p_1^2,m_1^2,m_2^2) \lambda(p_3^2,m_1^2,m_3^2)},
\end{align}
in the form of logarithmic branch points.

The main task is to determine whether either of these singularities $t_\pm$ can appear on the principal sheet. Starting with the assumption that all $y_{ij}$ and therefore also all $p_i^2$ are real, this implies that $p_1^2 < (m_1+m_2)^2$ and $p_3^2 < (m_1+m_3)^2$, i.e., $y_{13},y_{23} > -1$, due to the branch cuts in these variables. A singularity appearing on the principal sheet requires a solution with all $\alpha_i > 0$. Such a solution requires that at least two out of $y_{12}$, $y_{13}$, and $y_{23}$  be negative, so that all three of them fulfill $\abs{y_{ij}} < 1$. Physically, this means that all the particles are stable as $(m_1-m_2)^2 < p_1^2 < (m_1+m_2)^2$, $(m_1-m_3)^2 < p_3^2 < (m_1+m_3)^2$, and $(m_2-m_3)^2 < t < (m_2+m_3)^2$.

\begin{figure}[tb]
	\centering
	\includegraphics[width=0.37\textwidth]{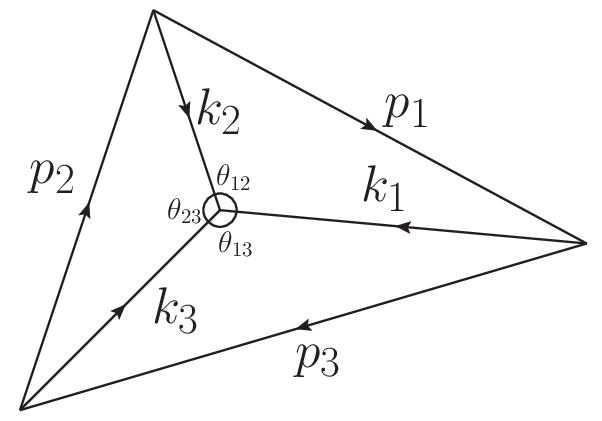}
	\caption{Dual diagram of the triangle diagram in Fig.~\ref{fig:triangle_diagram}.}
	\label{fig:triangle_dual_diagram}
\end{figure}

Under these circumstances, it is useful to consider the dual graph in Fig.~\ref{fig:triangle_dual_diagram}. Due to energy-momentum conservation both the three external momenta $p_i$ and the respective three momenta at each vertex can form closed triangles. As $\sum \alpha_i k_i = 0$ the $k_i$ are linearly dependent, which means that all triangles of the dual graph lie in the same plane. $\abs{y_{ij}} < 1$ implies that the $p_i^2>0$ and $m_i^2>0$ can be interpreted as the squares of the side lengths of the respective edges and $y_{ij} \equiv \cos \theta_{ij}$ as the angles between the $k_i$. All $\alpha_i > 0$ is equivalent to the middle point lying inside the outer triangle, i.e., $\theta_{12} + \theta_{13} + \theta_{23} = 2\pi$ and all $0 < \theta_{ij} < \pi$. This means
\begin{align}
	y_{23} &= \cos\theta_{23} = \cos(2\pi - \theta_{23}) = \cos(\theta_{12} + \theta_{13}) \\
	&= \cos\theta_{12} \cos\theta_{13} - \sin\theta_{12} \sin\theta_{13} = y_{12} y_{13} - \sqrt{(1-y_{12}^2)(1-y_{13}^2)},\notag
\end{align}
and $\theta_{12} + \theta_{13} > \pi$, which is equivalent to
\begin{equation}
	y_{12} + y_{13} = \cos\theta_{12} + \cos\theta_{13} < \cos\theta_{12} + \cos(\pi - \theta_{12}) = 0.
\end{equation}
These two equations imply that only $t_+$ from Eq.~\eqref{eq:triangle_singularities} can be a singularity on the principal sheet and this is the case precisely if
\begin{equation}\label{eq:triangle_singularity_condition}
	m_3 p_1^2 + m_2 p_3^2 > (m_2 + m_3)(m_1^2 + m_2 m_3).
\end{equation}

\subsection{Anomalous thresholds}

To see how $t_+$ moves from the second Riemann sheet to the principal one, we assign an infinitesimal imaginary part to the mass $p_1^2 \to p_1^2 + \iu \delta$ and then increase its value until the inequality in Eq.~\eqref{eq:triangle_singularity_condition} is fulfilled. For $y_{23}$ this corresponds to the replacement $y_{12} \to y_{12} - \iu \delta /2 m_1 m_2$. As the crossing happens when $y_{12} = - y_{13}$, so that $y_{23} = -1$, we can expand in $\delta$ to obtain the crossing point
\begin{equation}
	y_{23}^\text{c} = -1 -\delta^2 \frac{1}{4 m_1^2 m_2^2 (1-y_{12}^2)} + \order{\delta^4} = -1 +\delta^2 \frac{1}{\lambda(p_1^2,m_1^2,m_2^2)} + \order{\delta^4},
\end{equation}
or
\begin{equation} \label{eq:triangle_singularity_crossing_point}
	t_\text{c} = t_\text{thr} - \delta^2\frac{2 m_2 m_3}{\lambda(p_1^2,m_1^2,m_2^2)} + \order{\delta^4} > t_\text{thr},
\end{equation}
where in the last step we used that the stability condition implies $\lambda(p_1^2,m_1^2,m_2^2) < 0$. Thus, when increasing $p_1^2$, the singularity at $t_+$ approaches $t_\text{thr}$ from below on the second sheet, then circles around $t_\text{thr}$ clockwise moving through the unitarity cut onto the principal sheet at $t_\text{c}$, and then moves to the left-hand side of $t_\text{thr}$ again.

Since the singularity at $t_+$ is a logarithmic branch point, its corresponding branch cut extends towards $t_-$. As the latter is still on the second sheet, the branch cut goes from $t_+$ to $t_\text{thr}$ on the principal sheet and from $t_\text{thr}$ to $t_-$ on the second sheet. This introduces an additional discontinuity starting even below the unitarity cut, which needs to be taken into account when performing a dispersive analysis. For this reason, the triangle singularity $t_+$ is often referred to as anomalous threshold in the literature.

As a next step, we need to perform the analytic continuation beyond 
 $\abs{y_{ij}}<1$, to be able to treat the general case  
 with an arbitrary mass configuration (without loss of generality, we take $p_1^2 \geq p_3^2$). We start with $m_1^2$, $m_2^2$, and $m_3^2$ fixed at their physical values, but analytically continue in $p_1^2$ and $p_3^2$ to values such that $\abs{y_{ij}}<1$. After doing so, we are back at the special case discussed above, where we understand exactly whether the triangle singularity is on the principal sheet or not. Now we can reverse the argument and analytically continue back to the physical values, first in $p_1^2$ and then in $p_3^2$. By keeping track of the positions of $t_\pm$ during this process, we can therefore examine every possible mass configuration by starting from the special case.

\begin{table}[tb]
	\centering
	\renewcommand{\arraystretch}{1.3}
	\begin{tabular}{ccccccc}
	\toprule 
		$p_3^2$ & \multicolumn{2}{c}{$M_{K^{*\pm}}^2$} & $M_{K^\pm}^2$ & \multicolumn{2}{c}{$M_{\rho^{\pm}}^2$} & $M_{\pi^\pm}^2$ \\
		$m_1^2$ & $M_{K^0}^2$ & $M_{K^{*0}}^2$ & $M_{K^{*0}}^2$ & $M_{\pi^0}^2$ & $M_{\omega}^2$ & $M_{\rho^0}^2$ \\\midrule
		Case & A & B & C & A & B & C \\
		$t_+$ $\big[\text{GeV}^2\big]$ & $-57.8$ & $0.48-4.19\iu$ & $18.6$ & $-859$ & $0.73-4.81\iu$ & $26.4$ \\
		$t_-$ $\big[\text{GeV}^2\big]$ & $-1.0$ & $0.48+4.19\iu$ & $1.0$ & $-0.93$ & $0.73+4.81\iu$ & $0.95$ \\
		\bottomrule
	\end{tabular}
	\renewcommand{\arraystretch}{1.0}
	\caption{Triangle singularities for different mass configurations, where all have $p_1^2 = M_{B^\pm}^2$ and $m_2^2=m_3^2=M_{\pi^\pm}^2$, using particle masses from Ref.~\cite{ParticleDataGroup:2022pth}.}
	\label{tab:triangle_singularities_Bdecays}
\end{table}

To illustrate this procedure, let us consider some special cases listed in Table~\ref{tab:triangle_singularities_Bdecays}, with the general case summarized in App.~\ref{sec:triangle_position_general_case}. All three cases $A$, $B$, and $C$ have in common that $y_{12}<-1$. We therefore continue $p_1^2 > (m_1+m_2)^2$ towards its physical value, while keeping the small imaginary part $p_1^2 + \iu\delta$ as discussed above. In this case the discriminant of the square-root in Eq.~\eqref{eq:triangle_singularity_condition} becomes negative, which means that the square-root becomes imaginary and there are two branches from which to choose. However, the small imaginary part $\iu\delta$ singles out the correct branch and we find
\begin{align}\label{eq:triangle_singularities_analytic_continuation}
	t_\pm&=p_1^2 \frac{m_1^2 + m_3^2}{2m_1^2} + p_3^2 \frac{m_1^2+m_2^2}{2m_1^2} - \frac{p_1^2 p_3^2}{2m_1^2} - \frac{(m_1^2-m_2^2)(m_1^2-m_3^2)}{2m_1^2} \notag\\
	&\mp \frac{1}{2m_1^2} \lambda^{\frac12}(p_1^2,m_1^2,m_2^2) \lambda^{\frac12}(p_3^2,m_1^2,m_3^2)
\end{align}
to be the correct analytic continuation of the locations of the triangle singularity, where $\lambda^{\frac12}(a^2,b^2,c^2) \equiv \sqrt{a^2-(b+c)^2}\sqrt{a^2-(b-c)^2}$. In particular, the sign in the second line changes compared to Eq.~\eqref{eq:triangle_singularities}.

Now that we have the analytic continuation of $t_\pm$ at hand, let us trace the position of $t_+$ in the complex plane upon changing first $p_1^2$ and then $p_3^2$. As soon as $p_1^2$ grows bigger than $(m_1+m_2)^2$, the singularity $t_+$ moves into the lower complex half-plane ($\Im t_+ < 0$) of the principal Riemann sheet. At first glance this should be alarming as its counterpart $t_-$ on the upper complex half-plane ($\Im t_- > 0$) is still located on the second sheet and therefore it seems like the Schwarz reflection principle is violated. However, by analytically continuing $p_1^2$ across the branch point $p_1^2 = (m_1+m_2)^2$ and choosing one of the branches $p_1^2 + \iu\delta$, we unavoidably introduced an imaginary part to $C_0(t)$ on the entirety of the real axis. Thus, the Schwarz reflection principle does not apply anymore and there is no contradiction.\footnote{Had we chosen a negative imaginary part $p_1^2 - \iu\delta$ instead, then $t_+$ would have moved into the upper half-plane instead, which would lead to a conflict with the causality condition that all amplitudes be analytic in the upper-half plane of the physical sheet. This is part of the reason why $p_1^2+\iu\delta$ is the correct choice.}

\begin{figure}[tb]
	\centering
	\begin{subfigure}{0.48\textwidth}
		\centering
		\includegraphics[width=0.9\textwidth]{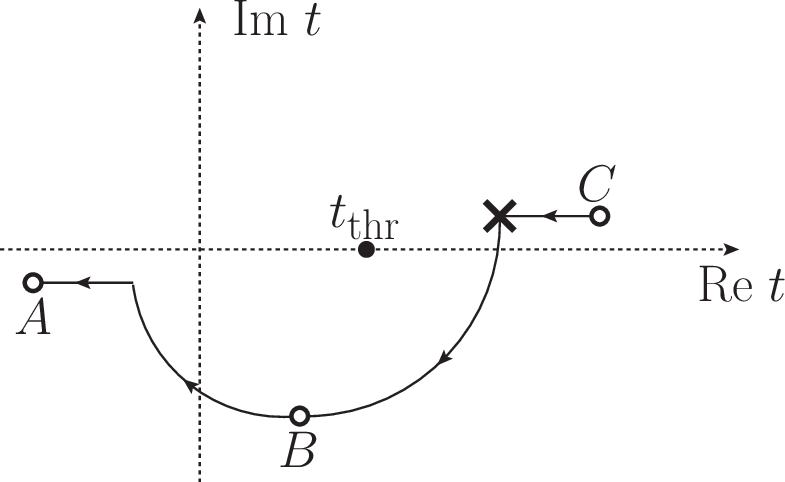}
		\caption{}
		\label{fig:triangle_position}
	\end{subfigure}
	\begin{subfigure}{0.48\textwidth}
		\centering
		\includegraphics[width=0.9\textwidth]{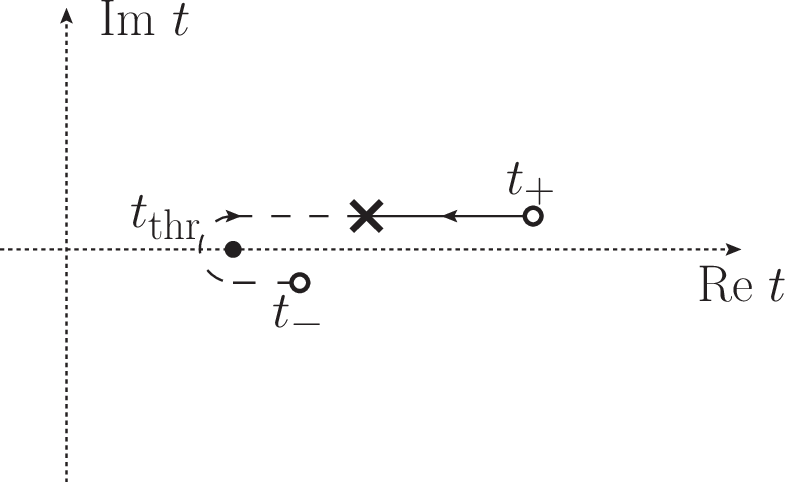}
		\caption{}
		\label{fig:triangle_position_cut}
	\end{subfigure}
	\caption{(a) The solid line indicates the path traced by $t_+$ in the complex plane upon increasing $p_3^2$ for the mass configurations listed in Table~\ref{tab:triangle_singularities_Bdecays}. The positions of $t_+$ for cases $A$, $B$, and $C$ are marked by the hollow circles. (b) The solid (dashed) line indicates the path traced by $t_+$ ($t_-$) in the complex plane on the second sheet upon further decreasing $p_3^2$ starting at case $C$ from Table~\ref{tab:triangle_singularities_Bdecays}. The dashed and solid lines meet at the point where the path in (a) crossed the unitarity cut, indicated by a cross in both figures, see main text for details.}
\end{figure}

This situation now corresponds to case $B$, i.e., $y_{12}<-1$, $|y_{13}|<1$, and therefore $t_+$ lies in the lower half-plane of the physical sheet (see Fig.~\ref{fig:triangle_position}). In the next step we can continue in $p_3^2$ towards its physical value. For case $A$ we have $y_{13}<-1$, i.e., $p_3^2>(m_1+m_3)^2$, which means we introduce another imaginary part $p_3^2 \to p_3^2+\iu\delta$. This leads to $t_+$ moving back onto the real axis to $t_+ = \Re t_+-\iu\delta$ with $\Re t_+<(m_2-m_3)^2$ (see Fig.~\ref{fig:triangle_position}). For case $C$ we instead have $y_{13}>1$, i.e., $p_3^2<(m_1-m_3)^2$. For these particular mass configurations in Table~\ref{tab:triangle_singularities_Bdecays} we further have $m_1>m_3$, which means that $t_+$ moves onto the unitarity cut from below and back onto the second sheet towards $t_+ = \Re t_+ + \iu\delta$ with $t_\text{thr} < \Re t_+ < t_\text{ps}$ (see Fig.~\ref{fig:triangle_position}), where we introduced the pseudothreshold $t_\text{ps} = \big(\sqrt{p_1^2}-\sqrt{p_3^2}\big)^2$.\footnote{From now on, the term ``pseudothreshold'' always refers to $t_\text{ps}$ unless stated otherwise. Other singularities often referred to as pseudothresholds, e.g., at $t=(m_2-m_3)^2$, are still present on the second Riemann sheet, but will not play a role in the following.} At the same time $t_-$ also moves onto the cut with $t_- = \Re t_- + \iu\delta$ and $t_\text{thr} < \Re t_- \leq \Re t_+ < t_\text{ps}$. Although both of them lie on the unitarity cut, they are not on the physical boundary $t = \Re t + \iu\epsilon$ with $\Re t > t_\text{thr}$, which is instead connected to the lower half plane $t = \Re t - \iu\epsilon$ of the second sheet.

\begin{figure}[tb]
	\centering
	\includegraphics[width=0.7\textwidth]{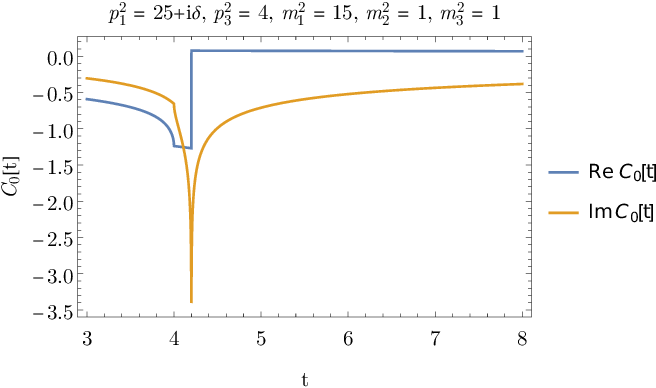}
	\caption{Example of a resonance-like peak in the triangle function $C_0(t+\iu\epsilon)$, caused by the triangle singularity at $t_- = 4.2-\iu\delta$ on the second sheet, located closely above the two-particle threshold $t_\text{thr}=4$. The other triangle singularity sits at $t_+=7+\iu\delta$ on the second sheet and does not cause such a peak. The mass configuration is given in the figure.}
	\label{fig:c0_resonance}
\end{figure}

However, when $p_3^2$ is further reduced beyond the point at which Eq.~\eqref{eq:triangle_singularity_condition} is not longer fulfilled, the second triangle singularity $t_-$ moves anticlockwise around $t_\text{thr}$ on the second sheet and onto the physical boundary $t_- = \Re t_- - \iu\delta$ with $\Re t_- > t_\text{thr}$ (see Fig.~\ref{fig:triangle_position_cut}). The crossing point $t_c$ in this case is located at
\begin{equation}
	t_\text{c} = t_\text{thr} - \delta^2\frac{2 m_2 m_3}{\lambda(p_1^2,m_1^2,m_2^2)} + \order{\delta^4} < t_\text{thr},
\end{equation}
which can be computed similarly to Eq.~\eqref{eq:triangle_singularity_crossing_point} before. The example in Fig.~\ref{fig:c0_resonance} illustrates that as soon as this crossing of $t_-$ onto the physical boundary has happened it causes a resonance-like peak.

We show in App.~\ref{sec:triangle_position_general_case} that this phenomenon can occur in general whenever $y_{12} + y_{13}>0$, with either $y_{12}<-1$ and $y_{13}>1$, or $y_{12}>1$ and $y_{13}<-1$. As  discussed recently in Refs.~\cite{Bayar:2016ftu,Guo:2019twa}, these conditions are in accordance with the Coleman--Norton theorem~\cite{Coleman:1965xm}, stating that the leading Landau singularity of a graph lies in the physical region if and only if all vertices of the graph describe energy- and momentum-conserving classical processes of real particles on their positive-energy mass-shell moving forward in time. The resulting resonance-like peak can of course imitate a seemingly new hadronic state in the measured data. One such example appears to be the $a_1(1420)$ resonance, which can be measured in the decay channel $a_1(1420) \to f_0(980) \pi$~\cite{COMPASS:2015kdx,COMPASS:2018uzl,Wagner:2024syk}. Recent evidence supports the hypothesis that it is not actually a hadronic state but instead caused by a triangle singularity with $p_1^2=M_{f_0(980)}^2$, $p_3^2=M_{\pi^\pm}^2$, $m_1^2=m_2^2=M_{K^\pm}^2$, and $m_3^2=M_{K^{*0}}^2$~\cite{Mikhasenko:2015oxp,Aceti:2016yeb,COMPASS:2020yhb}.\footnote{The nature of the $a_1(1420)$ is currently under study at Belle in the reaction $\tau\to 3\pi\nu_\tau$~\cite{Rabusov:2023otv,Rabusov:2024koz}, to help clarify whether it constitutes a genuine resonance or indeed the manifestation of a triangle singularity.}  A table with more (suspected) examples is provided in Ref.~\cite{Guo:2019twa}.

We also see in App.~\ref{sec:triangle_position_general_case} for the general case that $t_-$ can never move onto the physical sheet and that $t_+$ can only do so if $y_{12} + y_{13} < 0$. If either both $y_{12},y_{13}<-1$ or $|y_{12}|,|y_{13}|<1$, then $t_+$ sits on the real axis below threshold $t_+<t_\text{thr}$. If either $y_{12}<-1$ and $|y_{13}|<1$, or $|y_{12}|<1$ and $y_{13}<-1$, then $t_+$ is located in the lower complex half-plane. Else, if $|y_{12}|,|y_{13}|>1$ with $y_{12}  y_{13}<-1$, then $t_+$ as well as $t_-$ are located on the unitarity cut, but not on the physical boundary.

\subsection{Dispersion relation for the scalar triangle function} \label{sec:triangle_function_dispersion}

Based on the discussion in the previous section, we are now in the position to derive dispersion relations for $C_0(t)$ in the various possible mass configurations. We study this loop function in some detail, both as a test case in which numerical results can be easily cross-checked via standard packages such as \texttt{LoopTools}~\cite{Hahn:1998yk} and because the conclusions will  
 generalize to more complicated amplitudes with triangle topology.

\begin{figure}[tb]
	\centering
	\includegraphics[width=0.6\textwidth]{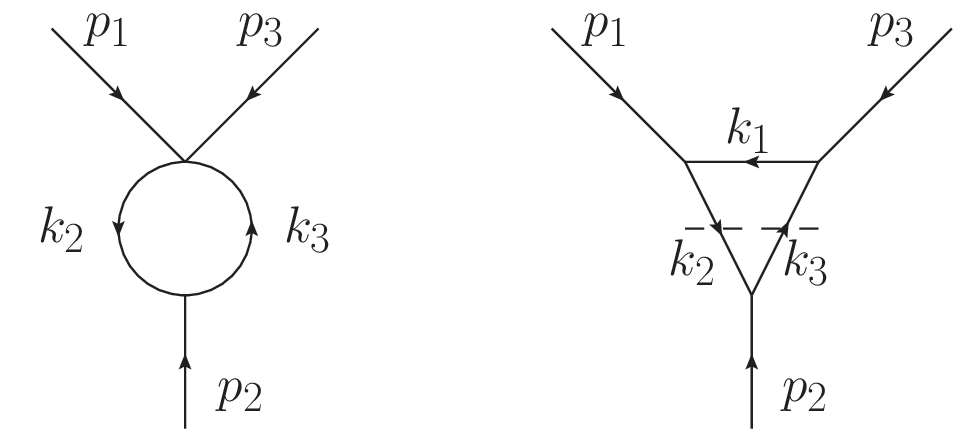}
	\caption{Left: Reduced triangle graph with $\alpha_1=0$ corresponding to the two-particle threshold. Right: Cut triangle graph yielding the discontinuity across the unitarity cut.}
	\label{fig:triangle_cut}
\end{figure}

Let us first derive the unitarity relation for the triangle function $C_0(t)$ as defined in Eq.~\eqref{eq:triangle_function}, using the short-hand notation
\begin{equation}
 t_1 \equiv t_\text{thr} = (m_2+m_3)^2,\qquad 
 t_2 \equiv t_\text{ps} = \Big(\sqrt{p_1^2}-\sqrt{p_3^2}\Big)^2,\qquad t_3 \equiv \Big(\sqrt{p_1^2}+\sqrt{p_3^2}\Big)^2.
\end{equation}
If we start in the kinematic region in which all particles are stable, i.e., $(m_1-m_2)^2 < p_1^2 < (m_1+m_2)^2$ and $(m_1-m_3)^2 < p_3^2 < (m_1+m_3)^2$, and where the triangle singularity does not yet occur on the physical sheet, i.e., $m_3 p_1^2 + m_2 p_3^2 < (m_2 + m_3)(m_1^2 + m_2 m_3)$, see Eq.~\eqref{eq:triangle_singularity_condition}, then the only singularity on the physical sheet is the two-particle threshold $t_1$ giving rise to the right-hand unitarity cut. We can compute the discontinuity across this cut via Cutkosky rules~\cite{Cutkosky:1960sp}, and thereby set up a dispersion relation for $C_0(t)$. Afterwards, we will then analytically continue in $p_1^2$ and $p_3^2$ back to the actual physical value of the masses.

\begin{figure}[tb]
	\centering
	\begin{subfigure}{0.48\textwidth}
		\centering
		\includegraphics[width=\textwidth]{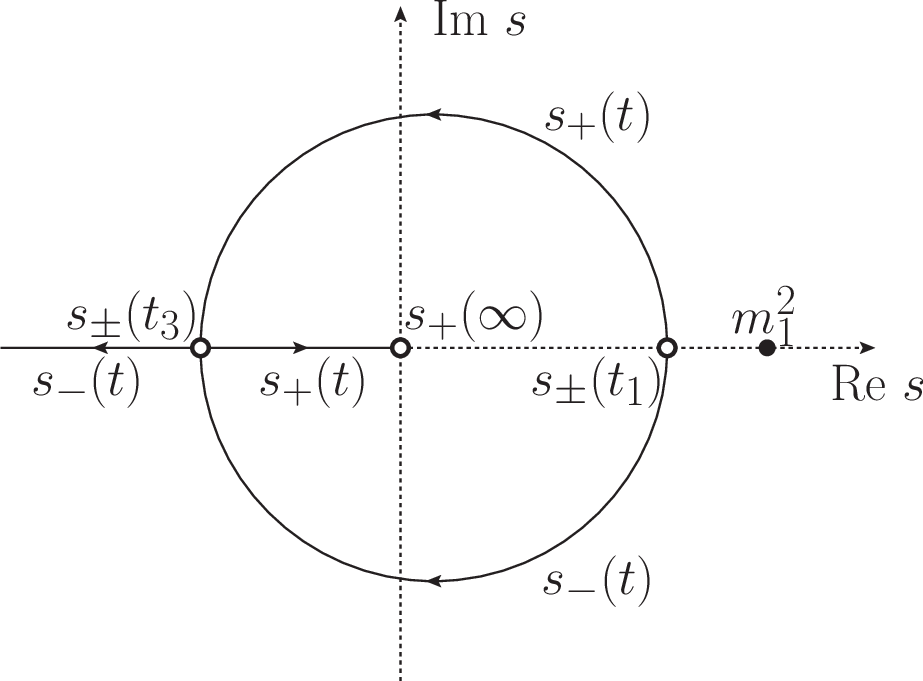}
		\caption{}
		\label{fig:pinocchio_stable}
	\end{subfigure}
	\begin{subfigure}{0.48\textwidth}
		\centering
		\includegraphics[width=\textwidth]{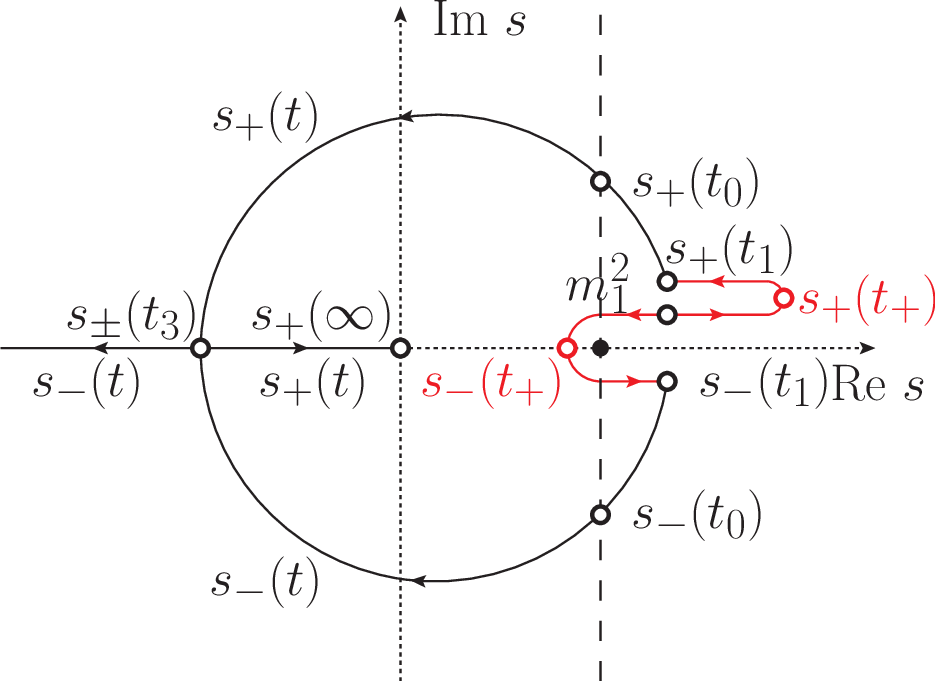}
		\caption{}
		\label{fig:pinocchio_stable_continued}
	\end{subfigure}
	\caption{The black solid lines show the paths traced by $s_+(t)$ and $s_-(t)$ upon increasing $t$ starting from $t=t_1$. The mass configuration is such that all particles are stable. (a) The triangle singularity $t_+$ is still located on the second sheet. (b) The triangle singularity $t_+$ has moved onto the physical sheet. The additional paths of $s_\pm(t\pm\iu\epsilon)$ for $t_+ < t < t_1$ are shown with the red solid lines.}
\end{figure}

From Cutkosky rules, the discontinuity becomes 
\begin{align}
	\disc C_0(t) &= 4\iu \int \dd[4]{q} \frac{1}{q^2-m_1^2} \delta((q+p_1)^2-m_2^2) \, \theta(-q^0-p_1^0) \, \delta((q-p_3)^2-m_3^2) \, \theta(q^0-p_3^0)\notag\\
	&= 2\pi\iu \theta(t-t_1) \lambda^{\frac12}(t,m_2^2,m_3^2) \frac{1}{2t} \int_{-1}^{+1} \dd{z} \frac{1}{s(t,z)-m_1^2},
\end{align}
where we introduced the Mandelstam variables $s = (p_1-k_2)^2 = (p_3+k_3)^2$, $t=p_2^2 = (p_1+p_3)^2 = (k_2-k_3)^2$, and $u = (p_1+k_3)^2 = (p_3-k_2)^2$, related by
\begin{equation} \label{eq:triangle_mandelstam}
	\begin{Bmatrix}s(t,z)\\u(t,z)\end{Bmatrix} = \frac{p_1^2+p_3^2+m_2^2+m_3^2-t}{2} \mp \frac{(p_1^2-p_3^2)(m_2^2-m_3^2)}{2t} \pm z \frac{\lambda^{\frac12}(t,p_1^2,p_3^2) \lambda^{\frac12}(t,m_2^2,m_3^2)}{2t}.
\end{equation}
The remaining integral
\begin{equation}
	\frac{1}{2t} \int_{-1}^{+1} \dd{z} \frac{1}{s(t,z)-m_1^2} = \int_{-1}^{+1} \dd{z} \frac{P_0(z)}{-Y(t)+\kappa(t)z} = -\frac{1}{\kappa(t)} \log\frac{Y(t)+\kappa(t)}{Y(t)-\kappa(t)},
\end{equation}
with 
\begin{align}
\label{Y_def}
Y(t) &\equiv t^2 - t(p_1^2+p_3^2+m_2^2+m_3^2-2m_1^2) + (p_1^2-p_3^2)(m_2^2-m_3^2), \notag\\
\kappa(t) &\equiv \lambda^{\frac12}(t,p_1^2,p_3^2) \lambda^{\frac12}(t,m_2^2,m_3^2), 
\end{align}
and the zeroth Legendre polynomial $P_0(z)=1$, is the $S$-wave projection of the $s$-channel Born amplitude $(s-m_1^2)^{-1}$ up to the prefactor of $1/t$. In the evaluation of the integral we used that $[\kappa(t)]^2 = [Y(t)]^2$ is solved exactly by the triangle singularities $t_\pm$ from Eq.~\eqref{eq:triangle_singularities_analytic_continuation}, reflecting the fact that $t_\pm$ are logarithmic branch points of $C_0(t)$. In the case in which all particles are stable the above integral is well-defined and does not have any singularities, as can be seen by rewriting it as the following contour integral
\begin{equation} \label{eq:pinocchio_integral_stable}
	\frac{1}{2t} \int_{-1}^{+1} \dd{z} \frac{1}{s(t,z)-m_1^2} = \frac{1}{\kappa(t)} \int_{\mathcal{C}_t} \dd{s} \frac{1}{s-m_1^2},
\end{equation}
with a straight path between $s_\pm(t) \equiv s(t,\pm 1)$ as the integration contour $\mathcal{C}_t$. In this case,  $s_\pm(t)$ trace the path shown in Fig.~\ref{fig:pinocchio_stable}. Due to the condition~\eqref{eq:triangle_singularity_condition}, the singularity $s=m_1^2$ of the integrand never lies inside the integration region, so that the integral is free of singularities for all $t\geq t_1$.
The discontinuity of $C_0(t)$ therefore takes the form
\begin{equation}\label{eq:triangle_function_discontinuity}
	\disc C_0(t) = - \frac{2\pi\iu \theta(t-t_1)}{\lambda^{\frac12}(t,p_1^2,p_3^2)} \log\frac{Y(t)+\kappa(t)}{Y(t)-\kappa(t)},
\end{equation}
which has a high-energy behavior $ \order{t^{-1}}$ and does not possess any zeros or singularities along the unitarity cut, except for the expected square-root singularity at threshold. We can therefore set up an unsubtracted dispersion relation
\begin{equation}
	C_0(t) = \frac{1}{2\pi\iu} \int_{t_\text{thr}}^{\infty} \dd{t^\prime} \frac{\disc C_0(t^\prime)}{t^\prime-t-\iu\epsilon}.
\end{equation}

\begin{figure}[tb]
	\centering
	\includegraphics[width=0.7\textwidth]{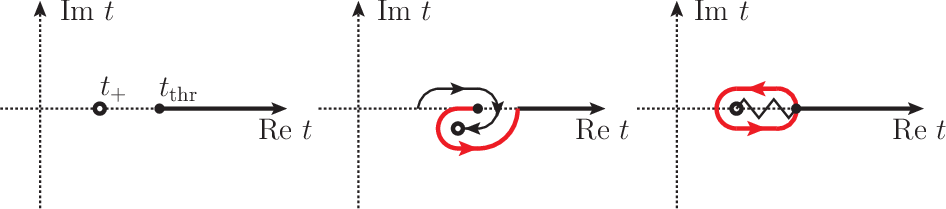}
	\caption{Left: positions of $t_\text{thr}$ and $t_+$ in the complex $t$-plane before the analytic continuation in $p_1^2$. The integration path is indicated by the thick solid line. Middle: integration path gets distorted when $t_+$ moves through the unitarity cut as indicated by the thin solid line with arrows. The additional path arising from that is highlighted in red. Right: deformed integration path around the anomalous branch cut, indicated by the zig-zag line, after the analytic continuation.}
	\label{fig:anomalous_discontinuity}
\end{figure}

When analytically continuing in the masses $p_1^2$ and $p_3^2$ in the next step, it is critical to not only analytically continue the integrand, but also 
observe how the aforementioned singularities move through the complex plane and how this might affect the integration contour. First of all, as soon as Eq.~\eqref{eq:triangle_singularity_condition} is fulfilled, the triangle singularity $t_+$ moves through the unitarity cut---and therefore the integration range of the dispersion integral. We therefore have to deform the integration contour  as shown in Fig.~\ref{fig:anomalous_discontinuity}, picking up an additional, so-called anomalous integral around the logarithmic branch cut. 

\begin{figure}[tb]
	\centering
	\begin{subfigure}[b]{0.2\textwidth}
		\centering
		\includegraphics[width=0.9\textwidth]{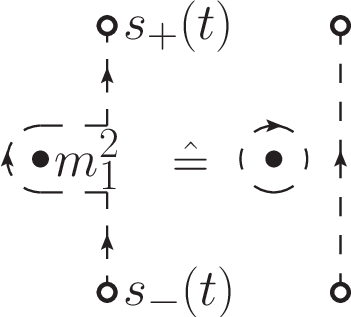}
		\caption{}
		\label{fig:pinocchio_scontour_deformed}
	\end{subfigure}
	\hspace{0.1\textwidth}
	\begin{subfigure}[b]{0.6\textwidth}
		\centering
		\includegraphics[width=0.9\textwidth]{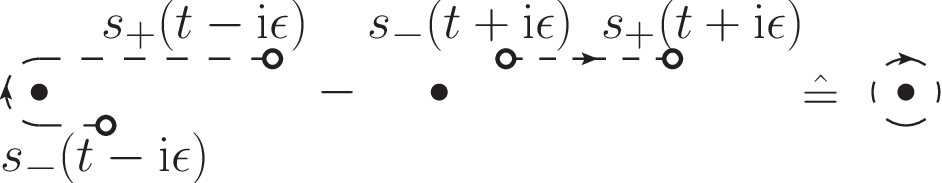}
		\caption{}
		\label{fig:pinocchio_scontour_anomalous}
	\end{subfigure}
	\caption{(a) Deformed integration contour $\mathcal{C}_t^\prime$ of the integral in Eq.~\eqref{eq:pinocchio_integral_stable} for $t_1<t<t_0$. Due to Cauchy's theorem it can be reduced to the original straight-line contour $\mathcal{C}_t$ and a residue at $s=m_1^2$. (b) Difference between the integration contour $\mathcal{C}_t^\prime$ of the integral in Eq.~\eqref{eq:pinocchio_integral_stable} for $t = \Re t \mp \iu\epsilon$ with $t_+<t<t_1$. Due to Cauchy's theorem this difference reduces to a residue at $s=m_1^2$.}
\end{figure}

To compute both the analytic continuation of the discontinuity $\disc C_0(t)$ in Eq.~\eqref{eq:triangle_function_discontinuity} and the discontinuity across the anomalous logarithmic branch cut, we consider Fig.~\ref{fig:pinocchio_stable_continued} and study what happens to $s_\pm(t)$ as soon as Eq.~\eqref{eq:triangle_singularity_condition} is fulfilled. The singularity $s=m_1^2$ of the integrand has moved through the integration region for all $t_1 < t < t_0$, where $t_0$ is defined as the positive root of $Y(t)=0$,
\begin{equation} \label{eq:disc_C0_t0}
	t_0 = \frac{p_1^2+p_3^2+m_2^2+m_3^2-2m_1^2}{2} + \sqrt{\left(\frac{p_1^2+p_3^2+m_2^2+m_3^2-2m_1^2}{2}\right)^2-(p_1^2-p_3^2)(m_2^2-m_3^2)}.
\end{equation}
Therefore, we need to deform the integration contour $\mathcal{C}_t$ of the integral in Eq.~\eqref{eq:pinocchio_integral_stable} around this singularity for all $t_1 < t < t_0$ in order to analytically continue $\disc C_0(t)$ correctly. The deformed integration path $\mathcal{C}_t^\prime$ is shown in Fig.~\ref{fig:pinocchio_scontour_deformed}. Due to Cauchy's theorem the integral along the deformed contour reduces to a residue at $s=m_1^2$ and the integral along the original contour $\mathcal{C}_t$, in such a way that we obtain for the discontinuity
\begin{align}
	\disc C_0(t) &= 2\pi\iu \theta(t-t_1) \frac{\lambda^{\frac12}(t,m_2^2,m_3^2)}{\kappa(t)} \int_{\mathcal{C}_t^\prime} \dd{s} \frac{1}{s-m_1^2} \notag\\
	&= 2\pi\iu \theta(t-t_1) \frac{\lambda^{\frac12}(t,m_2^2,m_3^2)}{\kappa(t)} \left( -2\pi\iu\Res[\frac{1}{s-m_1^2}]\theta(t_0-t) + \int_{\mathcal{C}_t} \dd{s} \frac{1}{s-m_1^2} \right) \notag\\
	&= 2\pi\iu \theta(t-t_1) \frac{\lambda^{\frac12}(t,m_2^2,m_3^2)}{\kappa(t)} \left( -2\pi\iu\theta(t_0-t) - \log\frac{Y(t)+\kappa(t)}{Y(t)-\kappa(t)} \right).
\end{align}
Introducing 
\begin{equation}
\label{kappapm_def}
\kappa_\pm(t) \equiv \sqrt{\pm\lambda(t,m_2^2,m_3^2)\lambda(t,p_1^2,p_3^2)}, 
\end{equation}
this can be rewritten as
\begin{equation} \label{eq:disc_C0_stable}
	\disc C_0(t) = \theta(t-t_1) \begin{cases}
		\frac{-2\pi\iu}{\sqrt{\lambda(t,p_1^2,p_3^2)}} \log\frac{Y(t)+\kappa_+(t)}{Y(t)-\kappa_+(t)}, & t>t_3, \\
		\frac{-4\pi\iu}{\sqrt{-\lambda(t,p_1^2,p_3^2)}} \left[ \arctan\frac{\kappa_-(t)}{Y(t)} + \pi \theta(t_0-t) \right], & t<t_3.
	\end{cases}
\end{equation}
Similarly, from the difference between the integration paths shown in Fig.~\ref{fig:pinocchio_scontour_anomalous} we can see that the anomalous discontinuity across the logarithmic branch cut amounts to
\begin{equation} \label{eq:disc_anom_C0_stable}
	\disca C_0(t) = -\frac{4\pi^2}{\lambda^{\frac12}(t,p_1^2+\iu\delta,p_3^2)} = -\frac{4\pi^2\iu}{\sqrt{-\lambda(t,p_1^2,p_3^2)}}.
\end{equation}
In the presence of the anomalous threshold $t_+$ on the physical sheet the dispersive representation of $C_0(t)$ therefore modifies to
\begin{equation} \label{eq:triangle_function_dispersion_relation}
	C_0(t) = \frac{1}{2\pi\iu} \int_{t_\text{thr}}^{\infty} \dd{t^\prime} \frac{\disc C_0(t^\prime)}{t^\prime-t-\iu\epsilon} + \frac{1}{2\pi\iu} \int_0^1 \dd{x} \pdv{t_x}{x} \frac{\disca C_0(t_x)}{t_x-t-\iu\epsilon},
\end{equation}
where $t_x \equiv x t_\text{thr} + (1-x)t_+$ and where $\disc C_0(t)$ and $\disca C_0(t)$ are given in Eqs.~\eqref{eq:disc_C0_stable} and~\eqref{eq:disc_anom_C0_stable}.

\begin{figure}[tb]
	\centering
	\includegraphics[width=0.5\textwidth]{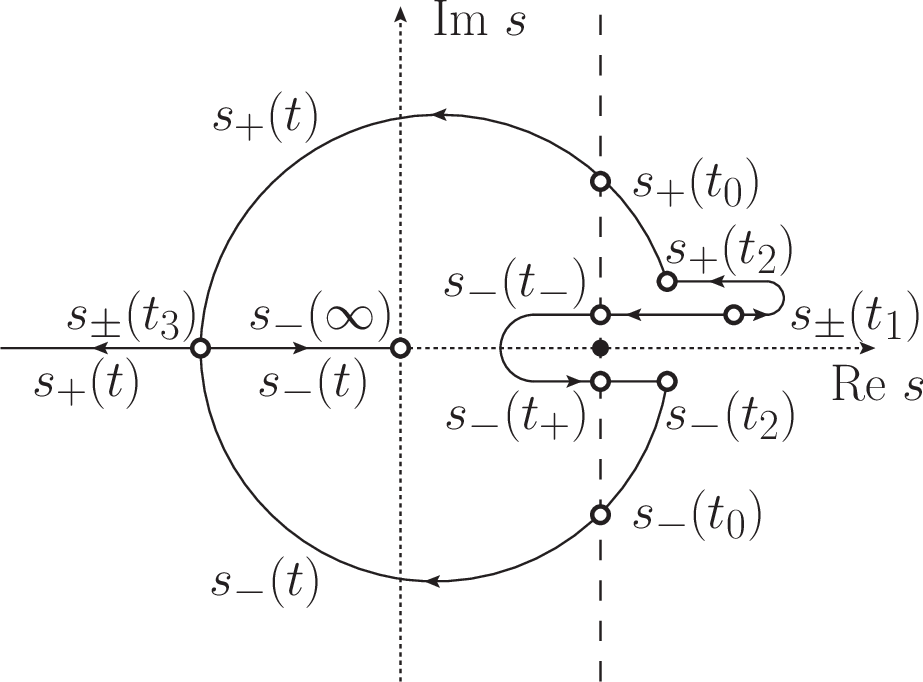}
	\caption{The black solid lines show the paths traced by $s_+(t)$ and $s_-(t)$ upon increasing $t$ starting from $t=t_1$ for case $C$ of the mass configurations in Table~\ref{tab:triangle_singularities_Bdecays}. The black dot indicates the position of the singularity at $s=m_1^2$.}
	\label{fig:pinocchio_case_c}
\end{figure}

Next, we perform the analytic continuation of this dispersive representation towards the mass configurations in Table~\ref{tab:triangle_singularities_Bdecays}, starting with $p_1^2 > (m_1+m_2)^2$ via the usual $p_1^2+\iu\delta$ prescription. In doing so, we lift the Schwarz reflection principle, and $t_+$ picks up an imaginary part, moving into the lower complex half-plane. As the anomalous discontinuity in the form of $\disca C_0(t) = -4\pi^2 / \lambda^{\frac12}(t,p_1^2,p_3^2)$ is analytic in the lower half-plane, we can analytically continue the anomalous integral simply by plugging in the new value of $t_+$ into the integration path $t_x$. However, the normal discontinuity changes because of the modified  positions of $t_0$ and $t_2$, leading to
\begin{equation}\label{eq:disc_C0_AB}
	\disc C_0(t) = \theta(t-t_1) \begin{cases}
		\frac{-2\pi\iu}{\sqrt{\lambda(t,p_1^2,p_3^2)}} \left[ \log\frac{Y(t)+\kappa_+(t)}{Y(t)-\kappa_+(t)}  + 2\pi\iu \theta(t_2 - t) \right], & \lambda(t,p_1^2,p_3^2)>0, \\
		\frac{-4\pi\iu}{\sqrt{-\lambda(t,p_1^2,p_3^2)}} \left[ \arctan\frac{\kappa_-(t)}{Y(t)} + \pi \theta(t_0-t) \right], & \lambda(t,p_1^2,p_3^2)<0,
	\end{cases}
\end{equation}
for both case $A$ and $B$ in Table~\ref{tab:triangle_singularities_Bdecays}. In particular, at the pseudothreshold a singularity that behaves as $\disc C_0(t) \sim {(t_2-t)^{-1/2}}$ has appeared due to the fact that the logarithm moves onto another sheet as soon as $t<t_0$. In order to numerically integrate the dispersion relation properly, one therefore has to treat this new pseudothreshold singularity as described in App.~\ref{sec:numerical_treatment_pseudothreshold}. For the integration path of the anomalous integral in Eq.~\eqref{eq:triangle_function_dispersion_relation} we replace the integration path by 
\begin{equation}
t_x = \frac{t_\text{thr}+t_+}{2} + \frac{t_\text{thr} - t_+}{2} e^{-\iu\pi x}
\end{equation}
in case $A$ to avoid the branch points of the integrand.

For case $C$ the analytic continuation is complicated by the fact that both $t_+$ and $t_-$ move onto the unitarity cut in the limit $\delta \to 0$. This means that we do not need the anomalous integral anymore, but in turn we obtain two additional logarithmic singularities at $t_\pm$ for the integrand. For the analytic continuation we again look at the paths traced by the endpoints $s_\pm(t)$ of the integration contour in Eq.~\eqref{eq:pinocchio_integral_stable}. By starting at a large value of $m_1^2$ and then decreasing it to its physical value we see in Fig.~\ref{fig:pinocchio_case_c} that we again have to deform the integration paths. This leads to the new integrand
\begin{equation}\label{eq:disc_C0_C}
	\disc C_0(t) = \theta(t-t_1) \begin{cases}
		\frac{-2\pi\iu}{\sqrt{\lambda(t,p_1^2,p_3^2)}} \left[ \log\frac{Y(t)+\kappa_+(t)}{Y(t)-\kappa_+(t)} + \pi\iu \big[\theta(t-t_+) + \theta(t-t_-) \big] \right], & t<t_2, \\
		\frac{-4\pi\iu}{\sqrt{-\lambda(t,p_1^2,p_3^2)}} \left[ \arctan\frac{\kappa_-(t)}{Y(t)} + \pi \theta(t_0-t) \right], & t_2<t<t_3, \\
		\frac{-2\pi\iu}{\sqrt{\lambda(t,p_1^2,p_3^2)}} \log\frac{Y(t)+\kappa_+(t)}{Y(t)-\kappa_+(t)}, & t>t_3.
	\end{cases}
\end{equation}
The singularity at $t_2$ is again  treated numerically as described in App.~\ref{sec:numerical_treatment_pseudothreshold}. In contrast, the logarithmic singularities are so mild that they can be dealt with by simply increasing the number of sampling points in their vicinity.

All of these analytic continuations can be cross-checked by comparing the numerical results of the dispersion relation with the triangle function from \texttt{LoopTools}~\cite{Hahn:1998yk} and they coincide up to a very high precision.\footnote{\texttt{LoopTools} has known numerical issues in certain kinematic regions, which can be prevented by replacing $C_0(p_1^2,p_2^2,p_3^2,m_1^2,m_2^2,m_3^2) \to -[C_0(-p_1^2,-p_2^2,-p_3^2,-m_1^2,-m_2^2,-m_3^2)]^*$.} In this section, we concentrated on the cases that will prove relevant for the $B\to(P,V)\,\gamma^*$ form factors, but there are
other mass configurations with even different analytic continuations than the ones presented above. The analytic continuation for all of these general mass configurations can be found in App.~\ref{sec:triangle_position_general_case}.

\subsection{General triangle topologies}\label{sec:general_triangle_topologies}
In a more general case of a triangle diagram whose intermediate state is in a higher partial wave $l$, the unitarity relation can be of the form
\begin{equation}
	\disc A(t) = 2\pi\iu\, \theta(t-(m_2+m_3)^2)\, \frac{a(t)}{\kappa(t)^l} \int_{-1}^{+1} \dd{z} \frac{P_l(z)}{-Y(t)+\kappa(t)z}
\end{equation}
or
\begin{equation}
	\disc \tilde{A}(t) = 2\pi\iu\, \theta(t-(m_2+m_3)^2)\, \frac{a(t)}{\kappa(t)^{l-1}} \int_{-1}^{+1} \dd{z} \frac{P_{l-1}(z)-P_{l+1}(z)}{-Y(t)+\kappa(t)z},
\end{equation}
where $a(t)$ is some analytic prefactor that vanishes like $(t-t_\text{thr})^{(2l+1)/2}$ for $t\to t_\text{thr}$ and $P_l(z)$ denotes the Legendre polynomials.\footnote{As a special case we recover $\disc C_0(t)$ from $\disc A(t)$ for $l=0$ and $a(t) = \lambda^{1/2}(t,m_2^2,m_3^2)$.} Therefore, we will need to evaluate the integral
\begin{equation}\label{eq:higher_partial_waves}
	b_l(t) \equiv \frac{1}{\kappa(t)^l} \int_{-1}^{+1} \dd{z} \frac{P_l(z)}{-Y(t)+\kappa(t)z} = -\frac{2}{\kappa(t)^{l+1}} Q_l\left(\frac{Y(t)}{\kappa(t)}\right),
\end{equation}
for any $l\in\mathbb{N}$, with the Legendre function of the second kind $Q_l(x)$. Here, the integral representation
\begin{equation} \label{eq:legendre_second_kind_integral_representation}
	Q_l(x) = \frac12 \int_{-1}^{+1} \dd{z} \frac{P_l(z)}{x-z}
\end{equation}
was used, which holds as long as $Y(t)/\kappa(t) \notin [-1,+1]$.  The prefactors have been chosen in such a way that $b_l(t)$ behaves like $b_l(t) \sim Y(t)^{-(l+1)} + \order{[\kappa(t)]^2}$ for $\kappa(t) \to 0$ and is thus free of kinematic singularities and zeros. This implies that both discontinuities $\disc A(t)$ and $\disc \tilde{A}(t)$ are free of kinematic singularities as well, but have a kinematic zero behaving like $\sim(t-t_\text{thr})^{(2l+1)/2}$ at threshold $t_\text{thr}$, introducing the usual square-root cusp.

Moreover, since all $Q_l(x)$ can be expressed as
\begin{equation}
	Q_l(x) = \frac12 P_l(x) \log\frac{x+1}{x-1} + W_{l-1}(x),
\end{equation}
where $W_{l-1}(x)$ is a polynomial of degree $l-1$ in $x$, they include the exact same logarithm as the discontinuity of the triangle function in Eq.~\eqref{eq:triangle_function_discontinuity}, so the analytic continuation of the dispersion relation of $A(t)$ will follow exactly the same procedure as with the triangle function $C_0(t)$. In this case, the pseudothreshold singularity occurring when the logarithm goes to another sheet is of the form $\disc A(t) \sim (t_\text{ps}-t)^{-(2l+1)/2}$. Therefore, we again need the procedure described in App.~\ref{sec:numerical_treatment_pseudothreshold} to treat this singularity numerically.


\section{Light-quark loop}
\label{sec:light}

The main objects of interest in this work are the non-local $B\to(P,V)\,\gamma^*$ transition form factors. 
Following the conventions from Ref.~\cite{Gubernari:2020eft}, we decompose the $B\to(P,V)\,\ell^+\ell^-$ amplitude in the following way
\begin{align} \label{eq:BtoPVll_amplitude_decomposition}
	\mathcal{A}\bigg[\bar{B}(q+k) \to \begin{Bmatrix}
		K^{(*)}(k) \\ \pi(\rho)(k)
	\end{Bmatrix} &\ell^+(q_1^\ell)\ell^-(q_2^\ell)\bigg] = \frac{G_F \alpha_\text{em}}{\sqrt{2}\pi} \begin{Bmatrix}
		V_{ts}^*V_{tb} \\ V_{td}^*V_{tb}
	\end{Bmatrix} \\ 
	&\times \left[ (C_{9} L_V^\mu + C_{10} L_A^\mu) \mathcal{F}_\mu - \frac{L_V^\mu}{q^2} (2 \iu m_b C_{7} \mathcal{F}_{T,\mu} + 16\pi^2 \mathcal{H}_\mu) \right], \notag
\end{align}
with $L_V^\mu = \bar{u}_\ell(q_1^\ell) \gamma^\mu v_\ell(q_2^\ell)$ and $L_A^\mu = \bar{u}_\ell(q_1^\ell) \gamma^\mu \gamma_5 v_\ell(q_2^\ell)$. Here, $C_{7}$, $C_{9}$, and $C_{10}$ are Wilson coefficients of the corresponding semileptonic operators $\mathcal{O}_{7}$, $\mathcal{O}_{9}$, and $\mathcal{O}_{10}$ from the Weak Effective Theory~\cite{Buchalla:1995vs,Aebischer:2017gaw}, $V_{ij}$ the standard CKM matrix elements, $G_F$ the Fermi constant, and $\alpha_\text{em}=e^2/(4\pi)$. The local vector and tensor form factors $\mathcal{F}_\mu$ and $\mathcal{F}_{T,\mu}$ are defined as
\begin{align} \label{eq:local_FF}
	\mathcal{F}_\mu^{B \to (P,V)}(k,q) &= \bigg\langle (P,V)(k) \, \bigg| \begin{Bmatrix}
		\bar{s} \\ \bar{d}
	\end{Bmatrix} \gamma_\mu b_L \bigg| \, \bar{B}(q+k) \bigg\rangle, \notag\\
	\mathcal{F}_{T,\mu}^{B \to (P,V)}(k,q) &= \bigg\langle (P,V)(k) \, \bigg| \begin{Bmatrix}
		\bar{s} \\ \bar{d}
	\end{Bmatrix} \sigma_{\mu\nu} q^\nu b_R \bigg| \, \bar{B}(q+k) \bigg\rangle,
\end{align}
and the non-local form factors $\mathcal{H}_\mu$ by
\begin{equation} \label{eq:non_local_FF}
	\mathcal{H}_\mu^{B \to (P,V)}(k,q) = \iu \int \dd[4]{x} \eu^{\iu q \cdot x} \bigg\langle (P,V)(k) \, \bigg| T\bigg\{ j_\mu^\text{em}(x),\sum_{i,q=u,c} C_i^q \mathcal{O}_i^q(0) \bigg\} \bigg| \, \bar{B}(q+k) \bigg\rangle,
\end{equation}
with $j_\mu^\text{em} = \sum_q Q_q \bar{q} \gamma_\mu q$ and the local operators
\begin{equation} \label{eq:operators_O1_O2}
	\mathcal{O}_1^q = \begin{Bmatrix}
		\bar{s} \\ \bar{d}
	\end{Bmatrix} \gamma_\mu T^a q_L \bar{q} \gamma^\mu T^a b_L, \hspace{30pt} \mathcal{O}_2^q = \begin{Bmatrix}
	\bar{s} \\ \bar{d}
	\end{Bmatrix} \gamma_\mu q_L \bar{q} \gamma^\mu b_L.
\end{equation}
The upper/lower entries in Eqs.~\eqref{eq:local_FF} and~\eqref{eq:operators_O1_O2} refer to $b\to s$/$b\to d$ transitions, respectively, and we have not displayed other operators suppressed by small Wilson coefficients or subleading CKM matrix elements. While ultimately the prime phenomenological interest concerns the charm loop, we will focus on the $u$-quark in this work, in which case the hadronization at low energies is dominated by $\pi\pi$ intermediate states, with a clear energy gap to inelastic corrections. 


\subsection{BTT decomposition and helicity amplitudes} \label{sec:BTT_decompostition}

As a first step, we need to decompose the non-local form factors $\mathcal{H}_\mu$ further into invariant Lorentz structures. For this decomposition we follow the BTT procedure~\cite{Bardeen:1968ebo,Tarrach:1975tu}, so that the scalar coefficient functions are automatically free of kinematic singularities and zeros, thus suitable to set up dispersion relations.

For the pseudoscalar meson case we only have one Lorentz structure
\begin{equation}\label{eq:BTT_structure_pseudoscalar}
	T_\mu = q^2 k_\mu - (q \cdot k) q_\mu,
\end{equation}
as there is only one helicity component. However, for the vector meson case there are three helicity components and, correspondingly, three independent Lorentz structures~\cite{Gasser:1974wd,Gasser:2015dwa,Colangelo:2015ama}
\begin{align}\label{eq:BTT_structures_vector}
	T_{1\alpha\mu} &= \epsilon_{\alpha\mu\beta\gamma} k^\beta q^\gamma, \notag\\
	T_{2\alpha\mu} &= (q \cdot k) g_{\alpha\mu} - q_\alpha k_\mu, \notag\\
	T_{3\alpha\mu} &= k^2 q^2 g_{\alpha\mu} + (q \cdot k) k_\alpha q_\mu - k^2 q_\alpha q_\mu - q^2 k_\alpha k_\mu.
\end{align}
With that the decomposition of $\mathcal{H}_\mu$ into these invariant structures looks like
\begin{equation}\label{eq:BTT_decomposition}
	\mathcal{H}_\mu^{B \to P} = M_B^2 T_\mu \Pi_P, \hspace{30pt} \mathcal{H}_\mu^{B \to V} = M_B^2 \eta_{V,\lambda}^{*\alpha} \sum_{i=1}^{3} T_{i\alpha\mu} \Pi_i,
\end{equation}
where $\eta_{V,\lambda}^\alpha$ is the vector meson's polarization vector. Choosing the following linear combinations (compare to the helicity basis found, e.g., in Ref.~\cite{Gubernari:2020eft})
\begin{equation}\label{eq:FF_helicity_basis}
	\Pi_\perp = \Pi_1, \hspace{20pt} \Pi_\parallel = (q \cdot k) \, \Pi_2 + q^2 k^2 \, \Pi_3, \hspace{20pt} \Pi_0 = \Pi_2 + (q \cdot k) \, \Pi_3,
\end{equation}
we arrive at the proper helicity components $\Pi_\lambda$ with $\lambda = \perp,\parallel$ labeling the transversal polarizations and $\lambda=0$ the longitudinal one. This can be seen by contracting the sum $\sum_{i=1}^{3} T_{i\alpha\mu} \Pi_i$ with polarization vectors for the vector meson and the virtual photon and factoring out kinematic zeros, see, e.g., Ref.~\cite{Hoferichter:2020lap}. Most important among these zeros is a factor of $q^2$ in front of the longitudinal form factor $\Pi_0(t)$, which ensures that the longitudinal component vanishes when the photon becomes real.


\begin{figure}[tb]
	\centering
	\includegraphics[width=0.24\textwidth]{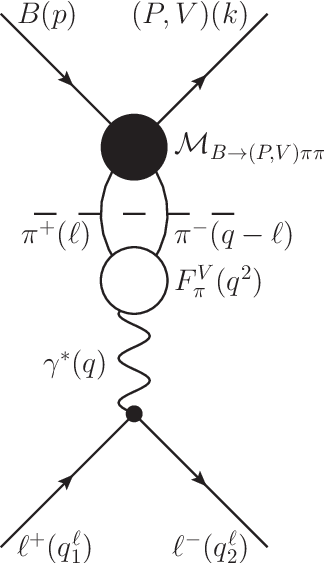}
	\caption{$\pi^+\pi^-$ cut in the $B \to (P,V)\,\ell^+\ell^-$ amplitude. The arrows of $B$ and $(P,V)$ indicate the momentum flow. Both $\ell^\pm$ have their momenta flowing outward. The filled black circle represents the $B \to (P,V)\,\pi^+\pi^-$ amplitude $\mathcal{M}$, the empty circle stands for the pion vector form factor $F_\pi^V(q^2)$.}
	\label{fig:BtoPVll_unitarity}
\end{figure}

\subsection[Dispersion relation for $\pi\pi$ intermediate states]{Dispersion relation for \texorpdfstring{$\boldsymbol{\pi\pi}$}{} intermediate states}
\label{sec:FF_unitarity_pipi}

In the next step, we consider $\pi\pi$ intermediate states and use the BTT decomposition from Eq.~\eqref{eq:BTT_decomposition} to find the corresponding unitarity relations for the scalar form factors $\Pi_{P,\lambda}$ (shorthand for $\Pi_{P}$ or $\Pi_{\lambda}$). Introducing the pion vector form factor $F_\pi^V(q^2)$ via
\begin{equation}
	\big\langle \, \pi^+(q_1) \, \pi^-(q_2) \, \big| \, j_\mu(0) \, \big| \, 0 \, \big\rangle = \iu (q_1-q_2)^\mu F_\pi^V(q^2), \hspace{30pt} q^2 = (q_1+q_2)^2,
\end{equation}
and the $B \to (P,V) \, \pi^+\pi^-$ decay amplitude $\mathcal{M}\big[B(p) \to (P,V)(k) \, \pi^+(q_1) \, \pi^-(q_2)\big]$ we can use Cutkosky rules to obtain the following unitarity relation corresponding to Fig.~\ref{fig:BtoPVll_unitarity} for the $B \to (P,V)\,\ell^+\ell^-$ amplitude from Eq.~\eqref{eq:BtoPVll_amplitude_decomposition},
\begin{align}\label{eq:BtoPVll_unitarity}
	\text{disc}_{\pi\pi}\, \mathcal{A} &= \mathcal{N} e^2 \, \frac{L_V^\mu}{q^2} \, \text{disc}_{\pi\pi}\, \mathcal{H}_\mu \notag\\
	&= -\iu e^2 \, \frac{L_V^\mu}{q^2} \int \frac{\dd[4]{\ell}}{(2\pi)^4} (2\ell-q)_\mu \, \mathcal{M}\big[B(p) \to (P,V)(k) \, \pi^+(\ell) \, \pi^-(q-\ell)\big] [F_\pi^V(q^2)]^* \notag\\ &\times (2\pi\iu)\,\delta(\ell^2-\mpi^2) \, (2\pi\iu)\,\delta\big((q-\ell)^2-\mpi^2\big),
\end{align}
where we introduced the normalization constant
\begin{equation}
	\mathcal{N} = \frac{-4 G_F M_B^2}{\sqrt{2}} \begin{Bmatrix}
		V_{ts}^* V_{tb} \\ V_{td}^* V_{tb}
	\end{Bmatrix},
\end{equation}
and used the fact that only the non-local form factor $\mathcal{H}_\mu$ contains the $\pi\pi$ cut. For the $B \to (P,V) \, \pi^+\pi^-$ amplitude $\mathcal{M}\big[B(p) \to (P,V)(k) \,\pi^+(q_1)\, \pi^-(q_2)\big] = \mathcal{M}_{B \to (P,V)\pi\pi}(s,t,u)$ we further introduce the Mandelstam variables
\begin{equation}
	s = (p-q_1)^2, \hspace{20pt} t = q^2 = (q_1 + q_2)^2, \hspace{20pt} u = (p-q_2)^2,
\end{equation}
which fulfill $s+t+u=M_B^2+M_{(P,V)}^2+2\mpi^2$, as well as the scattering angle
\begin{equation}
	z_t = \cos \theta_t = \frac{s-u}{\sigma_\pi(t) \lambda_{B(P,V)}^{1/2}(t)}.
\end{equation}
Here we also introduced the short-hand notation $\lambda_{B(P,V)}(t) \equiv \lambda(t,M_B^2,M_{(P,V)}^2)$ and $\sigma_\pi(t) \equiv \sigma(t,\mpi^2) = \sqrt{1-4M_\pi^2/t}$. Then Eq.~\eqref{eq:BtoPVll_unitarity} reduces to a unitarity relation for $\mathcal{H}_\mu$,
\begin{equation} \label{eq:non_local_FF_unitarity}
	\mathcal{N} \, \text{disc}_{\pi\pi}\, \mathcal{H}_\mu (t) = \frac{i\big[F_\pi^V(t)\big]^*}{4\pi^2} \int \dd[4]{\ell} (2\ell - q)_\mu \, \mathcal{M}_{B \to (P,V)\pi\pi}(s,t,u) \, \delta(\ell^2-\mpi^2) \, \delta\big((q-\ell)^2-\mpi^2\big).
\end{equation}
To further simplify this into unitarity relations for the individual scalar form factors we need to treat the pseudoscalar and vector meson case separately. Since we always consider this cut, we drop the $\pi\pi$ subscript in the discontinuity in the following.

\subsubsection{Pseudoscalar meson case}\label{sec:F_unitarity_relation_pseudoscalar}

In the case of a pseudoscalar meson the amplitude is given by a scalar function
\begin{equation} \label{eq:BtoPpipi_amplitude}
	\mathcal{M}_{B \to P\pi\pi}(s,t,u) = \mathcal{F}_P(t,z_t),\qquad \overline{|\mathcal{M}|^2} = |\mathcal{F}_P|^2.
\end{equation}
Following Ref.~\cite{Jacob:1959at} we can write down its partial-wave expansion as
\begin{equation} \label{eq:partial_waves_pseudoscalar}
	\mathcal{F}_P(t,z_t) = \sum_{l=0}^\infty (2l+1) \, P_l(z_t) \, f_P^l(t), \hspace{30pt} f_P^l(t) = \frac12 \int_{-1}^{+1} \dd{z_t} P_l(z_t) \, \mathcal{F}_P(t,z_t).
\end{equation}
Plugging this expansion 
into the unitarity relation~\eqref{eq:non_local_FF_unitarity}, we obtain
\begin{align}
	\mathcal{N} \, \disc \Pi_P(t) &= 2\iu \frac{\sigma_\pi(t)}{\lambda_{BP}^{1/2}(t)} \frac{\big[F_\pi^V(t)\big]^*}{4\pi^2} \int \dd[4]{\ell} z_t \, \mathcal{F}_P(t,z_t)  \, \delta(\ell^2-\mpi^2) \, \delta\big((q-\ell)^2-\mpi^2\big) \notag \\
	&= 2\iu \, \theta(t-4\mpi^2) \, \frac{\sigma_\pi^2(t)}{\lambda_{BP}^{1/2}(t)} \, \frac{\big[F_\pi^V(t)\big]^*}{16\pi} \int _{-1}^{+1} \dd{z_t} z_t \, \mathcal{F}_P(t,z_t) \notag\\
	&= 2\iu \, \theta(t-4\mpi^2) \, \frac{\sigma_\pi^2(t)}{8\pi\lambda_{BP}^{1/2}(t)} \, f_P^1(t) \, \big[F_\pi^V(t)\big]^*.
\end{align}
In order to keep this $P$-wave free of kinematic zeros, see Sec.~\ref{sec:MO_representation}, we rescale it according to
\begin{equation} \label{eq:pwave_scalar_scaled}
	g_P(t) = \frac{1}{8\pi\sigma_\pi(t)\lambda_{BP}^{1/2}(t)} f_P^1(t),
\end{equation}
so that the unitarity relation reduces to (cf.~Ref.~\cite{Akdag:2023pwx})
\begin{equation} \label{eq:FF_unitarity_pseudoscalar}
	\mathcal{N} \, \disc \Pi_P(t) = 2\iu \, \theta(t-4\mpi^2) \, \sigma_\pi^3(t) \, g_P(t) \, \big[F_\pi^V(t)\big]^*.
\end{equation}
%

\subsubsection{Vector meson case}\label{sec:F_unitarity_relation_vector}
The vector meson case is slightly more complicated due to the three different helicity components. Thus, the amplitude needs to be decomposed into Lorentz structures,
\begin{equation} \label{eq:BtoVpipi_amplitude}
	\mathcal{M}_{B \to V\pi\pi}(s,t,u) = \epsilon_{\alpha\beta\gamma\delta} \eta_{V,\lambda}^{*\alpha} q_1^\beta q_2^\gamma p^\delta \mathcal{F}_1(t,z_t) + \eta_{V,\lambda}^{*\alpha} \sum_\pm [ k^2 q_{\pm\alpha} - (q_\pm \cdot k) k_\alpha ] \mathcal{F}_\pm(t,z_t),
\end{equation}
where $q_\pm \equiv q_1 \pm q_2$. Again, by contracting with the polarization vectors for each helicity and factoring out kinematic zeros, we see that we have to choose the following linear combinations as our helicity basis,
\begin{equation}
	\mathcal{F}_\perp = \mathcal{F}_1, \hspace{20pt} \mathcal{F}_\parallel = \mathcal{F}_-, \hspace{20pt} \mathcal{F}_0 = \mathcal{F}_+ - \frac{2\sigma_\pi(t)z_t}{\lambda_{BV}^{1/2}(t)} \, (q \cdot k) \, \mathcal{F}_-.
\end{equation}
Squaring the amplitude in Eq.~\eqref{eq:BtoVpipi_amplitude} and summing over the polarizations of $V$,
\begin{equation}\label{eq:polarization_sum}
	\sum_{\lambda} \eta_{V,\lambda}^{*\alpha} \eta_{V,\lambda}^\beta = -g^{\alpha\beta} + \frac{k^\alpha k^\beta}{M_V^2},
\end{equation}
we obtain
\begin{align} \label{eq:BtoVpipi_squared_amplitude}
	\overline{|\mathcal{M}|^2} &= \frac{M_V^2}{4} \lambda_{BV}(t) |\mathcal{F}_0(t,z_t)|^2 + \frac{t-4\mpi^2}{16} \lambda_{BV}(t) (1-z_t^2) |\mathcal{F}_\perp(t,z_t)|^2 \notag\\
	&+ M_V^4 (t - 4\mpi^2) (1-z_t^2) |\mathcal{F}_\parallel(t,z_t)|^2.
\end{align}
Following again Ref.~\cite{Jacob:1959at}, the partial-wave projections read
\begin{align} \label{eq:partial_waves_vector}
	\mathcal{F}_0(t,z_t) &= \sum\limits_{l=0}^\infty (2l+1) \, P_l(z_t) \, f_0^l(t), & f_0^l(t) &= \frac12 \int_{-1}^{+1} \dd{z_t} P_l(z_t) \, \mathcal{F}_0(t,z_t),\\
		\mathcal{F}_\perp(t,z_t) &= \sum\limits_{l=1}^\infty \, P_l^\prime(z_t) \, f_\perp^l(t), & f_\perp^l(t) &= \frac12 \int_{-1}^{+1} \dd{z_t} [ P_{l-1}(z_t) - P_{l+1}(z_t) ] \, \mathcal{F}_\perp(t,z_t),\notag\\
		\mathcal{F}_\parallel(t,z_t) &= \sum\limits_{l=1}^\infty \, P_l^\prime(z_t) \, f_\parallel^l(t), & f_\parallel^l(t) &= \frac12 \int_{-1}^{+1} \dd{z_t} [ P_{l-1}(z_t) - P_{l+1}(z_t) ] \, \mathcal{F}_\parallel(t,z_t),\notag
\end{align}
where $P_l^\prime(z_t)$ are the derivatives of the Legendre polynomials. 

Performing the tensor decomposition and projecting onto the corresponding Lorentz structures
yields the following three unitarity relations for the scalar form factors
\begin{align}
	\mathcal{N}\,\disc \Pi_1(t) &= 2\iu \, \theta(t-4\mpi^2) \, \frac{t\sigma_\pi^3(t)}{96\pi} \, f_\perp^1(t) \, \big[F_\pi^V(t)\big]^*, \\
	\mathcal{N}\,\disc \Pi_2(t) &= -2\iu \, \theta(t-4\mpi^2) \, \frac{M_V^2\sigma_\pi(t)}{16\pi} \left[ \frac{2t\sigma_\pi(t)}{\lambda_{BV}^{1/2}(t)} \, f_0^1(t) + \frac{4(q \cdot k)t\sigma_\pi^2(t)}{3\lambda_{BV}(t)} \, f_\parallel^1(t) \right] \big[F_\pi^V(t)\big]^*, \notag\\
	\mathcal{N}\,\disc \Pi_3(t) &= 2\iu \, \theta(t-4\mpi^2) \, \frac{\sigma_\pi(t)}{16\pi} \left[ \frac{2(q \cdot k)\sigma_\pi(t)}{\lambda_{BV}^{1/2}(t)} \, f_0^1(t) + \frac{4 M_V^2 t\sigma_\pi^2(t)}{3\lambda_{BV}(t)} \, f_\parallel^1(t) \right] \big[F_\pi^V(t)\big]^*. \notag
\end{align}
Similarly to the pseudoscalar case we rescale the $P$-waves as
\begin{equation} \label{eq:pwaves_vector_scaled}
	g_\perp(t) = \frac{1}{96\pi} f_\perp^1(t), \hspace{20pt} 
	g_\parallel(t) = -\frac{k^2}{48\pi} f_\parallel^1(t), \hspace{20pt} g_0(t) = \frac{\lambda_{BV}^{1/2}(t)}{32\pi\sigma_\pi(t)} f_0^1(t),
\end{equation}
to keep them free of kinematic zeros and absorb some prefactors. Switching to the helicity basis as defined in Eq.~\eqref{eq:FF_helicity_basis} disentangles the unitarity relations according to (cf.~Ref.~\cite{Schneider:2012ez} for $\lambda=\perp$)
\begin{align}
\label{eq:FF_unitarity_vector}
	\mathcal{N}\,\disc \Pi_{\perp,\parallel}(t) &= 2\iu \, \theta(t-4\mpi^2) \, t \, \sigma_\pi^3(t) \, g_{\perp,\parallel}(t) \, \big[F_\pi^V(t)\big]^*, \notag\\
	\mathcal{N}\,\disc \Pi_{0}(t) &= 2\iu \, \theta(t-4\mpi^2) \, \sigma_\pi^3(t) \, g_{0}(t) \, \big[F_\pi^V(t)\big]^*.
\end{align}
In general, finding a basis in which the unitarity relations become diagonal is an important step in the solution of the dispersion relations, see, e.g., Refs.~\cite{Hoferichter:2019nlq,Crivellin:2023ter}. Notice that the unitarity relation for the longitudinal form factor is of the same form as the one for the pseudoscalar form factor in Eq.~\eqref{eq:FF_unitarity_pseudoscalar}.


\subsection{Muskhelishvili--Omn\`es representation}
\label{sec:MO_representation}

\begin{figure}[tb]
	\centering
	\includegraphics[width=0.85\textwidth]{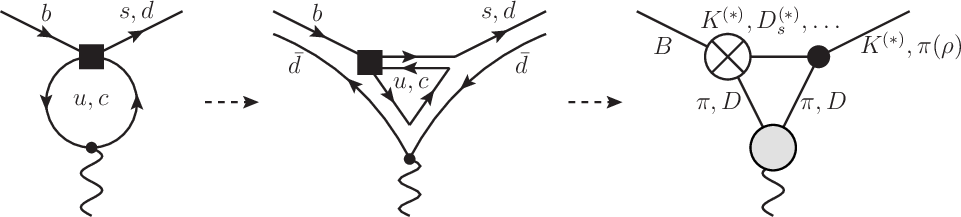}
	\caption{Triangle topology in the non-local $B \to (P,V)\, \gamma^*$ form factors $\mathcal{H}_\mu$ arising from $u$- or $c$-quark loops. Left: $b \to (s,d)\, \gamma^*$ via a $u$- or $c$-quark loop arising from the effective operators $\mathcal{O}_1^q$ and $\mathcal{O}_2^q$ depicted by the dark square. Middle: the same process including the spectator quark $\bar{d}$. Right: hadronized version of the process. The crossed circle stands for the weak effective vertex, the black circle for the strong vertex, and the gray circle is the pion vector form factor.}
	\label{fig:triangle_quark_level}
\end{figure}

In this section, we describe the $B\to(P,V)\,\pi\pi$ $P$-wave amplitudes with left-hand cuts induced by the $s$- and $u$-channel Born exchange of another pseudoscalar or vector meson. In all cases studied below only one of both channels contributes, but for the sake of generality we keep the sum of both in the derivations and just set one of the two couplings to zero in the end. On the quark level these diagrams come from $u$-quark loops, which arise from the effective operators $\mathcal{O}_1^u$ and $\mathcal{O}_2^u$ defined in Eq.~\eqref{eq:operators_O1_O2}. The quark loops hadronize as depicted in Fig.~\ref{fig:triangle_quark_level}, leading to triangle topologies in the non-local $B \to (P,V)\, \gamma^*$ form factors $\mathcal{H}_\mu$. For $c$-quark loops, this would be completely analogous with the $\pi\pi$ intermediate states replaced by $\bar DD$.

In the pseudoscalar case $B\to P\pi\pi$ we only consider a vector meson exchange as a strong $PP^\prime\pi$ vertex is forbidden by parity.\footnote{In contrast, a $PS^\prime\pi$ vertex and, thus, a scalar meson $S^\prime$ exchange would be allowed and may even yield sizable contributions. For example, the $s$-channel $K_0^*(1430)^0$ exchange in $B^+ \to K^+ \pi^+ \pi^-$ constitutes a significant effect~\cite{Belle:2005rpz,BaBar:2008lpx,BaBar:2015pwa}. However, we disregard that possibility for now due to the complicated lineshape of the $K_0^*(1430)^0$~\cite{VonDetten:2021rax} and its large mass.} In the vector case $B\to V\pi\pi$ both pseudoscalar and vector meson exchanges are possible. In particular, we consider $K^*$ exchange for $B\to K\pi\pi$, $K$ and $K^*$ exchange for $B\to K^*\pi\pi$, $\rho$ exchange for $B\to \pi\pi\pi$, and $\pi$ and $\omega$ exchange for $B\to \rho\pi\pi$. 

\subsubsection[Left-hand cuts from $P$ exchange]{Left-hand cuts from \texorpdfstring{$\boldsymbol{P}$}{} exchange} \label{sec:LHC_P_exchange}

Let us begin with the cases in which a pseudoscalar meson $P^\prime$ is exchanged. Due to parity conservation at the strong vertex such a diagram 
 will only contribute to the vector case $B\to V\pi\pi$. The two vertices can be expressed in terms of the amplitudes
\begin{align}
	\mathcal{M}\big[B(p) \to P^\prime(q_1) \pi(q_2)\big] &= g_{BP^\prime\pi}, \notag\\
	\mathcal{M}\big[V(k) \to P^\prime(q_1) \pi(q_2)\big] &= \eta_{V,\lambda}^\alpha (q_1-q_2)_\alpha g_{VP^\prime\pi},
 \label{P_amplitudes}
\end{align}
where another term proportional to $(q_1+q_2)_\alpha = k_\alpha$ would vanish when contracted with the polarization vector $\eta_{V,\lambda}^\alpha$. Combining this into
\begin{equation}
	\mathcal{M}\big[B(p) \to V(k) \pi^+(q_1) \pi^-(q_2)\big] =  g_{VP^\prime\pi}^* \eta_{V,\lambda}^{*\alpha} \left[ (q_+-q_-)_\alpha \frac{g_{BP^\prime\pi^+}}{s-M_{P^\prime}^2} + (q_++q_-)_\alpha \frac{g_{B\bar{P}^\prime\pi^-}}{u-M_{P^\prime}^2} \right], 
\end{equation}
where $q_\pm = q_1 \pm q_2$, yields
\begin{equation} \label{eq:LHC_P_exchange}
	\mathcal{F}_1(s,t,u) = 0,\qquad 
	\mathcal{F}_\pm(s,t,u) = \frac{g_{VP^\prime\pi}^*}{M_V^2} \left[ \pm\frac{g_{BP^\prime\pi^+}}{s-M_{P^\prime}^2} + \frac{g_{B\bar{P}^\prime\pi^-}}{u-M_{P^\prime}^2} \right],
\end{equation}
upon comparison with Eq.~\eqref{eq:BtoVpipi_amplitude}. To calculate the $P$-wave projections of these Born terms we use Eqs.~\eqref{eq:partial_waves_vector}, \eqref{eq:pwaves_vector_scaled}, and~\eqref{eq:higher_partial_waves} to obtain
\begin{align} \label{eq:born_pwaves_P}
	g_0^\text{Born}(t) &= \frac{g_{VP^\prime\pi}^* \Delta g_{BP^\prime\pi}}{16\pi\big[\kappa_V(t)\big]^2} \frac{M_V^2(M_V^2-t-M_B^2) + \Delta_{P^\prime\pi}(M_B^2-t-M_V^2)}{M_V^2}\notag\\
	&\quad\times\left( 2 - \frac{Y_{VP^\prime}(t)}{\kappa_V(t)} L_{VP^\prime}(t) \right), \notag\\
	g_\parallel^\text{Born}(t) &= -\frac{g_{VP^\prime\pi}^* \Delta g_{BP^\prime\pi}}{16\pi \big[\kappa_V(t)\big]^2} \left( Y_{VP^\prime}(t) - \frac{\big[Y_{VP^\prime}(t)\big]^2-\big[\kappa_V(t)\big]^2}{2\kappa_V(t)} L_{VP^\prime}(t) \right),
\end{align}
where $\Delta_{P^\prime\pi} = M_{P^\prime}^2-\mpi^2$, $\Delta g_{BP^\prime\pi} = g_{BP^\prime\pi^+} - g_{B\bar{P}^\prime\pi^-}$, $Y_{VP^\prime}(t) = t - M_B^2 - M_V^2 + 2\Delta_{P^\prime\pi}$, $\kappa_V(t) = \lambda_{BV}^{1/2}(t) \sigma_\pi(t)$, and
\begin{equation}
	L_{VP^\prime}(t) = \log\frac{Y_{VP^\prime}(t)+\kappa_V(t)}{Y_{VP^\prime}(t)-\kappa_V(t)}.
\end{equation}
%

\subsubsection[Left-hand cuts from $V$ exchange]{Left-hand cuts from \texorpdfstring{$\boldsymbol{V}$}{} exchange} 
\label{sec:LHC_V_exchange}

Next, we consider the vector meson $V^\prime$ exchange, which contributes to both the pseudoscalar case $B \to P\pi\pi$ and the vector case $B \to V\pi\pi$. The relevant amplitudes take the form
\begin{align}
	\mathcal{M}\big[B(p) \to V^\prime(q_1) \pi(q_2)\big]&= \eta_{V^\prime,\lambda^\prime}^\alpha (p+q_2)_\alpha g_{BV^\prime\pi}, \notag \\
	\mathcal{M}\big[P(k) \to V^\prime(q_1) \pi(q_2)\big] &= \eta_{V^\prime,\lambda^\prime}^\alpha (k+q_2)_\alpha g_{PV^\prime\pi}, \notag\\
	\mathcal{M}\big[V(k) \to V^\prime(q_1) \pi(q_2)\big] &= \eta_{V,\lambda}^\alpha \eta_{V^\prime,\lambda^\prime}^{*\beta} \epsilon_{\alpha\beta\gamma\delta} q_1^\gamma q_2^\delta g_{VV^\prime\pi},
 \label{V_amplitudes}
\end{align}
where the epsilon tensor is needed to ensure parity conservation at the strong $VV^\prime\pi$ vertex, which has odd intrinsic parity. Using the polarization sum from Eq.~\eqref{eq:polarization_sum} we can combine these into
\begin{align} \label{eq:LHC_V_exchange}
	\mathcal{F}_P(s,t,u) &= \mathcal{M}\big[B(p) \to P(k) \pi^+(q_1) \pi^-(q_2)\big]\notag\\
	&= g_{PV^\prime\pi}^* \frac{2M_{V^\prime}^2t + \Delta_{V^\prime\pi}^2+M_B^2M_P^2-(M_B^2+M_P^2)(\mpi^2+M_{V^\prime}^2)}{M_{V^\prime}^2} \notag\\&\quad\times\left[ \frac{g_{BV^\prime\pi^+}}{s-M_{V^\prime}^2} + \frac{g_{B\bar{V}^\prime\pi^-}}{u-M_{V^\prime}^2} \right], 
\end{align}
where constant non-pole terms have been dropped in the last step as they do not generate any left-hand cut contribution and can be absorbed into subtraction constants.
Similarly, we find
\begin{equation}
	\mathcal{M}\big[B(p) \to V(k) \pi^+(q_1) \pi^-(q_2)\big] 
	= 2g_{VV^\prime\pi}^* \, \eta_{V,\lambda}^{*\alpha} \epsilon_{\alpha\mu\beta\gamma} p^\mu q_1^\beta q_2^\gamma \left[ \frac{g_{BV^\prime\pi^+}}{s-M_{V^\prime}^2} - \frac{g_{B\bar{V}^\prime\pi^-}}{u-M_{V^\prime}^2} \right], 
\end{equation}
which yields
\begin{equation} \label{eq:LHC_V_exchange2}
	\mathcal{F}_1(s,t,u) = 2g_{VV^\prime\pi}^* \left[ \frac{g_{BV^\prime\pi^+}}{s-M_{V^\prime}^2} - \frac{g_{B\bar{V}^\prime\pi^-}}{u-M_{V^\prime}^2} \right],\qquad \mathcal{F}_\pm=0.
\end{equation}
 Using Eqs.~\eqref{eq:partial_waves_pseudoscalar}, \eqref{eq:pwave_scalar_scaled}, \eqref{eq:partial_waves_vector}, \eqref{eq:pwaves_vector_scaled}, and~\eqref{eq:higher_partial_waves} we find the $P$-wave projections
\begin{align} \label{eq:born_pwaves_V}
	g^\text{Born}_P(t) &= \frac{g_{PV^\prime\pi}^* \Delta g_{BV^\prime\pi}}{8\pi\big[\kappa_P(t)\big]^2} \frac{2M_{V^\prime}^2 t + M_P^2 M_B^2 - (M_B^2 + M_P^2)(\mpi^2 + M_{V^\prime}^2) + \Delta_{V^\prime\pi}^2}{M_{V^\prime}^2} \notag\\
	&\quad\times\left( 2 + \frac{Y_{PV^\prime}(t)}{\kappa_P(t)} L_{PV^\prime}(t) \right), \notag\\ 
	g^\text{Born}_\perp(t) &= -\frac{g_{VV^\prime\pi}^* \Delta g_{BV^\prime\pi}}{16\pi \big[\kappa_V(t)\big]^2} \left( Y_{VV^\prime}(t) + \frac{\big[Y_{VV^\prime}(t)\big]^2-\big[\kappa_V(t)\big]^2}{2\kappa_V(t)} L_{VV^\prime}(t) \right), 
\end{align}
where $\Delta_{V^\prime\pi} = M_{V^\prime}^2-\mpi^2$, $\Delta g_{BV^\prime\pi} = g_{BV^\prime\pi^+} - g_{B\bar{V}^\prime\pi^-}$, $Y_{(P,V)P^\prime}(t) = t - M_B^2 - M_{(P,V)}^2 + 2\Delta_{P^\prime\pi}$, $\kappa_{(P,V)}(t) = \lambda_{B(P,V)}^{1/2}(t) \sigma_\pi(t)$, and
\begin{equation}
	L_{(P,V)V^\prime}(t) = \log\frac{Y_{(P,V)V^\prime}(t)+\kappa_{(P,V)}(t)}{Y_{(P,V)V^\prime}(t)-\kappa_{(P,V)}(t)}.
\end{equation}
%

\subsubsection{Unitarization} \label{sec:unitarization_P_waves}

In the two previous sections we derived the $P$-wave $B \to (P,V)\,\pi\pi$ Born amplitudes $g^\text{Born}_{P,\lambda}(t)$ (shorthand for $g^\text{Born}_{P}(t)$ or $g^\text{Born}_{\lambda}(t)$ as in $\Pi_{P,\lambda}$ above) describing the $s$- or $u$-channel exchange of another pseudoscalar or vector meson. From the discussion in Sec.~\ref{sec:general_triangle_topologies} we know that they are free of kinematic singularities or zeros and exhibit the same logarithmic singularities $t_\pm$ as the scalar triangle function. However, these Born amplitudes do not yet fulfill the unitarity relation
\begin{equation} \label{eq:pwave_g_watson}
	\disc g_{P,\lambda}(t) = 2\iu  g_{P,\lambda}(t) \, \eu^{-\iu\delta(t)} \sin \delta(t) \, \theta(t-4\mpi^2),
\end{equation}
where $\delta(t)=\delta_1^1(t)$ is the $\pi\pi$ ($I=1$, $L=1$) scattering phase shift,\footnote{In the numerical analysis, we used the UFD Madrid phase~\cite{Garcia-Martin:2011iqs} as well as more recent determinations~\cite{Colangelo:2018mtw,Colangelo:2022prz}, but for the current purpose the details of the phase shift do not matter.} and thus violate Watson's theorem~\cite{Watson:1954uc}. 
To unitarize the amplitudes we therefore need to add an additional contribution $\tilde{g}_{P,\lambda}(t)$ with the correct right-hand cut that unitarizes the $P$-wave $g_{P,\lambda}(t) = g^\text{Born}_{P,\lambda}(t) + \tilde{g}_{P,\lambda}(t)$. For $\tilde{g}_{P,\lambda}(t)$ this leads to the unitarity relation 
\begin{equation}
	\disc \tilde{g}_{P,\lambda}(t) = 2\iu  \big[\tilde{g}_{P,\lambda}(t) + g^\text{Born}_{P,\lambda}(t)\big] \, \eu^{-\iu\delta(t)} \sin \delta(t) \, \theta(t-4\mpi^2),
\end{equation}
which constitutes an inhomogeneous MO problem~\cite{Muskhelisvili:1953,Omnes:1958hv}. Asymptotically, all Born amplitudes $g^\text{Born}_{P,\lambda}(t)$ from Eqs.~\eqref{eq:born_pwaves_P} and~\eqref{eq:born_pwaves_V} behave as $g^\text{Born}_{P,\lambda}(t) = \mathcal{O}(t^{-1})$. Therefore, the solution is given by the once-subtracted MO representation,
\begin{equation}
\label{MO_once_subtracted}
	\tilde{g}_{P,\lambda}(t) = \Omega(t) \left[ a_{P,\lambda} + \frac{t}{\pi} \int_{4\mpi^2}^{\infty} \frac{\dd{t^\prime}}{t^\prime} \frac{g^\text{Born}_{P,\lambda}(t^\prime) \sin \delta(t^\prime)}{\abs{\Omega(t^\prime)} \, (t^\prime-t)} \right],
\end{equation}
where $a_{P,\lambda}$ is a subtraction constant that remains to be determined and $\Omega(t)$ is the Omn\`es function~\cite{Omnes:1958hv}
\begin{equation} \label{eq:omnes_function}
	\Omega(t) = \exp \left[ \frac{t}{\pi} \int_{4\mpi^2}^{\infty} \dd{t^\prime} \frac{\delta(t^\prime)}{t^\prime(t^\prime-t)} \right].
\end{equation}

\begin{figure}[tb]
	\centering
	\includegraphics[width=0.65\textwidth]{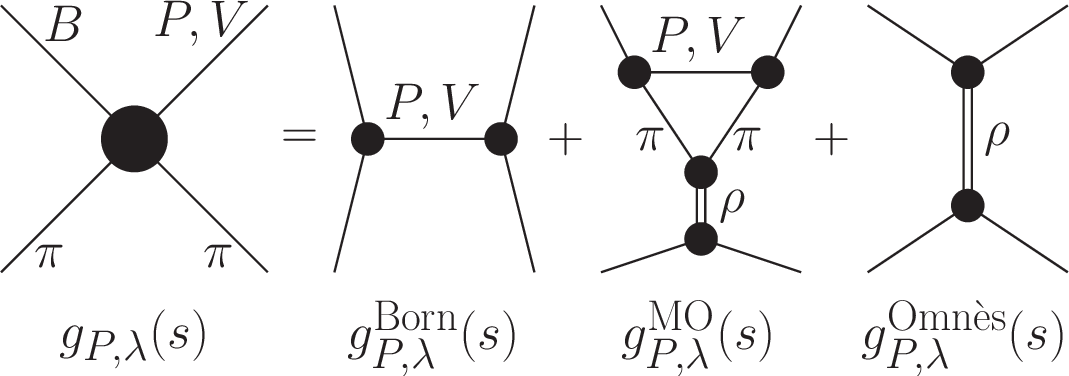}
	\caption{Diagrammatic representation of the unitarized $B\to(P,V)\,\pi\pi$ $P$-wave amplitudes given by the sum $g_{P,\lambda}(t) = g^\text{Born}_{P,\lambda}(t) + g^\text{MO}_{P,\lambda}(t)  + g^\text{Omn\`es}_{P,\lambda}(t)$. The MO contribution $g^\text{MO}_{P,\lambda}(t)$ contains a triangle topology.}
	\label{fig:MO_sum}
\end{figure}

In addition to the standard MO representation~\eqref{MO_once_subtracted}, the appearance of the triangle singularities $t_\pm$ in the discontinuities of $\tilde{g}_{P,\lambda}(t)$,
\begin{equation}
	\disc \frac{\tilde{g}_{P,\lambda}(t)}{\Omega(t)} = 2\iu \frac{g^\text{Born}_{P,\lambda}(t) \sin \delta(t)}{\abs{\Omega(t)}},
\end{equation}
potentially leads to  anomalous contributions, see  
Sec.~\ref{sec:general_triangle_topologies}. Therefore, the unitarized $P$-waves
\begin{equation}\label{eq:pwaves_unitarized}
	g_{P,\lambda}(t) = g^\text{Born}_{P,\lambda}(t) + g^\text{MO}_{P,\lambda}(t) + g^\text{Omn\`es}_{P,\lambda}(t)
\end{equation}
are the sum of the Born amplitudes, the scaled Omn\`es functions $g^\text{Omn\`es}_{P,\lambda}(t) = a_{P,\lambda} \, \Omega(t)$, and the MO contributions
\begin{equation} \label{eq:pwaves_MO_representation}
	g^\text{MO}_{P,\lambda}(t) = \Omega(t) \left[\frac{t}{\pi} \int_{4\mpi^2}^{\infty} \frac{\dd{t^\prime}}{t^\prime} \frac{g^\text{Born}_{P,\lambda}(t^\prime) \sin \delta(t^\prime)}{\abs{\Omega(t^\prime)} \, (t^\prime-t)} + \frac{t}{\pi} \int_{0}^{1} \frac{\dd{x}}{t_x} \pdv{t_x}{x} \frac{\disca g^\text{Born}_{P,\lambda}(t_x) \, \sigma_\pi(t_x) \, t_1^1(t_x)}{\Omega(t_x) \, (t_x-t)} \right],
\end{equation}
including the anomalous discontinuities
\begin{equation}
	\disca \frac{\tilde{g}_{P,\lambda}(t)}{\Omega(t)} = 2\iu \frac{\disca g^\text{Born}_{P,\lambda}(t) \sin \delta(t)}{\abs{\Omega(t)}} = 2\iu \frac{\disca g^\text{Born}_{P,\lambda}(t) \, \sigma_\pi(t) \, t_1^1(t)}{\Omega(t)},
\end{equation}
with
\begin{align} \label{eq:pwaves_anomalous_discontinuity}
	\disca g^\text{Born}_P(t) &= -\iu \frac{g_{PV^\prime\pi}^* \Delta g_{BV^\prime\pi}}{4\big[\kappa_P(t)\big]^2}\\
	&\quad\times\frac{2M_{V^\prime}^2 t + M_P^2 M_B^2 - (M_B^2 + M_P^2)(\mpi^2 + M_{V^\prime}^2) + \Delta_{V^\prime\pi}^2}{M_{V^\prime}^2}  \frac{Y_{PV^\prime}(t)}{\kappa_P(t)}, \notag\\ 
	\disca g^\text{Born}_\perp(t) &= \iu \frac{g_{VV^\prime\pi}^* \Delta g_{BV^\prime\pi}}{8 \big[\kappa_V(t)\big]^2} \frac{\big[Y_{VV^\prime}(t)\big]^2-\big[\kappa_V(t)\big]^2}{2\kappa_V(t)}, \notag\\
	\disca g^\text{Born}_\parallel(t) &= \iu \frac{g_{VP^\prime\pi}^* \Delta g_{BP^\prime\pi}}{8 \big[\kappa_V(t)\big]^2} \frac{\big[Y_{VP^\prime}(t)\big]^2-\big[\kappa_V(t)\big]^2}{2\kappa_V(t)}, \notag\\
	\disca g^\text{Born}_0(t) &= -\iu \frac{g_{VP^\prime\pi}^* \Delta g_{BP^\prime\pi}}{8 \big[\kappa_V(t)\big]^2} \frac{M_V^2(M_V^2-t-M_B^2) + \Delta_{P^\prime\pi}(M_B^2-t-M_V^2)}{M_V^2} \frac{Y_{VP^\prime}(t)}{\kappa_V(t)}. \notag
\end{align}
Diagrammatically, the unitarized $P$-waves can be represented as in Fig.~\ref{fig:MO_sum}. In practice, for the analytic continuation of the anomalous integrand into the complex plane we used the partial wave
\begin{equation}
    t_1^1(t) = \frac{1}{\sigma_\pi(t)} \eu^{\iu\delta(t)} \sin \delta(t)
\end{equation}
to make the replacement~\cite{Hoferichter:2013ama,Dax:2018rvs,Niehus:2021iin}
\begin{equation}
	\frac{\sin \delta(t)}{\abs{\Omega(t)}} = \frac{\sigma_\pi(t) \, t_1^1(t)}{\Omega(t)},
\end{equation}
in which form all constituents  have a well-defined analytic continuation. While in principle such a full analytic continuation  could be constructed using Roy equations~\cite{Roy:1971tc,Ananthanarayan:2000ht,Garcia-Martin:2011iqs,Caprini:2011ky}, for the present application it is sufficient to
 describe the $\pi\pi$ isovector $P$-wave amplitude $t_1^1(t)$ in terms of the unitarized next-to-leading-order ChPT  result, see, e.g., Ref.~\cite{Niehus:2020gmf}.


\section{Phenomenological estimates}
\label{sec:pheno}

For the numerical evaluation of unitarized $B\to(P,V)\,\pi\pi$ $P$-wave amplitudes we need to determine both the coupling constants $g$ and the subtraction constants $a_{P,\lambda}$ from experimental input. In particular, we are interested in $P=K,\pi$ and in $V=K^*,\rho$, and we use masses, decay widths, branching ratios, and polarization fractions from Ref.~\cite{ParticleDataGroup:2022pth}, see App.~\ref{app:coupling}, where we also collect the resulting couplings. To obtain these phenomenological estimates, the included left-hand cuts that give rise to anomalous contributions are as follows:\footnote{Since the remaining cases already prove representative of the possible configurations that can occur, we do not consider the neutral final states $P=\pi^0$ and $V=\rho^0$, as for those both $s$- and $u$-channel Born diagrams contribute, leading to more complicated combinatorics.}
\begin{enumerate}
 \item $P=K,\pi$: left-hand cuts from vector mesons $V'=K^*,\rho$, respectively, leading to case C in Table~\ref{tab:triangle_singularities_Bdecays}, i.e., an anomalous branch point on the unitarity cut.
 \item $V=K^*,\rho$: left-hand cuts from pseudoscalars $P'=K,\pi$ and vectors $V'=K^*,\omega$, respectively. The former/latter correspond to case A/case B in Table~\ref{tab:triangle_singularities_Bdecays}, i.e., an anomalous branch point on the negative real axis/in the lower complex half-plane.  
\end{enumerate}
In general, one thus needs to know the relative signs of the couplings that enter the two left-hand cuts in the vector case, which matter as soon as interferences of different helicity amplitudes are considered. In this work, however, we study the size of anomalous contributions in each helicity amplitude, and since pseudoscalar (vector) left-hand cuts only appear for $\lambda=0,\parallel$ ($\lambda=\perp$), no interference effects arise. This remains true for the determination of subtraction constants, since we only need the decay rate, in which all helicity components are squared separately.

\subsection{Determination of subtraction constants} \label{sec:subtraction_constants}
The unitarized $B\to(P,V)\,\pi\pi$ $P$-waves in Eq.~\eqref{eq:pwaves_unitarized} each contain a subtraction constant that needs to be determined. It yields a term directly proportional to an Omnès function and therefore includes the $t$-channel $\rho$-resonance contribution $B \to \rho(\to\pi\pi)\, (P,V)$ to the process in question. 
Accordingly, the subtraction constant can be inferred
 by demanding that the amplitude reproduce the branching ratio $\Br [B \to \rho\, (P,V)]$ as described in the following. 

The differential decay rate for these processes can be written as
\begin{equation}\label{eq:differential_decay_rate_BPVpipi}
	\dd{\Gamma}[B\to(P,V)\,\pi\pi] = \frac{1}{(2\pi)^3 32 M_B^3} |\mathcal{M}(s,t,u)|^2 \dd{t} \dd{s} = \frac{|\sigma_\pi(t) \lambda_{B(P,V)}^{1/2}(t)|}{(2\pi)^3 64 M_B^3} |\mathcal{M}(t,z_t)|^2 \dd{t} \dd{z_t},
\end{equation}
where we introduced the short-hand notation ${\mathcal{M} \equiv \mathcal{M}[B\to(P,V)\,\pi\pi]}$ and the squared amplitudes are given in Eqs.~\eqref{eq:BtoPpipi_amplitude} and~\eqref{eq:BtoVpipi_squared_amplitude}. If we only consider the $P$-wave part $\mathcal{M}(t,z)_P$ of the amplitude, the $P$-wave differential decay rates are therefore given by
\begin{align} \label{eq:BKpipi_differential_decay_rates}
	&\dd{\Gamma}[B\to P\pi\pi]_P = \frac{3|\sigma_\pi^3(t) \lambda_{BP}^{3/2}(t)|}{4 \pi M_B^3} |g_P(t)|^2 \dd{t} \equiv \phi_P(t)\, |g_P(t)|^2 \dd{t}, \notag \\
	&\dd{\Gamma}[B\to V\pi\pi]_P = \dd{\Gamma}[B\to V\pi\pi]_{0,P} + \dd{\Gamma}[B\to V\pi\pi]_{\perp,P} + \dd{\Gamma}[B\to V\pi\pi]_{\parallel,P}, \notag \\
	&\dd{\Gamma}[B\to V\pi\pi]_{0,P} = \frac{3 M_V^2 |\sigma_\pi^3(t) \lambda_{BV}^{1/2}(t)|}{\pi M_B^3} |g_0(t)|^2 \dd{t} \equiv \phi_0(t)\, |g_0(t)|^2 \dd{t}, \notag\\
	&\dd{\Gamma}[B\to V\pi\pi]_{\perp,P} = \frac{3t|\sigma_\pi^3(t) \lambda_{BV}^{3/2}(t)|}{2 \pi M_B^3} |g_\perp(t)|^2 \dd{t} \equiv \phi_\perp(t)\, |g_\perp(t)|^2 \dd{t}, \notag \\
	&\dd{\Gamma}[B\to V\pi\pi]_{\parallel,P} = \frac{6t |\sigma_\pi^3(t) \lambda_{BV}^{1/2}(t)|}{\pi M_B^3} |g_\parallel(t)|^2 \dd{t} \equiv \phi_\parallel(t)\, |g_\parallel(t)|^2 \dd{t},  
\end{align}
where the partial-wave decompositions from Eqs.~\eqref{eq:partial_waves_pseudoscalar}, \eqref{eq:partial_waves_vector}, \eqref{eq:pwave_scalar_scaled}, and~\eqref{eq:pwaves_vector_scaled} were used.

\begin{table}[tb]
	\centering
	\renewcommand{\arraystretch}{1.3}
	\begin{tabular}{llll}\toprule
		$(P,V)$ & $a_{P,\lambda}$ & Positive solution & Negative solution \\
		\midrule
		$K^+$ & $a_P$ & \SI{7.35e-11}{\giga\electronvolt\tothe{-2}} & (\SI{-7.72e-11}{\giga\electronvolt\tothe{-2}}) \\
		$K^0$ & $a_P$ & \SI{6.76e-11}{\giga\electronvolt\tothe{-2}} & (\SI{-7.12e-11}{\giga\electronvolt\tothe{-2}}) \\
		$\pi^+$ & $a_P$ & \SI{9.70e-11}{\giga\electronvolt\tothe{-2}} & \SI{-10.25e-11}{\giga\electronvolt\tothe{-2}} \\\midrule
		& $a_0$ & \SI{9.74e-10}{\giga\electronvolt\tothe{-1}} & \SI{-10.26e-10}{\giga\electronvolt\tothe{-1}} \\
		$K^{*+}$ & $a_\perp$ & \SI{2.03e-11}{\giga\electronvolt\tothe{-3}} & \SI{-2.34e-11}{\giga\electronvolt\tothe{-3}} \\
		& $a_\parallel$ & \SI{2.50e-10}{\giga\electronvolt\tothe{-1}} & \SI{-3.09e-10}{\giga\electronvolt\tothe{-1}} \\\midrule
		& $a_0$ & \SI{3.84e-10}{\giga\electronvolt\tothe{-1}} & \SI{-4.31e-10}{\giga\electronvolt\tothe{-1}} \\
		$K^{*0}$ & $a_\perp$ & \SI{3.70e-11}{\giga\electronvolt\tothe{-3}} & \SI{-3.98e-11}{\giga\electronvolt\tothe{-3}} \\
		& $a_\parallel$ & \SI{4.83e-10}{\giga\electronvolt\tothe{-1}} & \SI{-5.35e-10}{\giga\electronvolt\tothe{-1}} \\\midrule
		& $a_0$ & \SI{2.91e-9}{\giga\electronvolt\tothe{-1}} & \SI{-2.96e-9}{\giga\electronvolt\tothe{-1}} \\
		$\rho^+$ & $a_\perp$ & \SI{2.19e-11}{\giga\electronvolt\tothe{-3}} & \SI{-2.51e-11}{\giga\electronvolt\tothe{-3}} \\
		& $a_\parallel$ & \SI{2.87e-10}{\giga\electronvolt\tothe{-1}} & \SI{-3.22e-10}{\giga\electronvolt\tothe{-1}} \\\bottomrule
	\end{tabular}
	\renewcommand{\arraystretch}{1.0}
	\caption{Subtraction constants $a_{P,\lambda}$ for some phenomenologically interesting cases determined via Eq.~\eqref{eq:subtraction_constant_quadratic_equation}. For $K^+$ and $K^0$ the negative solution can be eliminated using information on relative phases from Dalitz plot analyses~\cite{Belle:2005rpz,BaBar:2008lpx}.}
	\label{tab:subtraction constants}
\end{table}

Assuming that in the vicinity of the $\rho$-band the $P$-wave part $\mathcal{M}(t,z)_P$ of the amplitude is dominated by the $\rho$-resonance, we can determine the subtraction constants by demanding
\begin{align}\label{eq:subtraction_constants_condition}
	\Gamma^{(\rho)}_P \equiv \Gamma[ B \to \rho P] &= \mathcal{N}_P^{-1} \int_{M_\rho^2-\Delta}^{M_\rho^2+\Delta} \dd{t} \dv{\Gamma[B\to P\pi\pi]_P}{t}, \notag\\
	\Gamma^{(\rho)}_\lambda \equiv f_\lambda\, \Gamma[ B \to \rho V] &= \mathcal{N}_\lambda^{-1} \int_{M_\rho^2-\Delta}^{M_\rho^2+\Delta} \dd{t} \dv{\Gamma[B\to P\pi\pi]_{\lambda,\text{P}}}{t},
\end{align}
where we integrate over the $\rho$-band of width $\Delta = 2M_\rho\Gamma_\rho$ normalized by the factor
\begin{equation}
	\mathcal{N}_{P,\lambda} = \frac{ \int_{M_\rho^2-\Delta}^{M_\rho^2+\Delta} \dd{t} \phi_{P,\lambda}(t) |\Omega(t)|^2 }{ \int_{4\mpi^2}^{(M_B-M_{P,V})^2} \dd{t} \phi_{P,\lambda}(t)  |\Omega(t)|^2},
\end{equation}
to make up for the fact that we integrated only over the $\rho$-band instead of the entire phase space. Moreover, $f_\text{L} \equiv f_0$, $f_\perp$, and $f_\parallel$ are the polarization fractions for $B \to \rho V$. For the cases where only $f_L$ is known experimentally, we split up the $f_\text{T} = 1- f_\text{L}$ component evenly between the two transversely polarized amplitudes, $f_\perp=f_\parallel=f_\text{T}/2$. This is justified by a hierarchy prediction for the helicity components established via QCD factorization in Ref.~\cite{Beneke:2006hg} giving the following scalings for QCD and electromagnetic corrections
\begin{equation}
	\mathcal{A}_0 : \mathcal{A}_- : \mathcal{A}_+ = 1 : \frac{\Lambda_\text{QCD}}{m_b} : \left(\frac{\Lambda_\text{QCD}}{m_b}\right)^2, \qquad \mathcal{A}_0^\gamma : \mathcal{A}_-^\gamma : \mathcal{A}_+^\gamma = 1 : \frac{\alpha_\text{em} m_b}{\Lambda_\text{QCD}} : \alpha_\text{em}.
\end{equation}
According to these scalings the linear combinations $\mathcal{A}_\parallel = (\mathcal{A}_+ + \mathcal{A}_-)/\sqrt{2}$ and $\mathcal{A}_\perp = (\mathcal{A}_+ - \mathcal{A}_-)/\sqrt{2}$, i.e., the two transversal polarizations, should be of roughly the same size. The experimental input quantities are again collected in App.~\ref{app:coupling}.

To determine the subtraction constants, we insert the decomposition of the unitarized partial waves $g_{P,\lambda}(t)$ from Eq.~\eqref{eq:pwaves_unitarized} into Eq.~\eqref{eq:subtraction_constants_condition} to obtain
\begin{align} \label{eq:subtraction_constant_quadratic_equation}
	\mathcal{N}_{P,\lambda}\Gamma^{(\rho)}_{P,\lambda} &= \int_{M_\rho^2-\Delta}^{M_\rho^2+\Delta} \dd{t} \phi_{P,\lambda}(t) \Big[\big|g^\text{Born}_{P,\lambda}(t) + g^\text{MO}_{P,\lambda}(t)\big|^2 + \big|a_{P,\lambda} \Omega(t)\big|^2 \notag\\
	&\qquad+2\Re\Big\{a_{P,\lambda}\big(g^\text{Born}_{P,\lambda}(t) + g^\text{MO}_{P,\lambda}(t)\big)\big[\Omega(t)\big]^*\Big\} \Big]\notag \\
	&\equiv a_{P,\lambda}^2 I_{P,\lambda}^{(2)} + a_{P,\lambda} I_{P,\lambda}^{(1)} + I_{P,\lambda}^{(0)}, 
\end{align}
under the assumption that the $a_{P,\lambda}$ are real. This equation determines $a_{P,\lambda}$ up to a two-fold ambiguity. For the case of $B \to K \pi\pi$ (both charged and neutral) we can choose the positive root as in the Dalitz plot analyses of these decays in Refs.~\cite{Belle:2005rpz,BaBar:2008lpx} the relative phase between the $B \to K^* \pi$ and the $B \to K \rho$ contribution was compatible with zero. In the other cases, we do not have such Dalitz plot data and therefore need to consider both solutions. The resulting subtraction constants $a_{P,\lambda}$ for the cases of interest are listed in Table~\ref{tab:subtraction constants}.

\begin{table}[tb]
	\centering
	\renewcommand{\arraystretch}{1.3}
	\begin{tabular}{llllll}\toprule
	$(P,V)$ & $B\to(P,V)\,\pi\pi$ & References & $B\to(P,V)\,\rho$& Born (full) & Narrow resonance\\\midrule
	$K^0$ & $4.97(18)$ &\cite{CLEO:2002jwu,Belle:2006ljg,BaBar:2009jov,LHCb:2017mlz} & $0.34(11)$ & $0.47$ & $0.50(3)$\\
	$K^{*0}$ & $5.5(5)$ & \cite{BaBar:2007nlh} & $0.39(13)$ & $2.6$ & --\\
	$K^+$ & $5.10(29)$ & \cite{Belle:2005rpz,BaBar:2008lpx} & $0.37(5)$ & $0.66$ &$0.67(3)$\\
	$K^{*+}$ & $7.5(1.0)$ & \cite{BaBar:2006qhm} & $0.46(11)$ & $3.9$ & --\\
	$\pi^+$ & $1.52(14)$ & \cite{BaBar:2009vfr} & $0.83(12)$ & $0.75$ &$0.83(12)$\\
	$\rho^+$ &  -- & -- & $2.40(19)$ & $2.0$ & --\\\bottomrule
	\end{tabular}
	\renewcommand{\arraystretch}{1.0}
	\caption{Branching fractions for $B\to (P,V)\,\pi\pi$, all in units of $10^{-5}$. The second column gives the experimental results for the full $\pi\pi$ channel where available and the fourth column the $P$-wave contributions from the $\rho$ region, see also Table~\ref{tab:BR_fL}, both compared to the branching fractions we obtain from the Born diagrams alone (prior to unitarization) in the last two columns for the full amplitude and a narrow-width approximation, respectively.}
	\label{tab:BR_pipi}
\end{table}

\subsection[Saturation of $B\to(P,V)\,\pi\pi$]{Saturation of \texorpdfstring{$\boldsymbol{B\to(P,V)\,\pi\pi}$}{}}

In the preceding section we determined the subtraction constants $a_{P,\lambda}$ by demanding that our dispersive representation reproduce the measured values of the $B\to(P,V)\,\rho$ branching fractions, see Table~\ref{tab:BR_fL}. In addition, it is instructive to compare to the full $\pi\pi$ final state, as summarized in Table~\ref{tab:BR_pipi}. First, we observe that the part of the spectrum that combines to the ($t$-channel) $\pi^+\pi^-\simeq \rho$ in general only gives a subdominant contribution to the entire branching fraction, suppressed by about an order of magnitude (the exception being the $B^+\to 3\pi$ decay). This indicates that other partial waves and/or decay mechanisms play an important role, and
it is for that reason that data for $B\to(P,V)\,\rho$ directly prove valuable to determine the free parameters in the dispersion relation.  

Table~\ref{tab:BR_pipi} also compares the measured branching fractions to the ones obtained by integrating the Born-term amplitudes over the entire phase space, according to
\begin{equation}
	\Gamma[B \to (P,V) \, \pi \pi] = \frac{1}{(2\pi)^3 32 M_B^3} \int_{t_\text{min}}^{t_\text{max}} \dd{t} \int_{s_\text{min}}^{s_\text{max}} \dd{s} \overline{|\mathcal{M}|^2},
\end{equation}
with integration boundaries
\begin{align}
	t_\text{min} &= 4 M_\pi^2,\hspace{10pt} t_\text{max} = (M_B - M_{P,V})^2,\notag \\
	s_\text{min/max} &= \frac12 \left[M_B^2 + 2 M_\pi^2 + M_{P,V}^2 - t \pm \sigma_\pi(t) \lambda_{B(P,V)}^{1/2}(t) \right],
\end{align}
and the squared amplitudes from Eqs.~\eqref{eq:BtoPpipi_amplitude}, \eqref{eq:BtoVpipi_squared_amplitude}, \eqref{eq:LHC_P_exchange}, \eqref{eq:LHC_V_exchange}, and~\eqref{eq:LHC_V_exchange2}. To avoid singularities in the integration region, we introduce a width to the $V^\prime = K^*,\rho,\omega$, propagators (the $P^\prime=K,\pi$ poles lie outside the integration region)
\begin{equation}
	\frac{1}{s - M_{V^\prime}^2} \longrightarrow \frac{1}{s - M_{V^\prime}^2 + \iu M_{V^\prime} \Gamma_{V^\prime}}.
\end{equation}
The resulting pseudoscalar branching fractions are similar in size to $\Br[B\to P\rho]$, but for the vector final states much bigger results are obtained, saturating about half of $\Br[B\to V\pi\pi]$. The last two columns in Table~\ref{tab:BR_pipi} also demonstrate how the Born-term results can be made plausible from a narrow-width approximation
\begin{align}
  \Br[B^0\to K^0\pi^+\pi^-]\big|_\text{Born}&\simeq \Br[B^0\to K^{*+}\pi^-]\Br[K^{*+}\to K^0\pi^+]\simeq \frac{2}{3} \Br[B^0\to K^{*+}\pi^-],\notag\\ 
  \Br[B^+\to K^+\pi^+\pi^-]\big|_\text{Born}&\simeq \Br[B^+\to K^{*0}\pi^+]\Br[K^{*0}\to K^+\pi^-]\simeq \frac{2}{3} \Br[B^+\to K^{*0}\pi^+],\notag\\
  \Br[B^+\to \pi^+\pi^+\pi^-]\big|_\text{Born}&\simeq \Br[B^+\to \rho^{0}\pi^+]\Br[\rho^{0}\to \pi^+\pi^-]\simeq \Br[B^+\to \rho^0\pi^+],
\end{align}
at least for those cases in which the on-shell decay can happen within the phase space. 

These considerations illustrate that the amplitude for $B\to(P,V)\,\pi\pi$ is a complicated object, and some care is required to extract the features relevant for the dispersive analysis of $B\to(P,V)\,\gamma^*$ form factors.  For the $\pi\pi$ intermediate states, this is most easily achieved by relying on the measured $B\to(P,V)\,\rho$ branching fractions.

\begin{figure}[tb]
	\centering
	\includegraphics[width=0.6\textwidth]{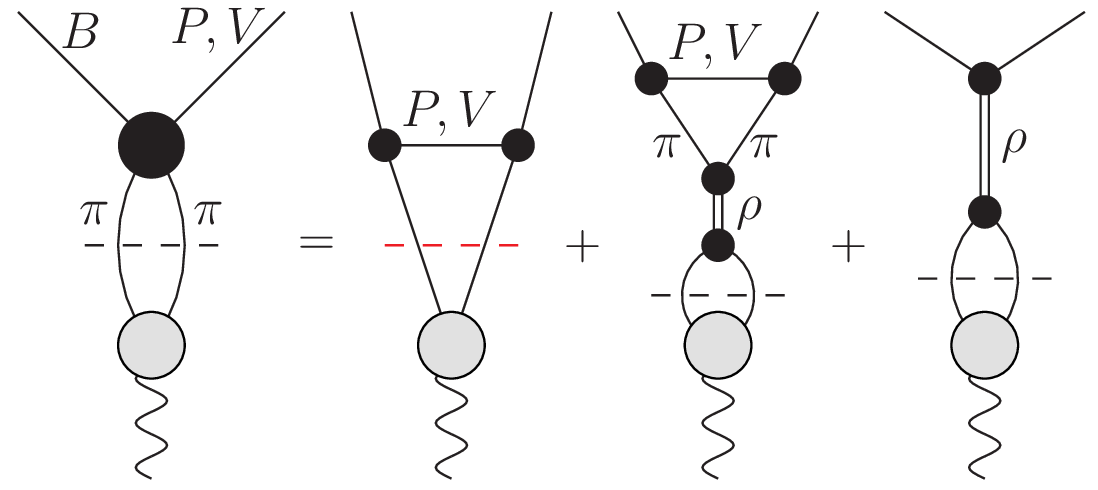}
	\caption{Diagrammatic representation of the $B\to(P,V)\,\gamma^*$ form factor unitarity relations with $\pi\pi$ intermediate states and the unitarized $B\to(P,V)\,\pi\pi$ amplitudes containing the three terms $g_{P,\lambda}(t) = g^\text{Born}_{P,\lambda}(t) + g^\text{MO}_{P,\lambda}(t)  + g^\text{Omnès}_{P,\lambda}(t)$. Only the Born term (marked in red) leads to an anomalous contribution.}
	\label{fig:FF_unitarity_sum}
\end{figure}

\subsection{Form factor dispersion relations} \label{sec:FF_dispersion_relations}

Having established the unitarized $B\to(P,V)\,\pi\pi$ $P$-waves and determined all parameters from experimental data as far as possible, we turn to setting up dispersion relations for the $B\to(P,V)\,\gamma^*$ form factors. 
In the helicity basis we can rewrite
the unitarity relations~\eqref{eq:FF_unitarity_pseudoscalar} and~\eqref{eq:FF_unitarity_vector} in the common form
\begin{equation}
	\mathcal{N} \, \disc \Pi_{P,\lambda}(t) = 2\iu \, \theta(t-4\mpi^2) \, \nu_{P,\lambda}(t)  g_{P,\lambda}(t) \, \big[\Omega(t)\big]^*,
\end{equation}
where we introduced $\nu_P(t)=\nu_0(t) \equiv \sigma_\pi^3(t)$, $\nu_\perp(t)=\nu_\parallel(t)\equiv t\sigma_\pi^3(t)$, and identified $F_\pi^V(t)$ with the Omn\`es factor for simplicity. 
Plugging in the unitarized $P$-waves from Eq.~\eqref{eq:pwaves_unitarized} into these form factor unitarity relations, we can see that $\Pi_P(t),\Pi_0(t)=\mathcal{O}(t^{-2})$ and $\Pi_1(t),\Pi_\parallel(t)=\mathcal{O}(t^{-1})$, so that no further subtractions are needed in the final dispersion relation
\begin{align} \label{eq:FF_dispersion_relation}
	\mathcal{N}\, \Pi_{P,\lambda}(t) &= \frac{1}{\pi}\int_{4\mpi^2}^{\infty} \dd{x} \frac{g_{P,\lambda}(x) \, \nu_{P,\lambda}(x) \, \big[\Omega(x)\big]^*}{(x-t)} \notag\\
	&\quad+ \frac{1}{\pi} \int_{0}^{1} \dd{x} \pdv{t_x}{x} \frac{\disca g_{P,\lambda}^\text{Born}(t_x) \, \nu_{P,\lambda}(t_x) \, \Omega(t_x)}{(t_x-t)} \notag\\
	&\equiv \mathcal{N}\, \Pi_{P,\lambda}^\text{norm}(t)  + \mathcal{N}\, \Pi_{P,\lambda}^\text{anom}(t). 
\end{align}
In the normal part $\Pi^\text{norm}(t)$ the pseudothreshold singularity may appear, to be treated as described in App.~\ref{sec:numerical_treatment_pseudothreshold}. Due to the form of the $P$-waves we again have three diagrams contributing to the form factors, see Fig.~\ref{fig:FF_unitarity_sum}. For the anomalous integrand we made use of the fact that only the first of these three terms (arising from $g_{P,\lambda}^\text{Born}(t)$) yields a triangle topology and, therefore, the anomalous discontinuity amounts to\footnote{Notice that compared to the normal discontinuity the complex conjugation of the Omnès function $\Omega(t)$ is not present anymore in the anomalous discontinuity. This is due to the fact that one needs to analytically continue the integrand into the lower complex half-plane and in the process of doing so needs to make use of $[\Omega(t+\iu\epsilon)]^* = \Omega(t-\iu\epsilon)$ along the unitarity cut.}
\begin{equation}
	\mathcal{N}\, \disca \Pi_{P,\lambda}(t) = 2\iu \, \nu_{P,\lambda}(t) \, \disca g_{P,\lambda}^\text{Born}(t) \, \Omega(t),
\end{equation}
where the expressions for $\disca g_{P,\lambda}^\text{Born}(t)$ are given in Eq.~\eqref{eq:pwaves_anomalous_discontinuity}.

\begin{figure}[t]
	\centering
	\begin{subfigure}{0.31\textwidth}
		\centering
		\includegraphics[width=\textwidth]{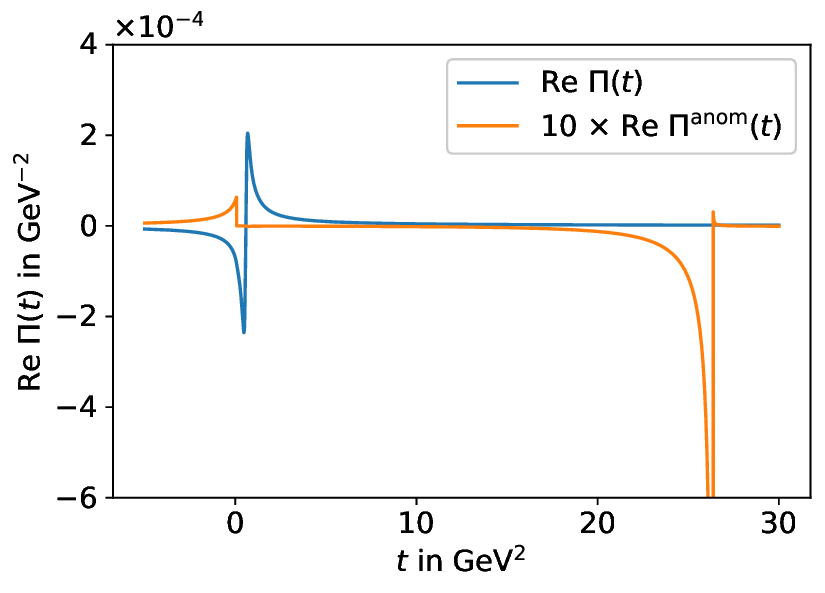}
	\end{subfigure}
	\begin{subfigure}{0.31\textwidth}
		\centering
		\includegraphics[width=\textwidth]{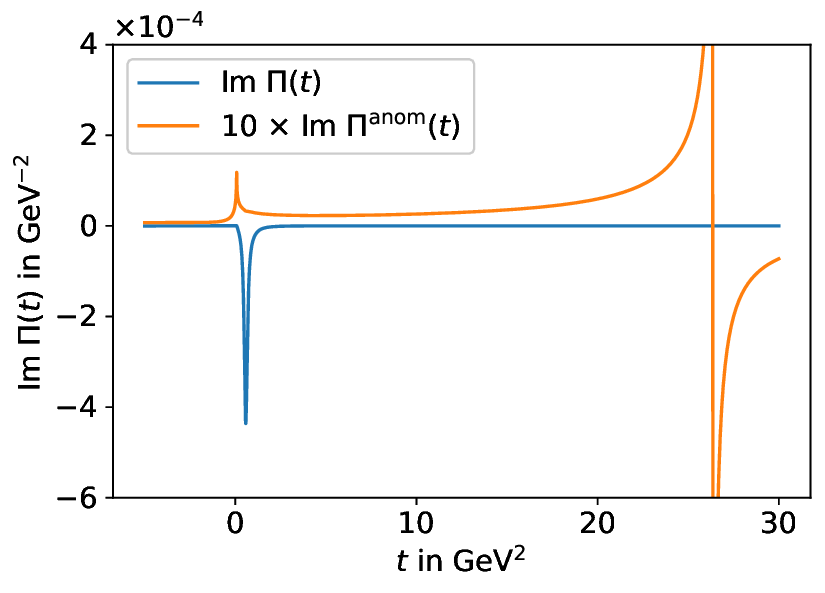}
	\end{subfigure}
	\begin{subfigure}{0.31\textwidth}
		\centering
		\includegraphics[width=\textwidth]{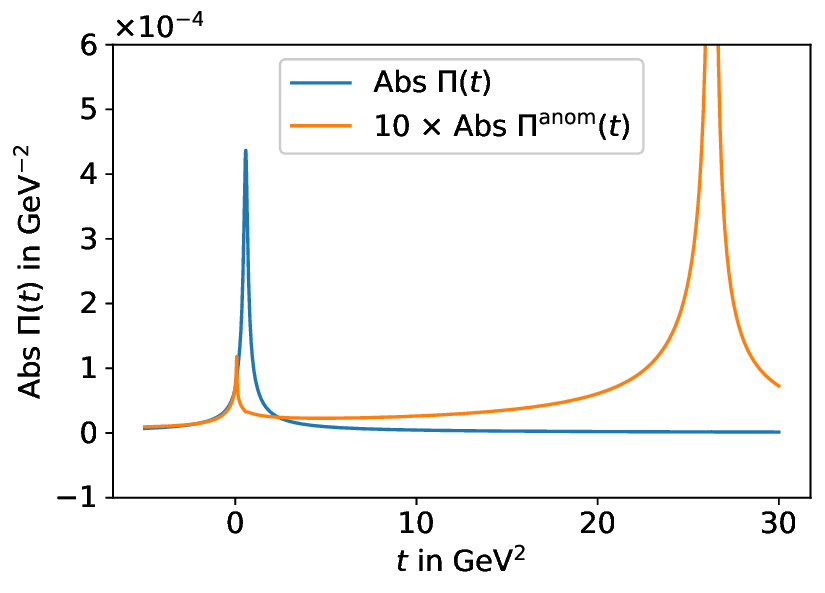}
	\end{subfigure}
	\caption{$B^+ \to \pi^+ \ell \ell$ form factors. Left: real part of the total form factor $\Pi(t)$ and its anomalous part $\Pi^\text{anom}(t)$, middle: imaginary part, right: absolute values.}
	\label{fig:Btopill_charged_FF}
\end{figure}

\begin{figure}[t]
	\centering
	\begin{subfigure}{0.31\textwidth}
		\centering
		\includegraphics[width=\textwidth]{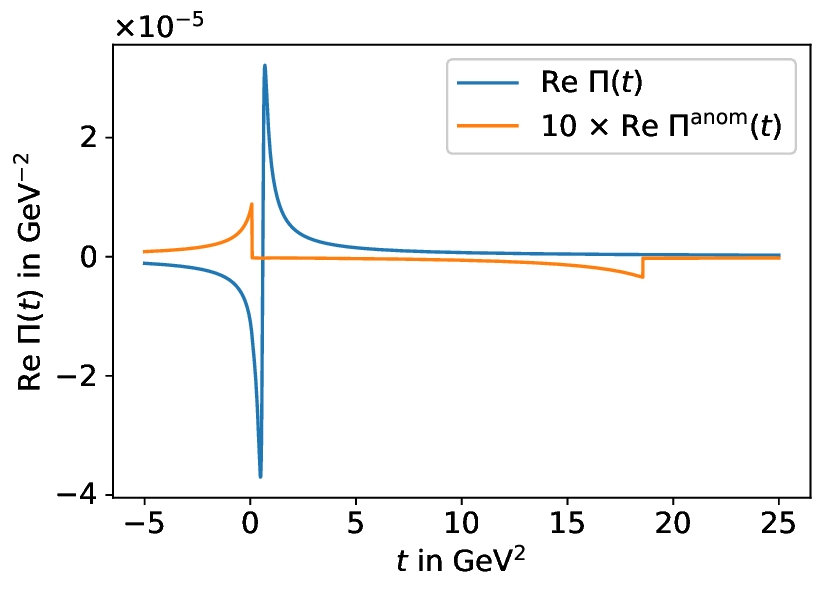}
	\end{subfigure}
	\begin{subfigure}{0.31\textwidth}
		\centering
		\includegraphics[width=\textwidth]{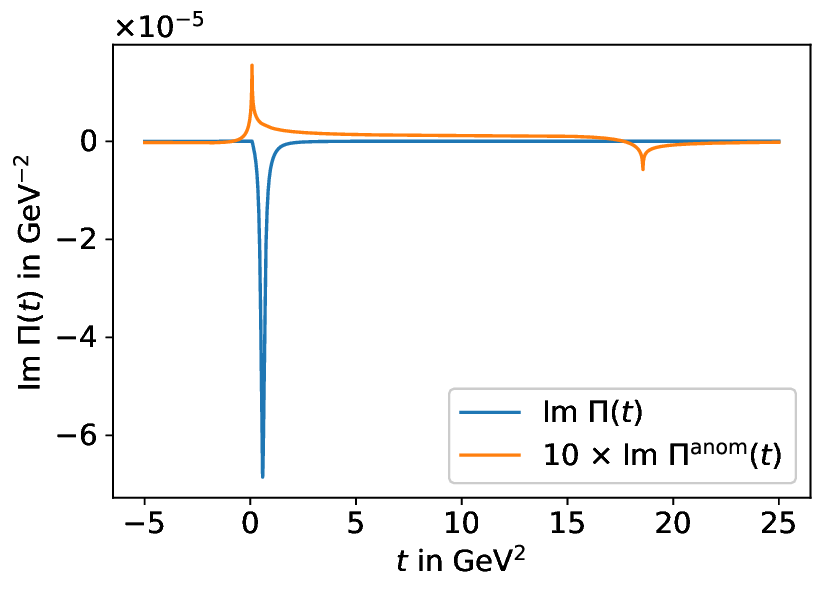}
	\end{subfigure}
	\begin{subfigure}{0.31\textwidth}
		\centering
		\includegraphics[width=\textwidth]{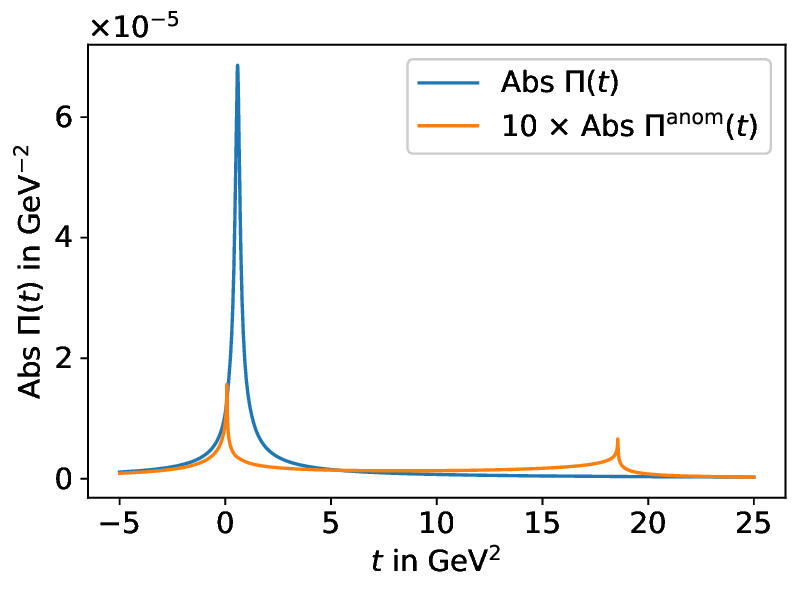}
	\end{subfigure}
	\caption{$B^+ \to K^+ \ell \ell$ form factors. Left: real part of the total form factor $\Pi(t)$ and its anomalous part $\Pi^\text{anom}(t)$, middle: imaginary part, right: absolute values.}
	\label{fig:BtoKll_charged_FF}
\end{figure}

\begin{figure}[t!]
	\centering
	\begin{subfigure}{0.31\textwidth}
		\centering
		\includegraphics[width=\textwidth]{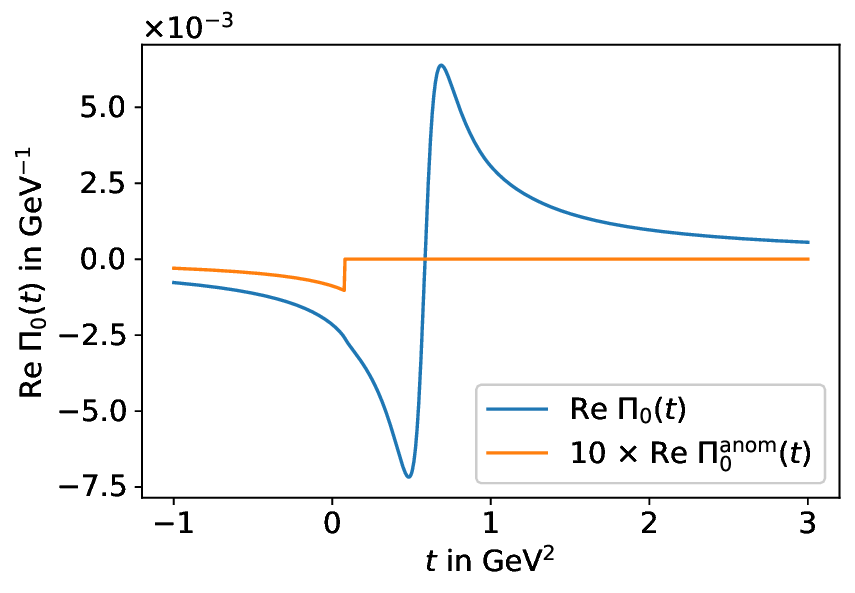}
	\end{subfigure}
	\begin{subfigure}{0.31\textwidth}
		\centering
		\includegraphics[width=\textwidth]{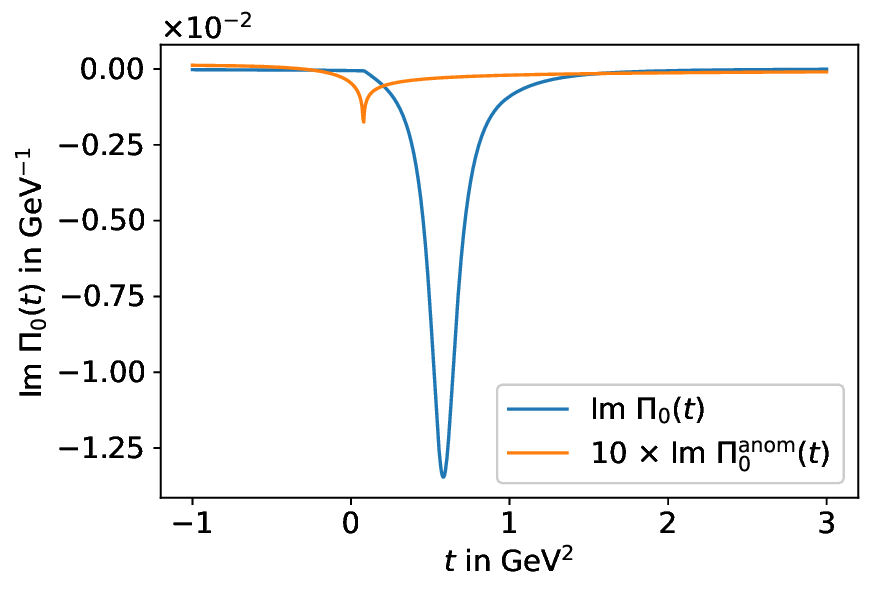}
	\end{subfigure}
	\begin{subfigure}{0.31\textwidth}
		\centering
		\includegraphics[width=\textwidth]{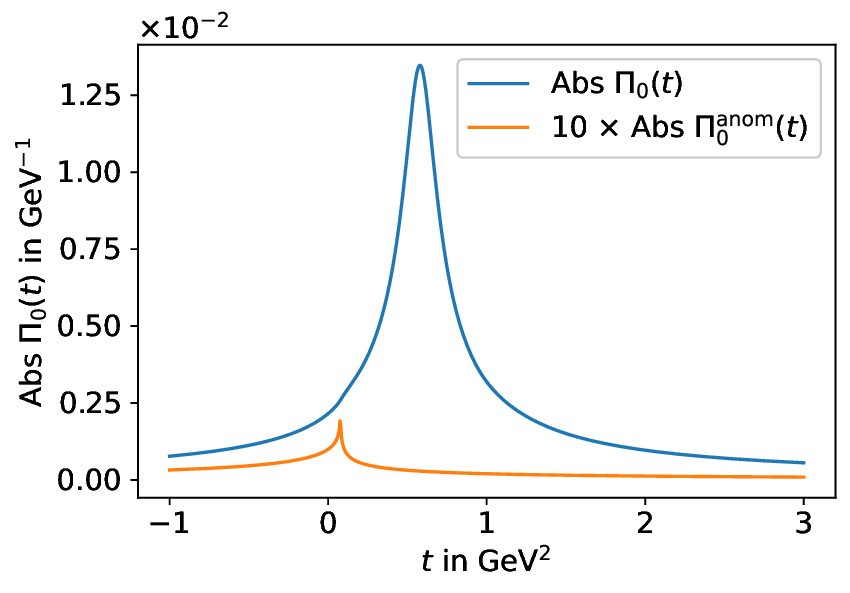}
	\end{subfigure}
	\begin{subfigure}{0.31\textwidth}
		\centering
		\includegraphics[width=\textwidth]{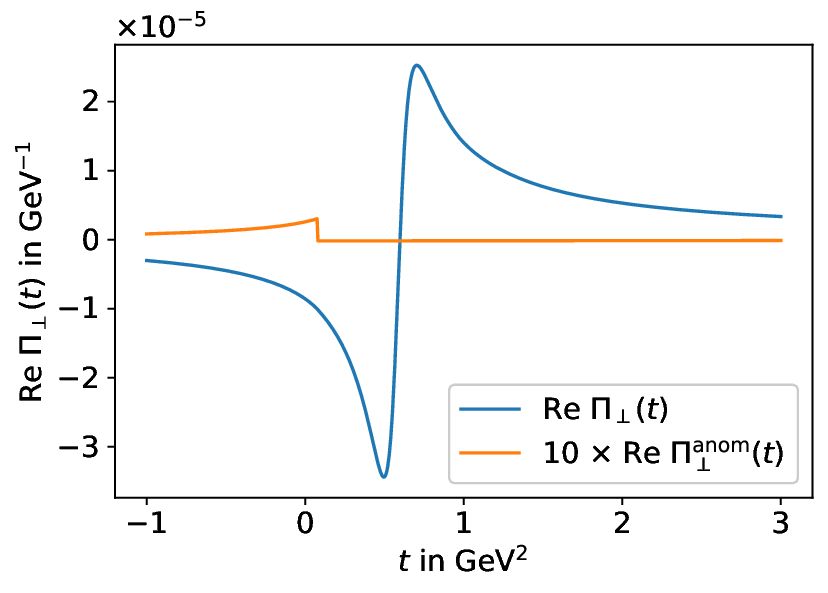}
	\end{subfigure}
	\begin{subfigure}{0.31\textwidth}
		\centering
		\includegraphics[width=\textwidth]{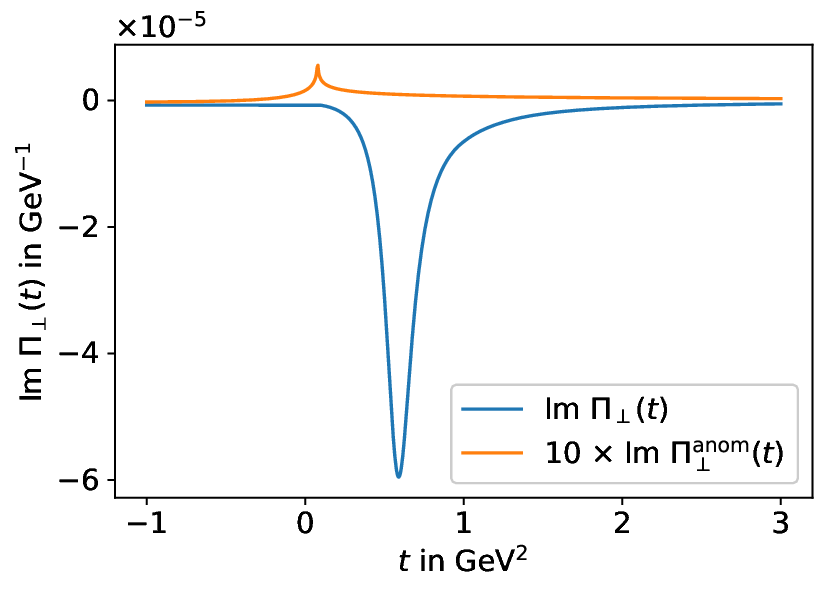}
	\end{subfigure}
	\begin{subfigure}{0.31\textwidth}
		\centering
		\includegraphics[width=\textwidth]{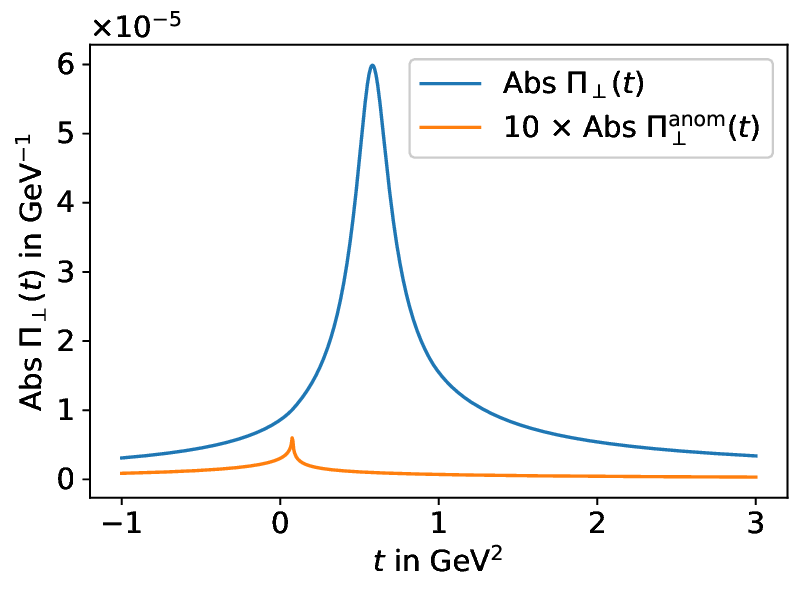}
	\end{subfigure}
	\begin{subfigure}{0.31\textwidth}
		\centering
		\includegraphics[width=\textwidth]{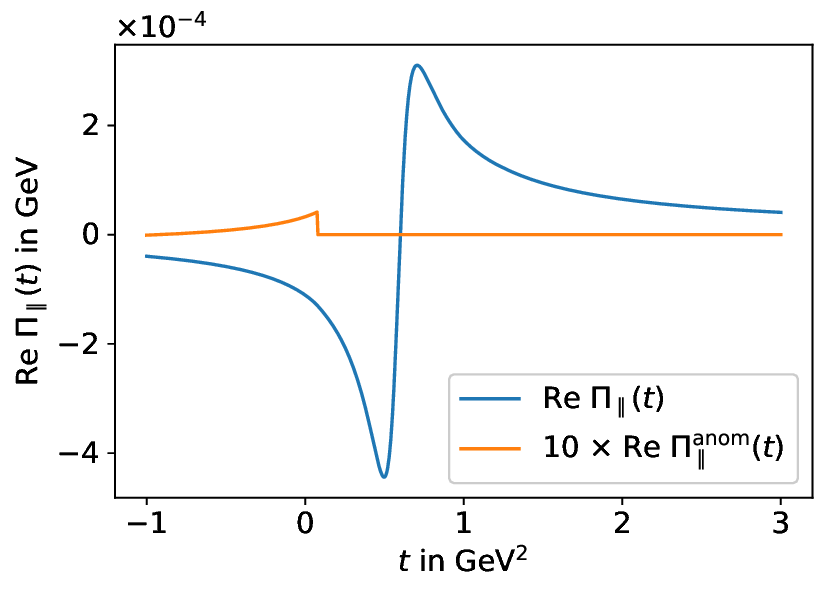}
	\end{subfigure}
	\begin{subfigure}{0.31\textwidth}
		\centering
		\includegraphics[width=\textwidth]{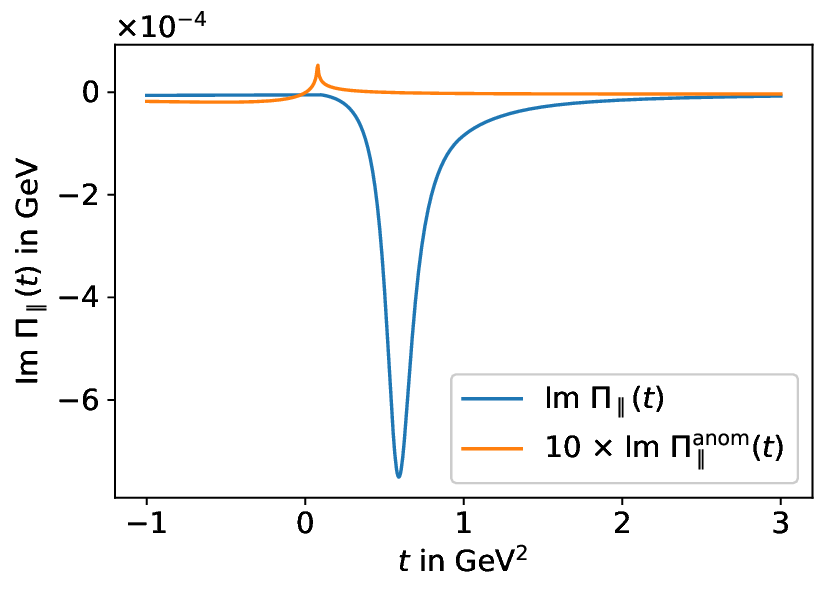}
	\end{subfigure}
	\begin{subfigure}{0.31\textwidth}
		\centering
		\includegraphics[width=\textwidth]{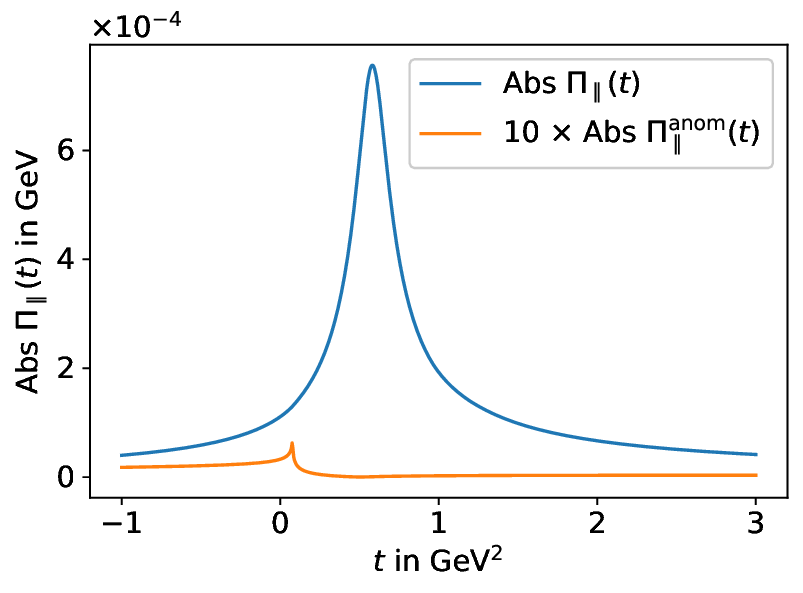}
	\end{subfigure}
	\caption{$B^+ \to \rho^+ \ell \ell$ form factors. First row: $\Pi_0(t)$, second row: $\Pi_\perp(t)$, third row: $\Pi_\parallel(t)$. First column: real part of the total form factor $\Pi_i(t)$ and its anomalous part $\Pi_i^\text{anom}(t)$, second column: imaginary part, third column: absolute value.}
	\label{fig:BtoRholl_charged_FF}
\end{figure}

\subsection{Anomalous contributions to form factors}\label{sec:FF_anom_phenomenological_estimates}

\begin{figure}[t!]
	\centering
	\begin{subfigure}{0.31\textwidth}
		\centering
		\includegraphics[width=\textwidth]{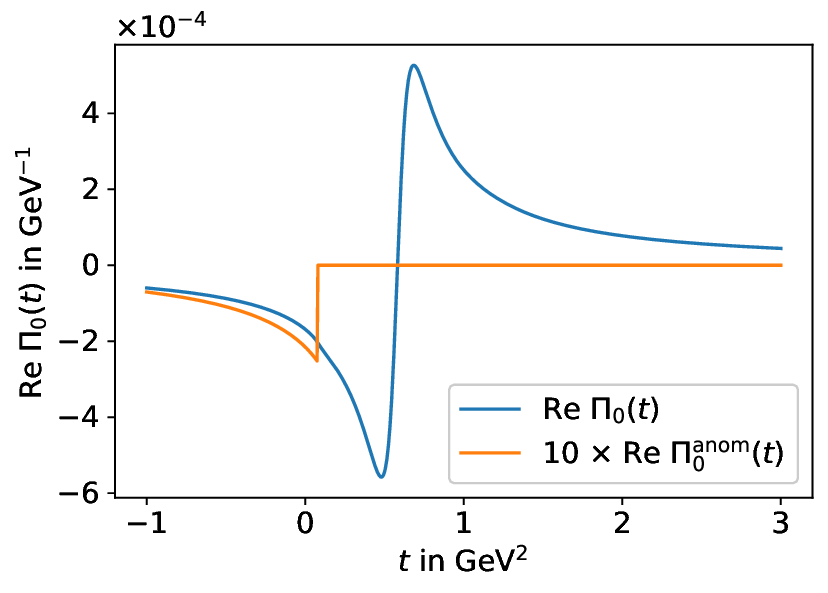}
	\end{subfigure}
	\begin{subfigure}{0.31\textwidth}
		\centering
		\includegraphics[width=\textwidth]{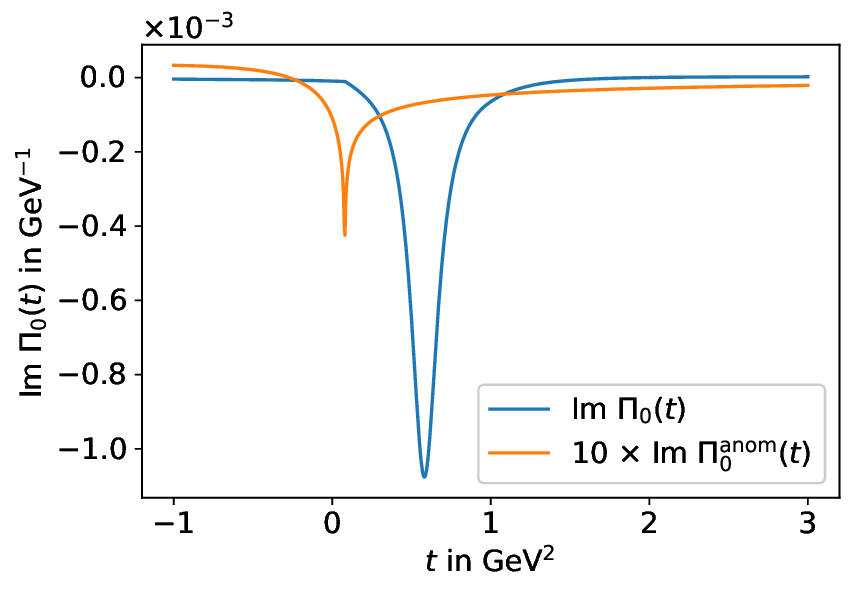}
	\end{subfigure}
	\begin{subfigure}{0.31\textwidth}
		\centering
		\includegraphics[width=\textwidth]{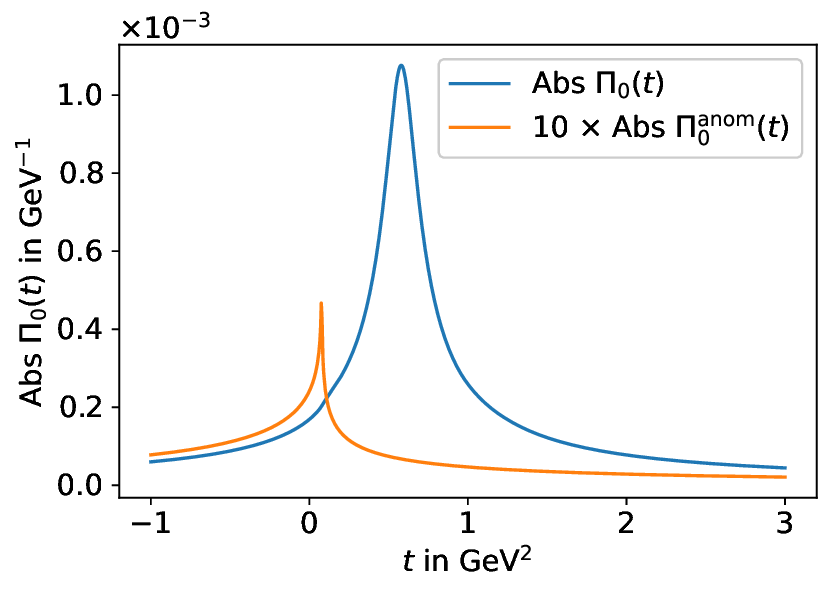}
	\end{subfigure}
	\begin{subfigure}{0.31\textwidth}
		\centering
		\includegraphics[width=\textwidth]{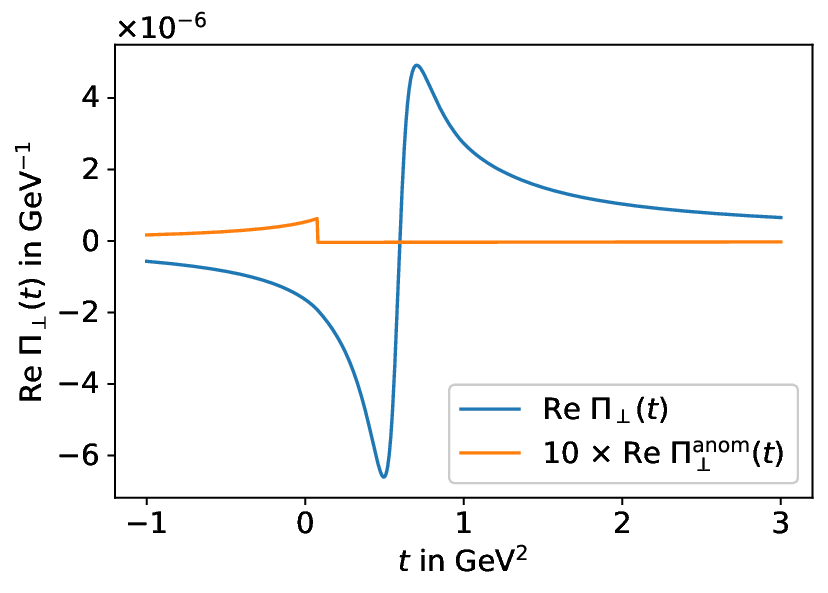}
	\end{subfigure}
	\begin{subfigure}{0.31\textwidth}
		\centering
		\includegraphics[width=\textwidth]{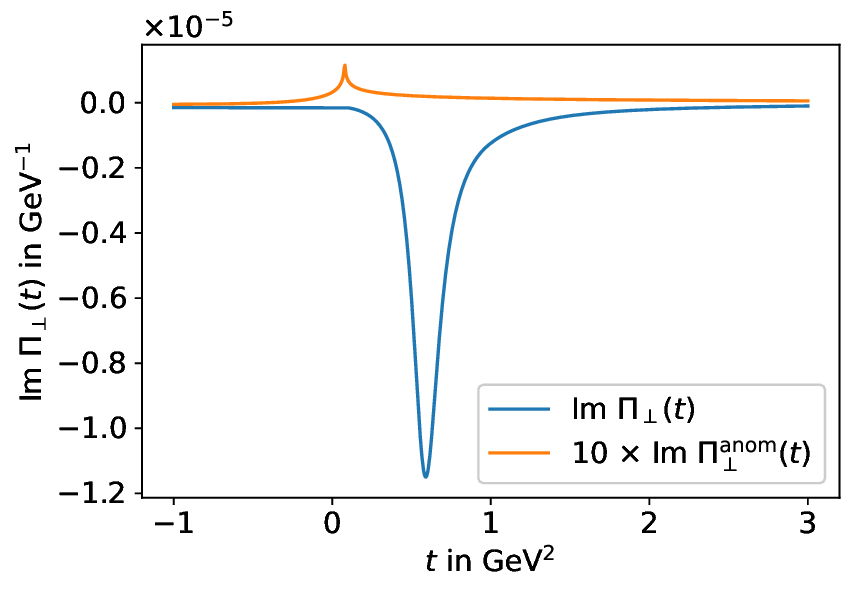}
	\end{subfigure}
	\begin{subfigure}{0.31\textwidth}
		\centering
		\includegraphics[width=\textwidth]{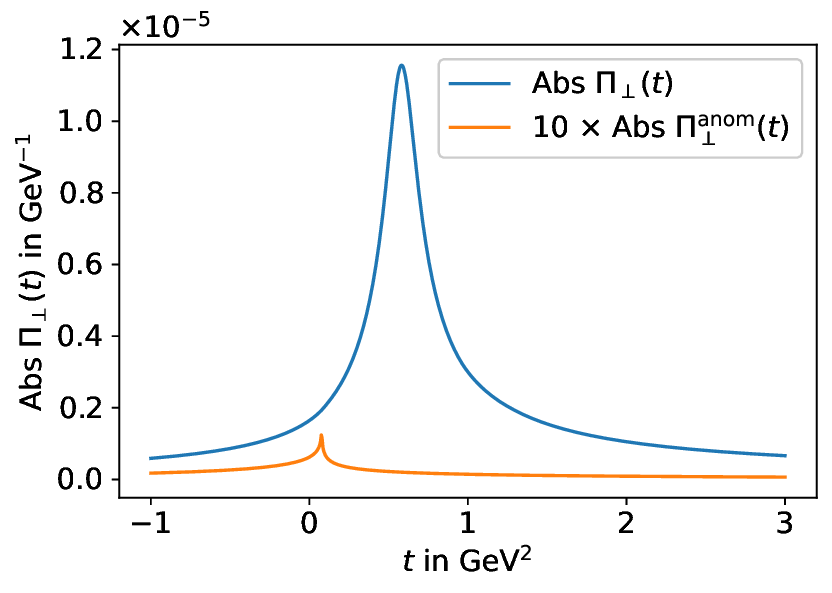}
	\end{subfigure}
	\begin{subfigure}{0.31\textwidth}
		\centering
		\includegraphics[width=\textwidth]{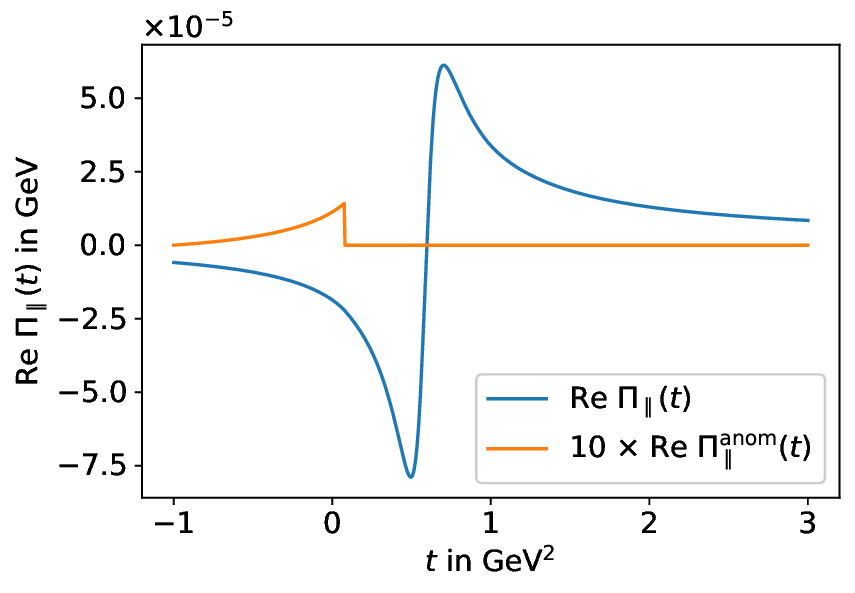}
	\end{subfigure}
	\begin{subfigure}{0.31\textwidth}
		\centering
		\includegraphics[width=\textwidth]{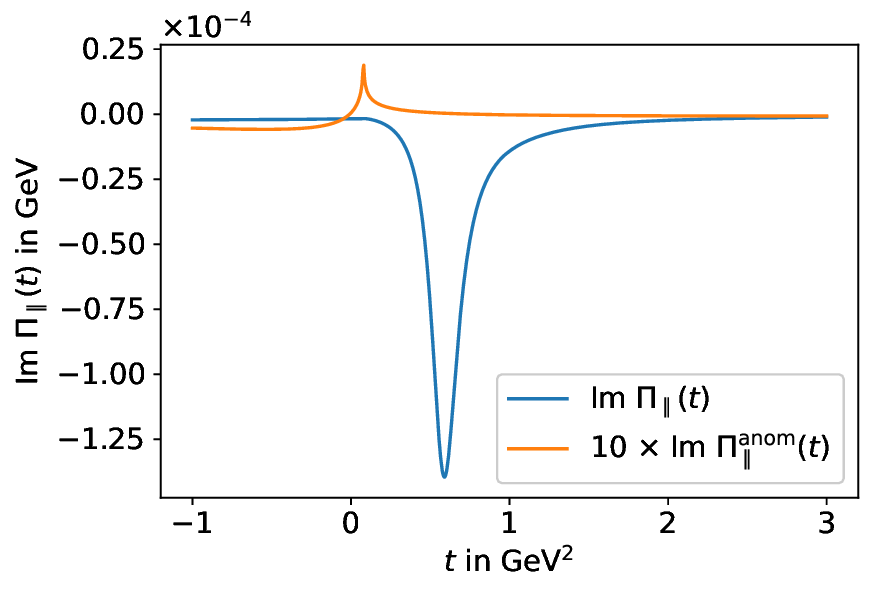}
	\end{subfigure}
	\begin{subfigure}{0.31\textwidth}
		\centering
		\includegraphics[width=\textwidth]{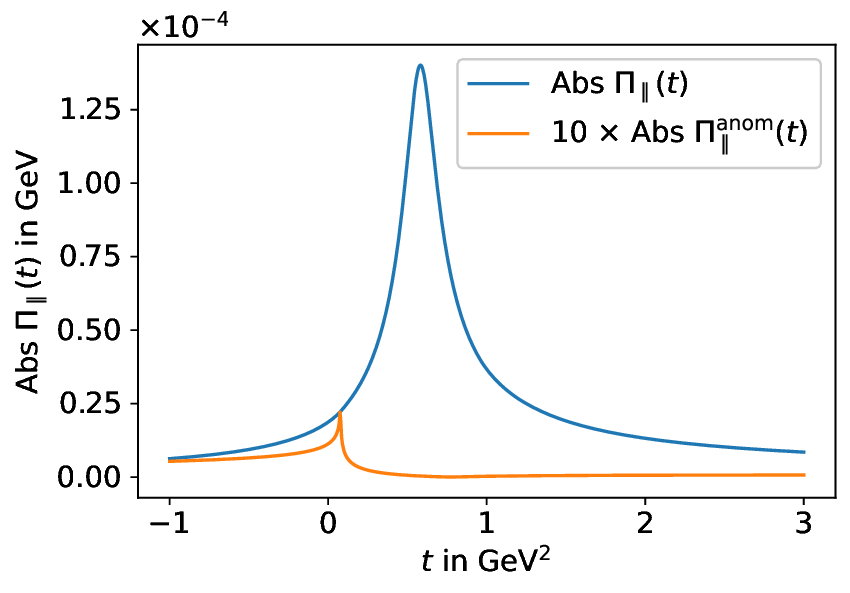}
	\end{subfigure}
	\caption{$B^+ \to K^{*+} \ell \ell$ form factors. First row: $\Pi_0(t)$, second row: $\Pi_\perp(t)$, third row: $\Pi_\parallel(t)$. First column: real part of the total form factor $\Pi_i(t)$ and its anomalous part $\Pi_i^\text{anom}(t)$, second column: imaginary part, third column: absolute value.}
	\label{fig:BtoKstarll_charged_FF}
\end{figure}

\begin{figure}[t!]
	\centering
	\begin{subfigure}{0.31\textwidth}
		\centering
		\includegraphics[width=\textwidth]{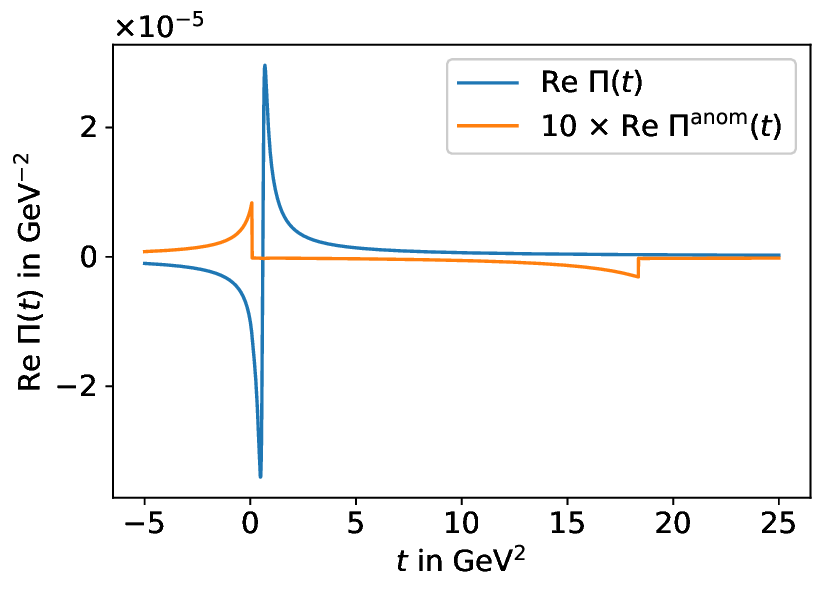}
	\end{subfigure}
	\begin{subfigure}{0.31\textwidth}
		\centering
		\includegraphics[width=\textwidth]{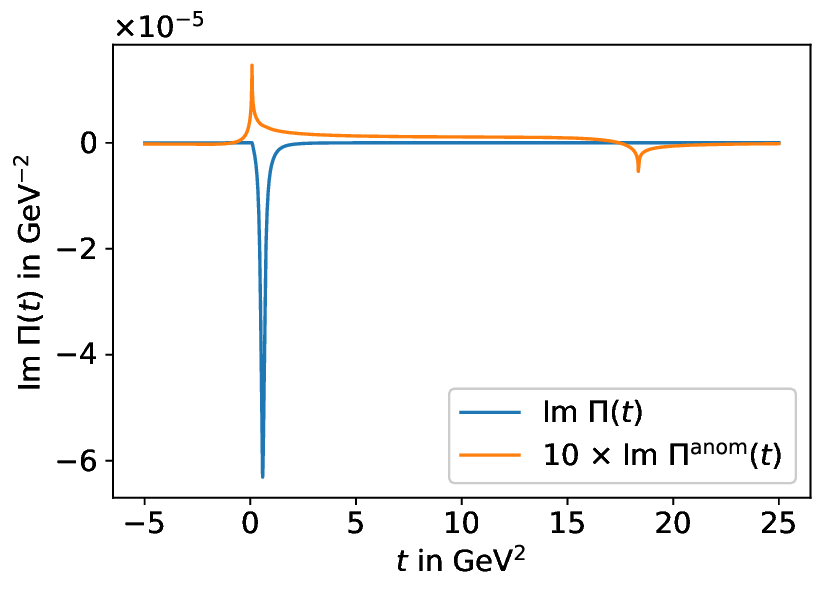}
	\end{subfigure}
	\begin{subfigure}{0.31\textwidth}
		\centering
		\includegraphics[width=\textwidth]{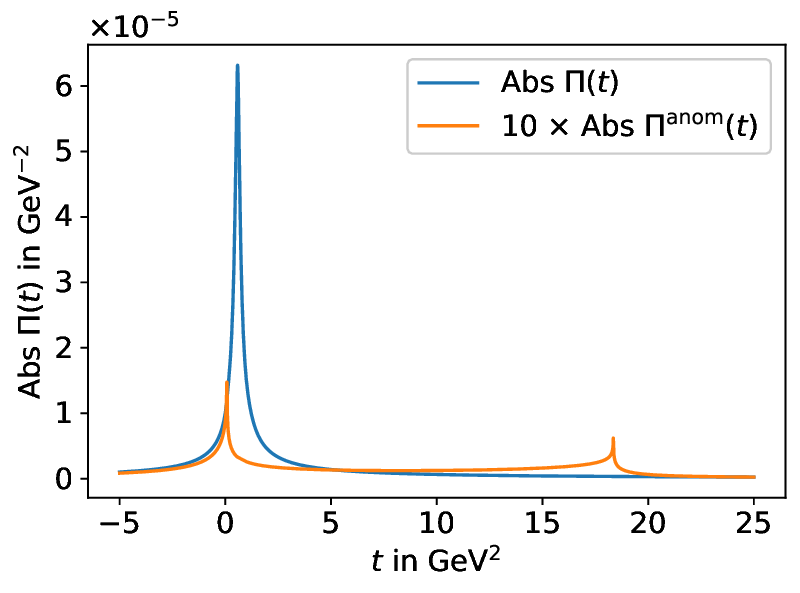}
	\end{subfigure}
	\caption{$B^0 \to K^0 \ell \ell$ form factors. Left: real part of the total form factor $\Pi(t)$ and its anomalous part $\Pi^\text{anom}(t)$, middle: imaginary part, right: absolute values.}
	\label{fig:BtoKll_neutral_FF}
\end{figure}

\begin{figure}[t!]
	\centering
	\begin{subfigure}{0.31\textwidth}
		\centering
		\includegraphics[width=\textwidth]{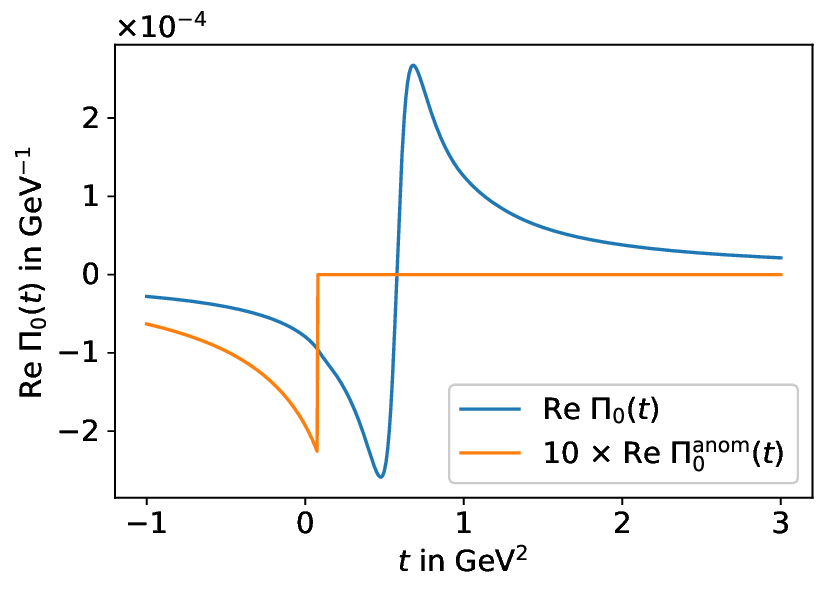}
	\end{subfigure}
	\begin{subfigure}{0.31\textwidth}
		\centering
		\includegraphics[width=\textwidth]{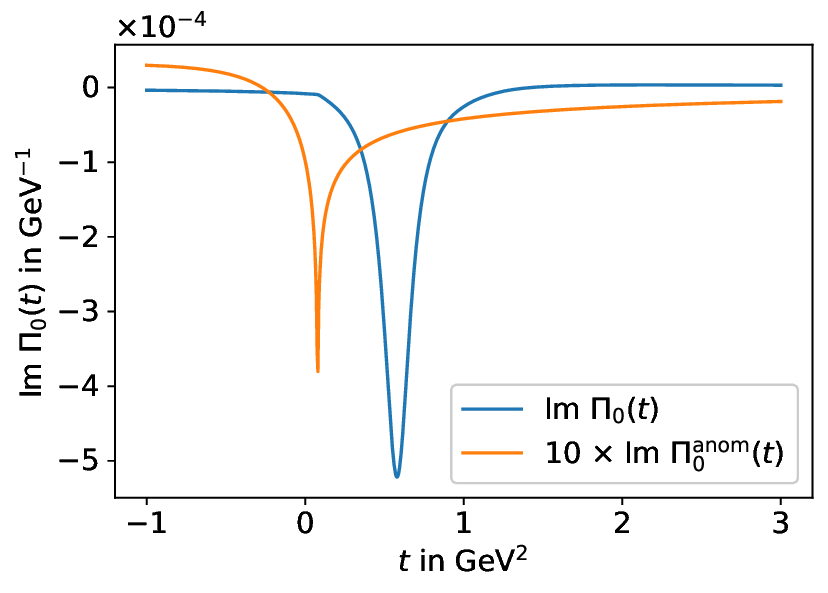}
	\end{subfigure}
	\begin{subfigure}{0.31\textwidth}
		\centering
		\includegraphics[width=\textwidth]{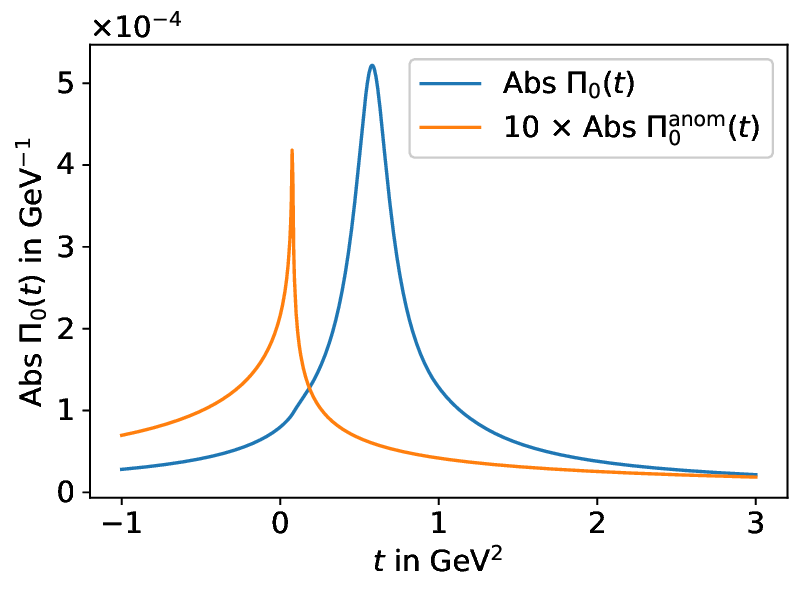}
	\end{subfigure}
	\begin{subfigure}{0.31\textwidth}
		\centering
		\includegraphics[width=\textwidth]{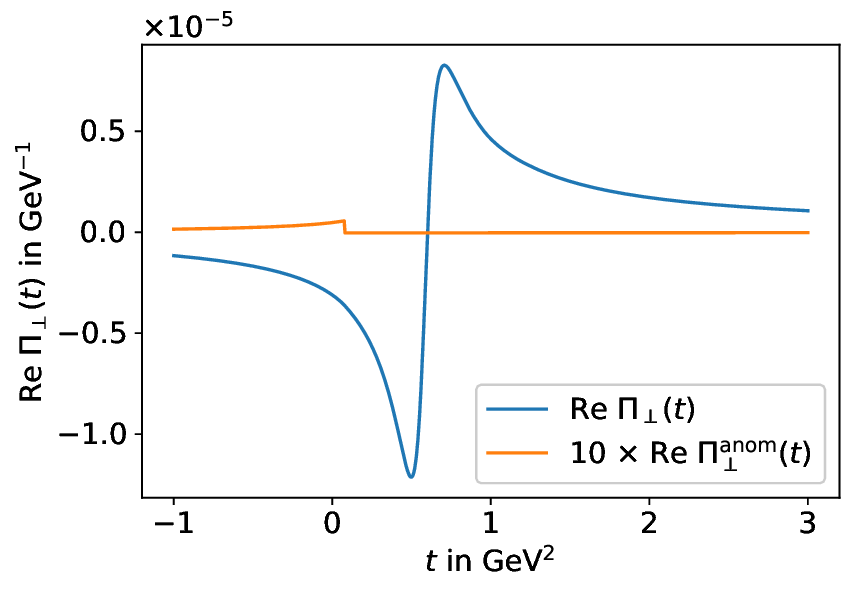}
	\end{subfigure}
	\begin{subfigure}{0.31\textwidth}
		\centering
		\includegraphics[width=\textwidth]{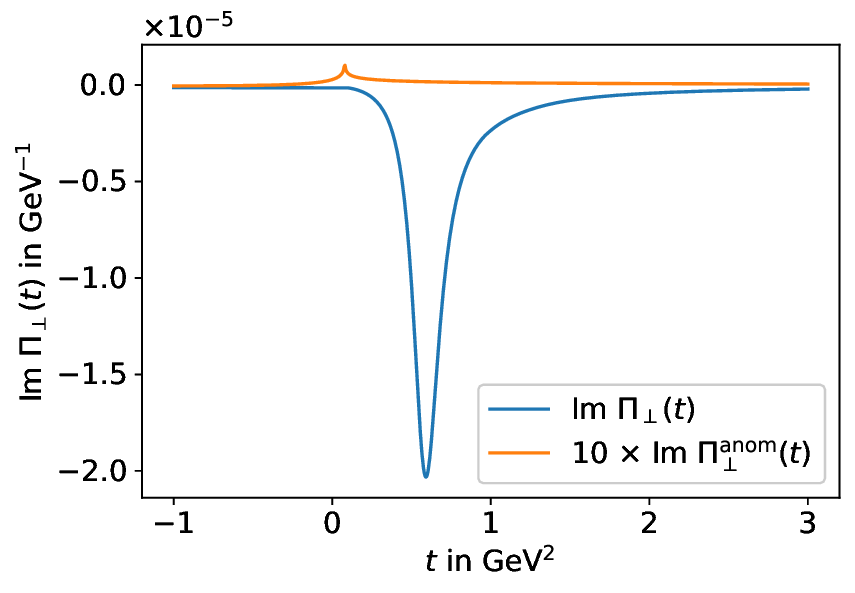}
	\end{subfigure}
	\begin{subfigure}{0.31\textwidth}
		\centering
		\includegraphics[width=\textwidth]{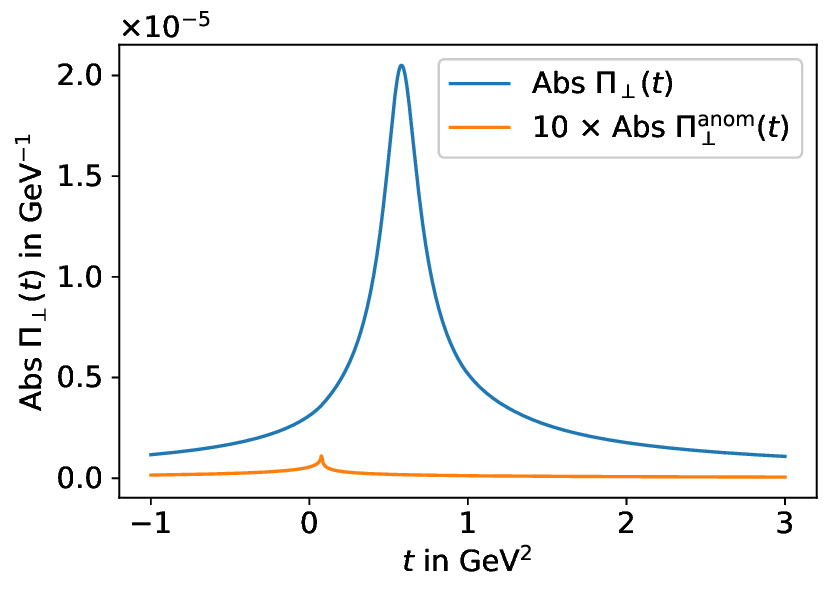}
	\end{subfigure}
	\begin{subfigure}{0.31\textwidth}
		\centering
		\includegraphics[width=\textwidth]{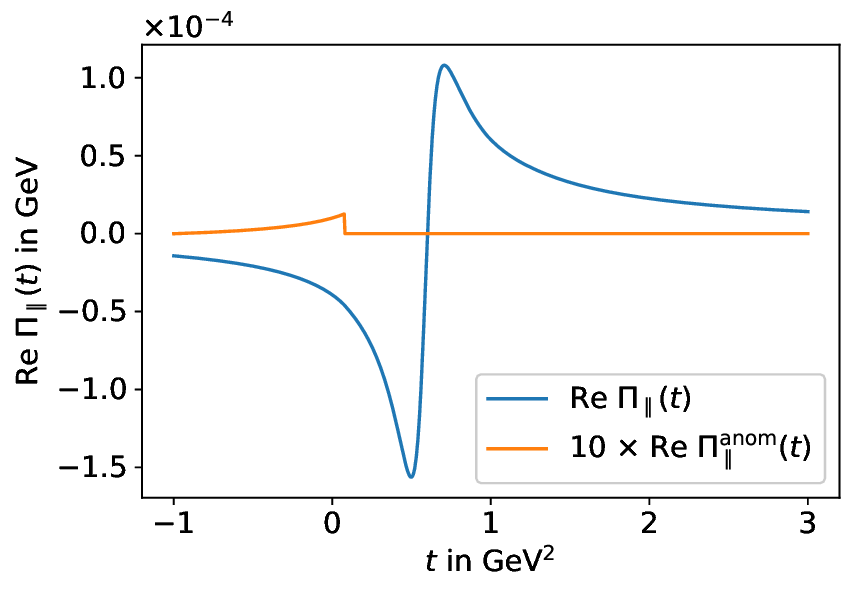}
	\end{subfigure}
	\begin{subfigure}{0.31\textwidth}
		\centering
		\includegraphics[width=\textwidth]{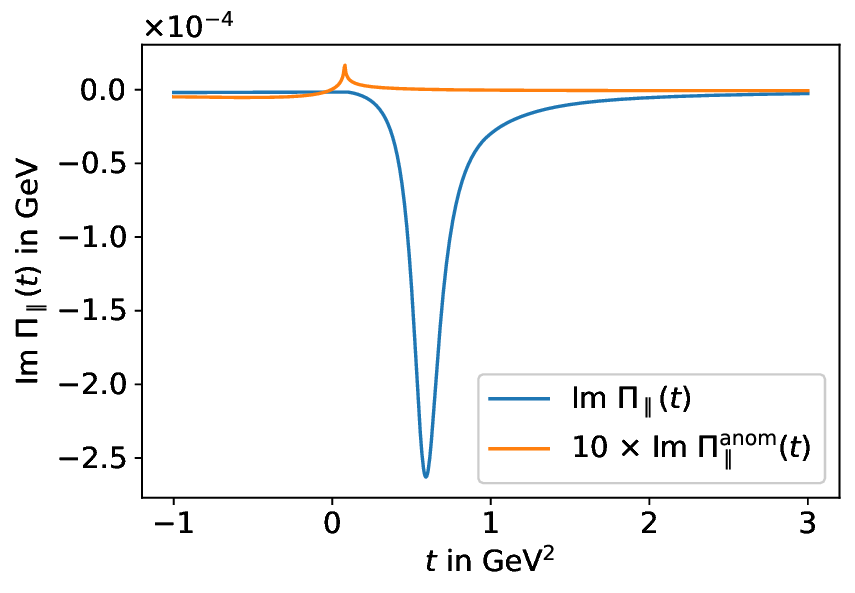}
	\end{subfigure}
	\begin{subfigure}{0.31\textwidth}
		\centering
		\includegraphics[width=\textwidth]{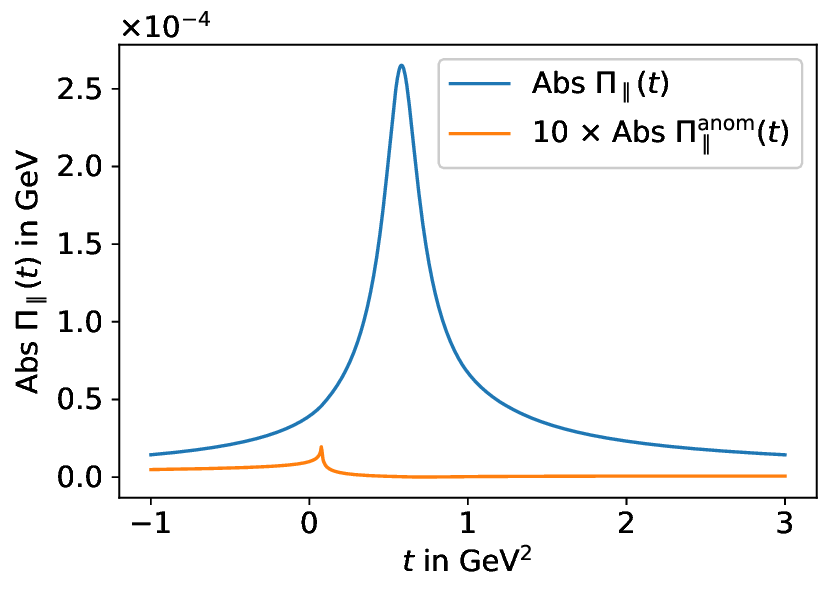}
	\end{subfigure}
	\caption{$B^0 \to K^{*0} \ell \ell$ form factors. First row: $\Pi_0(t)$, second row: $\Pi_\perp(t)$, third row: $\Pi_\parallel(t)$. First column: real part of the total form factor $\Pi_i(t)$ and its anomalous part $\Pi_i^\text{anom}(t)$, second column: imaginary part, third column: absolute value.}
	\label{fig:BtoKstarll_neutral_FF}
\end{figure}

Using the input parameters summarized in App.~\ref{app:coupling}, we can now numerically implement the form factor dispersion relations from Eq.~\eqref{eq:FF_dispersion_relation} for the cases of interest listed in Table~\ref{tab:triangle_singularities_Bdecays}. The results are shown in Figs.~\ref{fig:Btopill_charged_FF}--\ref{fig:BtoKstarll_neutral_FF}. 
All channels have in common that the $\rho$ peak at $t \simeq 0.6\GeV^2$ constitutes the dominant feature, and all results display the expected square-root cusp at $t_\text{thr}\simeq 0.08\GeV^2$. Likewise, the anomalous parts $\Pi_{P,\lambda}^\text{anom}(t)$ display a logarithmic singularity at $t_\text{thr}$. 

The three pseudoscalar cases $B \to P \gamma^*$, $P=(\pi^+,K^+,K^0)$, represent a mass configuration for which the triangle singularity $t_+$ lies on the unitarity cut and, thus, the anomalous part also has an additional logarithmic singularity at $t_+$. Initially,   the anomalous contribution was absorbed into the discontinuity of the normal dispersion integral in these cases, see App.~\ref{sec:triangle_position_general_case}, in such a way that there was no anomalous part for these form factors. However, in order to estimate the impact of the triangle singularity in these cases, we now split off the anomalous integral as in Eq.~\eqref{eq:FF_dispersion_relation} and calculate it separately.

\begin{figure}[t!]
	\centering
	\begin{subfigure}{0.31\textwidth}
		\centering
		\includegraphics[width=\textwidth]{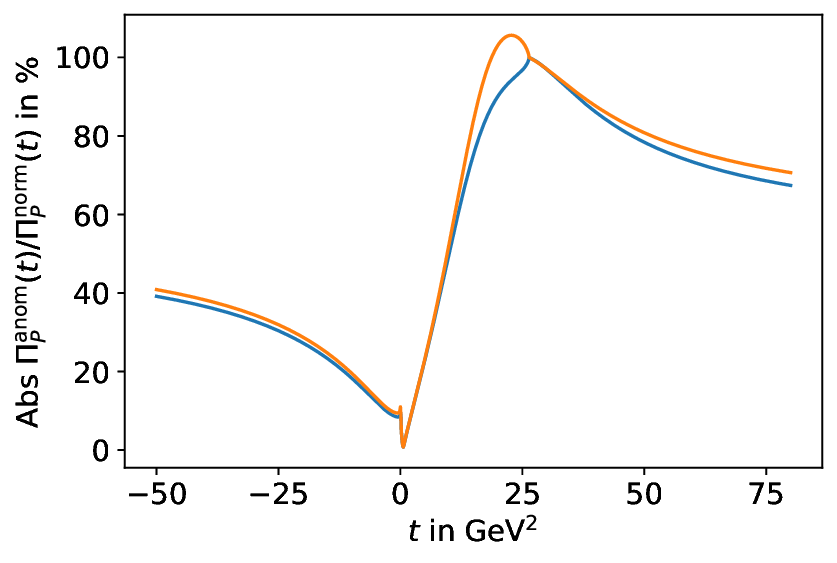}
	\end{subfigure}
	\begin{subfigure}{0.31\textwidth}
		\centering
		\includegraphics[width=\textwidth]{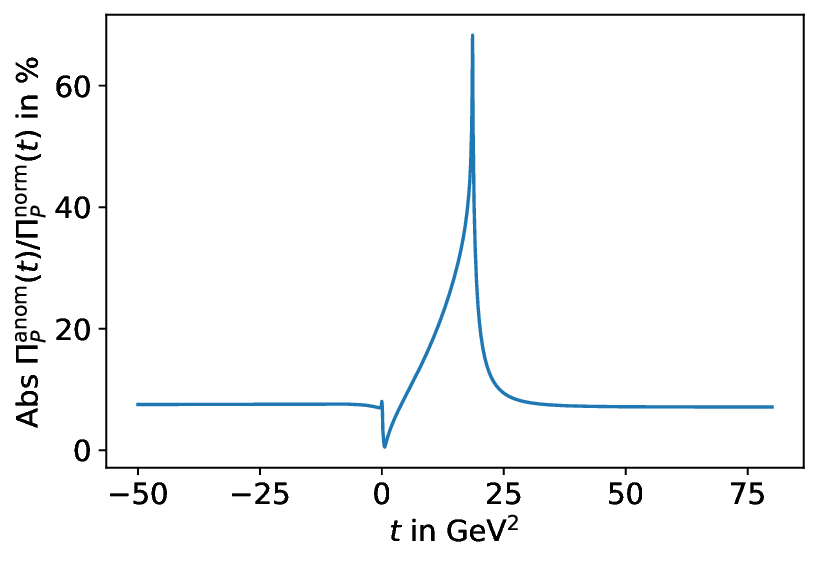}
	\end{subfigure}
	\begin{subfigure}{0.31\textwidth}
		\centering
		\includegraphics[width=\textwidth]{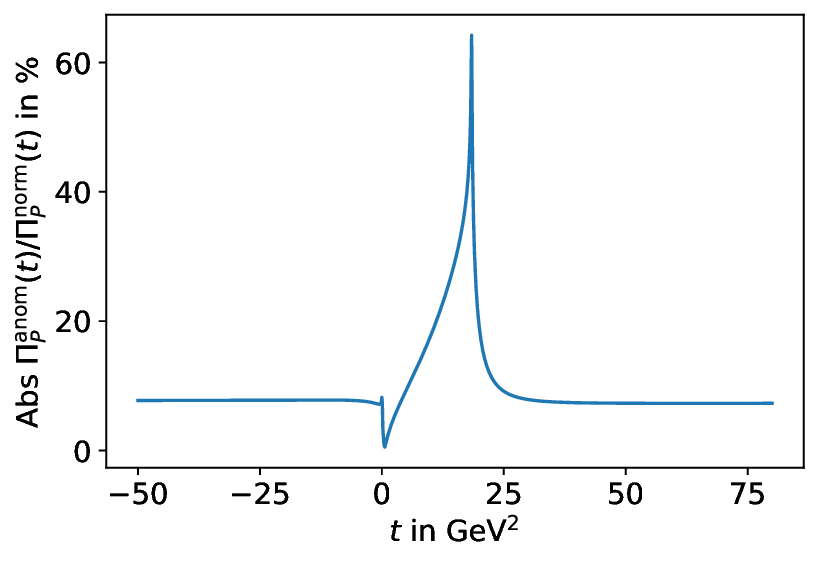}
	\end{subfigure}
	\caption{Anomalous fraction $\abs{\Pi_\lambda^\text{anom}/\Pi_\lambda^\text{norm}}$ for $B \to P \ell \ell$. Left: $P=\pi^+$, middle: $P=K^+$, right: $P=K^0$. Blue: solution for positive subtraction constant, orange: solution for negative subtraction constant. For $K^+$ and $K^0$ the negative solution can be excluded, see Sec.~\ref{sec:subtraction_constants}.}
	\label{fig:BtoPll_anomfrac}
\end{figure}

\begin{figure}[t]
	\centering
	\begin{subfigure}{0.31\textwidth}
		\centering
		\includegraphics[width=\textwidth]{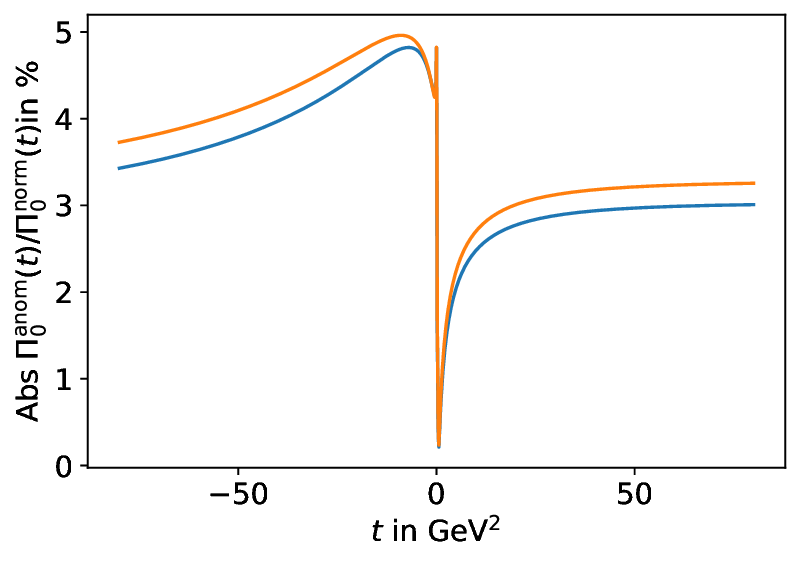}
	\end{subfigure}
	\begin{subfigure}{0.31\textwidth}
		\centering
		\includegraphics[width=\textwidth]{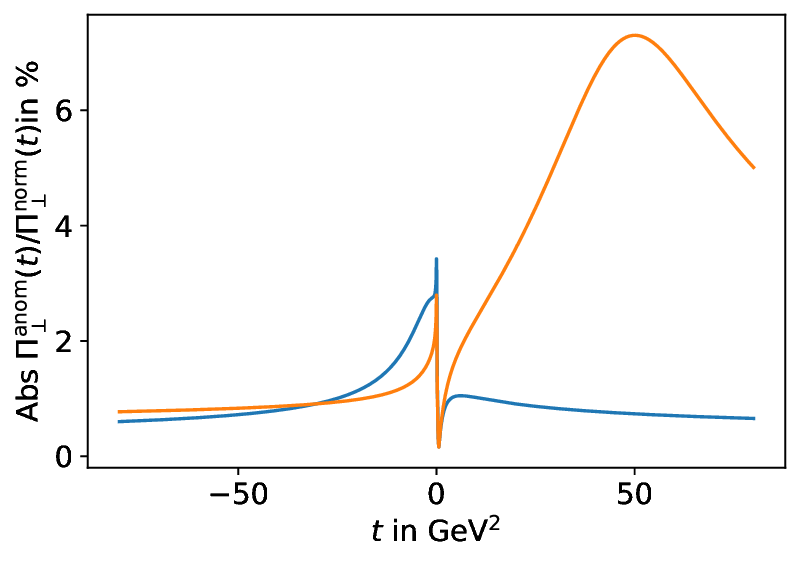}
	\end{subfigure}
	\begin{subfigure}{0.31\textwidth}
		\centering
		\includegraphics[width=\textwidth]{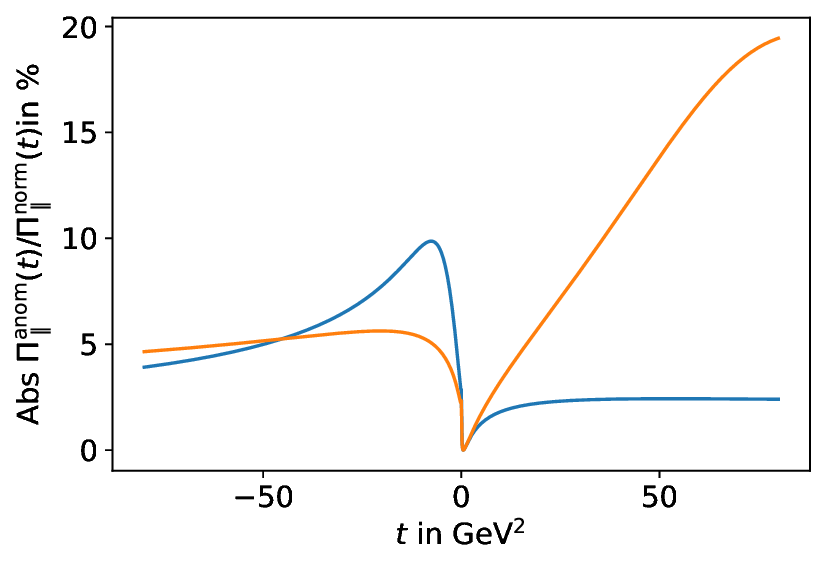}
	\end{subfigure}
	\caption{Anomalous fraction $\abs{\Pi_\lambda^\text{anom}/\Pi_\lambda^\text{norm}}$ for $B^+ \to \rho^+ \ell \ell$. Left: $\lambda=0$, middle: $\lambda=\perp$, right: $\lambda=\parallel$. Blue: solution for positive subtraction constant, orange: solution for negative subtraction constant.}
	\label{fig:Btorholl_charged_anomfrac}
\end{figure}

\begin{figure}[t!]
	\centering
	\begin{subfigure}{0.31\textwidth}
		\centering
		\includegraphics[width=\textwidth]{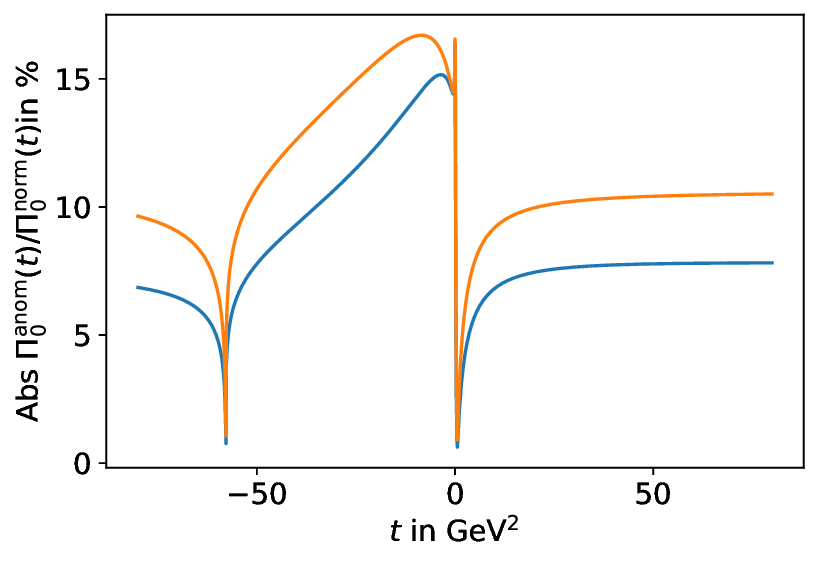}
	\end{subfigure}
	\begin{subfigure}{0.31\textwidth}
		\centering
		\includegraphics[width=\textwidth]{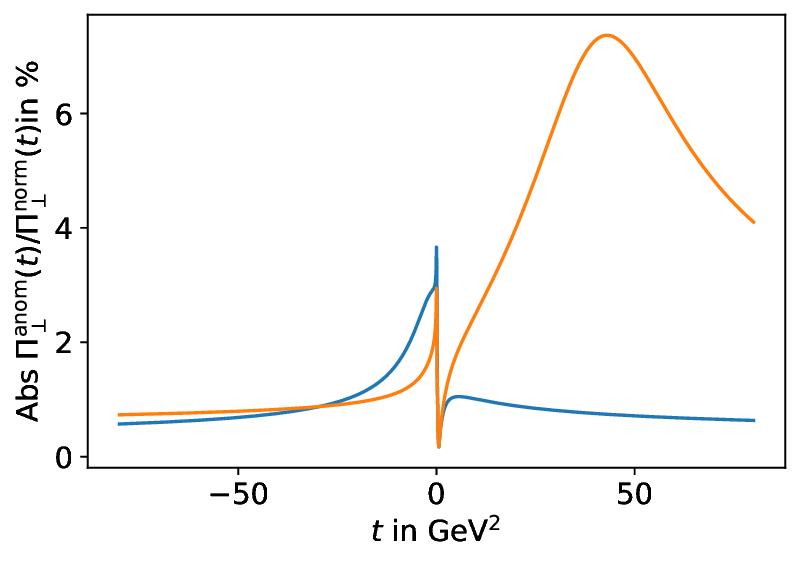}
	\end{subfigure}
	\begin{subfigure}{0.31\textwidth}
		\centering
		\includegraphics[width=\textwidth]{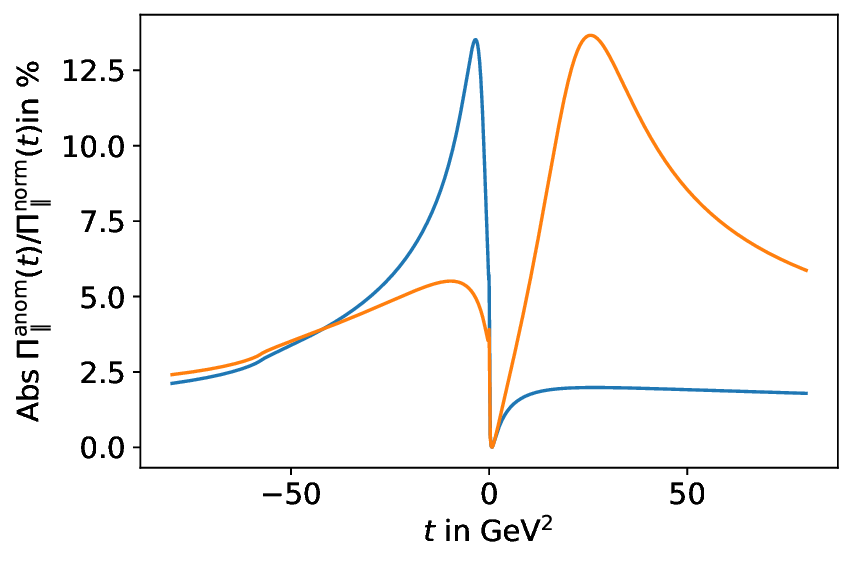}
	\end{subfigure}
	\caption{Anomalous fraction $\abs{\Pi_\lambda^\text{anom}/\Pi_\lambda^\text{norm}}$ for $B^+ \to K^{*+} \ell \ell$. Left: $\lambda=0$, middle: $\lambda=\perp$, right: $\lambda=\parallel$. Blue: solution for positive subtraction constant, orange: solution for negative subtraction constant.}
	\label{fig:BtoKstarll_charged_anomfrac}
\end{figure}

\begin{figure}[t]
	\centering
	\begin{subfigure}{0.31\textwidth}
		\centering
		\includegraphics[width=\textwidth]{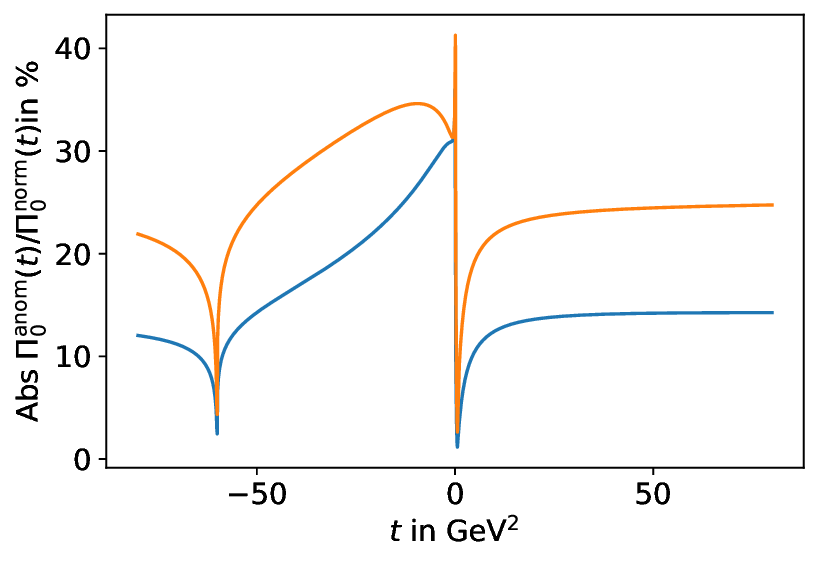}
	\end{subfigure}
	\begin{subfigure}{0.31\textwidth}
		\centering
		\includegraphics[width=\textwidth]{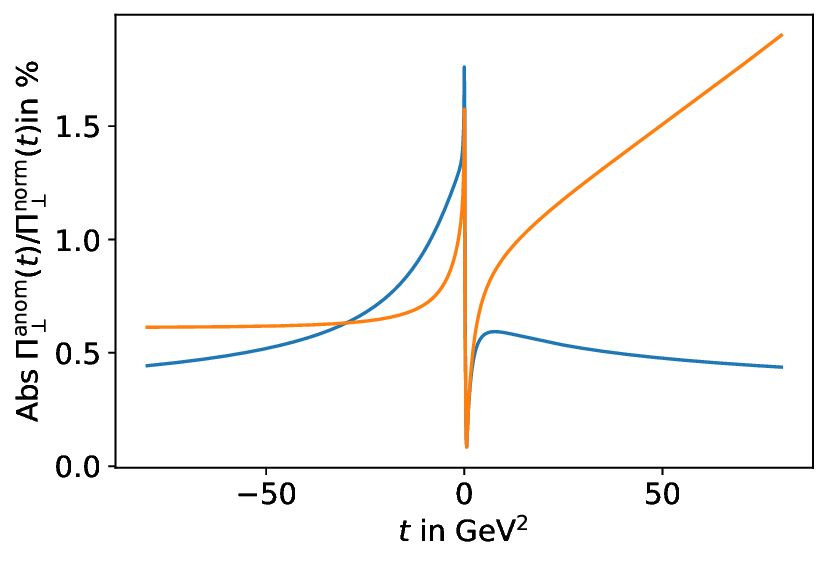}
	\end{subfigure}
	\begin{subfigure}{0.31\textwidth}
		\centering
		\includegraphics[width=\textwidth]{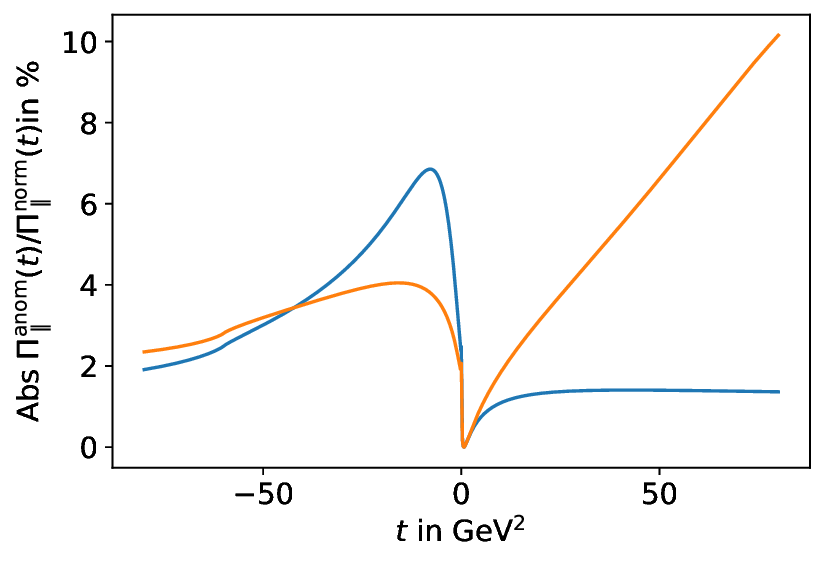}
	\end{subfigure}
	\caption{Anomalous fraction $\abs{\Pi_\lambda^\text{anom}/\Pi_\lambda^\text{norm}}$ for $B^0 \to K^{*0} \ell \ell$. Left: $\lambda=0$, middle: $\lambda=\perp$, right: $\lambda=\parallel$. Blue: solution for positive subtraction constant, orange: solution for negative subtraction constant.}
	\label{fig:BtoKstarll_neutral_anomfrac}
\end{figure}

In order to quantify the size of the contribution of the anomalous part $\Pi_{P,\lambda}^\text{anom}(t)$  to the total form factor $\Pi_{P,\lambda}(t)$ we further look at the anomalous fraction $\big|\Pi_{P,\lambda}^\text{anom}(t)/\Pi_{P,\lambda}^\text{norm}(t)\big|$ for each of the cases in Figs.~\ref{fig:BtoPll_anomfrac}--\ref{fig:BtoKstarll_neutral_anomfrac}. Due to the sign ambiguity of the subtraction constants $a_{P,\lambda}$ in most of the cases, see the discussion in Sec.~\ref{sec:subtraction_constants}, we computed the anomalous fractions for both of these solutions and included them in the figure for comparison. Qualitatively they exhibit the same behavior, but quantitatively they can differ more substantially as the interference between the three contributions in Fig.~\ref{fig:FF_unitarity_sum} changes. 

As expected, two common features shared by all results concern a suppression in the vicinity of the $\rho$ peak and a sharp spike near $t_\text{thr}$ due to the logarithmic divergence of both the normal and the anomalous integral. The latter phenomenon also appears at $t_+$ for the cases for which $t_+$ lies on the real axis.
These divergences cancel in the sum of normal and anomalous contributions, so that, at higher resolution, 
the anomalous fractions $|\Pi_{P,\lambda}^\text{anom}(t)/\Pi_{P,\lambda}^\text{norm}(t)|$ in Fig.~\ref{fig:BtoPll_anomfrac} tend to $100\%$ at $t_+$. Away from the thresholds $t_\text{thr}$ and $t_+$ the anomalous fractions eventually flatten and tend towards a constant value. While this value is of the order of a few percent for most of the cases, it can become as large as $(10\text{--}20)\,\%$  for the longitudinal $B^+ \to K^{*+} \gamma^*$ and $B^0 \to K^{*0} \gamma^*$ form factors and even 50\,\% for the $B^+ \to \pi^+ \gamma^*$ form factor. In the cases in which $t_+$ lies on the negative real axis, the anomalous fraction is further amplified between the two thresholds $t_\text{thr}$ and $t_+$ along the negative real axis.

The rather large discrepancy in the order of magnitude of the anomalous fraction between $B^+ \to \pi^+ \gamma^*$ on the one hand, and $B^+ \to K^+ \gamma^*$ and $B^0 \to K^0 \gamma^*$ on the other hand can be explained phenomenologically as follows. The anomalous integrand in Eq.~\eqref{eq:FF_dispersion_relation} contains a factor of~$[\lambda^{1/2}(t,M_B^2,M_{(P,V)}^2)]^{-3}$, which grows big in the vicinity of the pseudothreshold ${t_\text{ps}=(M_B - M_{(P,V)})^2}$. For $B^{+,0} \to K^{+,0} \gamma^*$ this does not have a large impact, as the anomalous integration path only runs up to $t_+ \simeq 18.6\GeV^2$ and $t_+ \simeq 18.4\GeV^2$, respectively, which is still quite distant from $t_\text{ps} \simeq 22.9\GeV^2$,
whereas for $B^+ \to \pi^+ \gamma^*$ a large amplification is to be expected as $t_+ \simeq 26.38\GeV^2$ and $t_\text{ps} \simeq 26.42\GeV^2$ are in close proximity.

From kinematic arguments one could similarly expect that the anomalous contributions for the $\lambda=0,\parallel$   form factors for $B^+ \to \rho^+ \gamma^*$ should be bigger than the ones for $B^+ \to K^{*+} \gamma^*$ and $B^0 \to K^{*0} \gamma^*$, since the anomalous integration path reaches far more into the negative-$q^2$ region, with $t_+ \simeq -859\GeV^2$ compared to $t_+ \simeq -57.8\GeV^2$ and $t_+ \simeq -60.0\GeV^2$, respectively. However, this expected hierarchy is compensated by comparably smaller coupling constants determining the strength of the left-hand singularity in $B^+ \to \rho^+ \gamma^*$. In this context, we also remark on the origin of scales as large as several hundred $\GeV^2$ in the position of $t_+$. Expanding Eq.~\eqref{eq:triangle_singularities_analytic_continuation} for $M_B\to\infty$ and $\mpi\to 0$, one has 
\begin{equation}
\label{eq:chiral_limit}
t_+\simeq -\frac{M_B^2 M_\rho^2}{\mpi^2}\simeq -860\GeV^2,\qquad 
t_-\simeq -\frac{M_B^2\mpi^2}{M_\rho^2}\simeq -0.9\GeV^2, 
\end{equation}
so that, in the chiral limit, $t_+$ indeed tends to minus infinity. Finally, for the $\lambda = \perp$ component we generally observe the smallest effects, depending on channel and relative signs the anomalous contribution is typically limited to a few percent.

\section{Generalizations}
\label{sec:generalizations}

The dispersive formalism developed in this work can be extended in a number of ways. First, we assumed that the widths of the vector mesons that give rise to the critical left-hand cuts can be neglected, while for a refined implementation one should consider a weighting with  spectral functions~\cite{Lomon:2012pn,Moussallam:2013una,Zanke:2021wiq,Crivellin:2022gfu}. Second, we have restricted the analysis to $\pi\pi$ intermediate states, but at higher energies also other isovector contributions such as $4\pi\simeq\pi\omega$ will play a role, as will isoscalar channels. Moreover, the study could be extended to $B_s$ decays, changing the phenomenology due to the additional strangeness. Most importantly, the question arises to what extent these considerations for the $u$-quark loop can be generalized to charm. In this section, we comment on several aspects of such future applications.  

\begin{table}[t!]
	\centering
	\renewcommand{\arraystretch}{1.3}
 \scalebox{0.96}{
	\begin{tabular}{ccccccc}
	\toprule 
		$p_3^2$ & \multicolumn{2}{c}{$M_{K^{*\pm}}^2$} & \multicolumn{2}{c}{$M_{K^\pm}^2$} & $M_{\rho^{\pm}}^2$ & $M_{\pi^\pm}^2$ \\
		$m_1^2$ & $M_{K^\pm}^2$ & $M_{K^{*\pm}}^2$ & $M_{K^\pm}^2$ & $M_{K^{*\pm}}^2$ & $M_{\pi^\pm}^2$  & $M_{\rho^\pm}^2$ \\\midrule
		Case & B & B & B & B & B & B \\
		$t_+$ $\big[\text{GeV}^2\big]$ & $4.1-43.6\iu$ & $11.1-21.3\iu$ & $35.3-26.7\iu$ & $20.4-13.1\iu$ & $23-154\iu$ & $27.7-4.9\iu$ \\
		$t_-$ $\big[\text{GeV}^2\big]$ & $4.1+43.6\iu$ & $11.1+21.3\iu$ & $35.3+26.7\iu$ & $20.4+13.1\iu$ & $23+154\iu$ & $27.7+4.9\iu$ \\
		\midrule
  $p_3^2$ & \multicolumn{2}{c}{$M_{K^{*\pm}}^2$} & \multicolumn{2}{c}{$M_{K^\pm}^2$} & $M_{\rho^{\pm}}^2$ & $M_{\pi^\pm}^2$ \\
		$m_1^2$ & $M_{K^\pm}^2$ & $M_{K^{*\pm}}^2$ & $M_{K^\pm}^2$ & $M_{K^{*\pm}}^2$ & $M_{\pi^\pm}^2$  & $M_{\rho^\pm}^2$ \\\midrule
		Case & A & B & -- & C & A & C \\
		$t_+$ $\big[\text{GeV}^2\big]$ & $-57.5$ & $0.93-3.99\iu$ & -- & $18.7$ & $-786$ & $26.4$ \\
		$t_-$ $\big[\text{GeV}^2\big]$ & $-0.32$ & $0.93+3.99\iu$ & -- & $1.53$ & $-0.26$ & $1.48$ \\
		\bottomrule
	\end{tabular}}
	\renewcommand{\arraystretch}{1.0}
	\caption{Triangle singularities for $p_1^2 = M_{B^\pm}^2$ and $m_2^2=M_{\pi^0}^2$, $m_3^2=M_\omega^2$ (upper panel) and $m_2^2=M_\omega^2$, $m_3^2=M_{\pi^0}^2$  (lower panel), using particle masses from Ref.~\cite{ParticleDataGroup:2022pth}. Isospin-breaking contributions are omitted.}
	\label{tab:triangle_singularities_Bdecays_piomega}
\end{table}

\subsection{Higher intermediate states, isoscalar decays, and strangeness}

In the isovector channel, inelastic contributions start, in principle, at the $4\pi$ threshold, but phenomenologically no significant inelasticities are observed below the $\pi\omega$ threshold, as can be demonstrated using the Eidelman--\L{}ukaszuk bound~\cite{Lukaszuk:1973jd,Eidelman:2003uh}.\footnote{Isospin-breaking $3\pi$ contributions start before, strongly localized at the $\omega$ resonance, but for simplicity we restrict the discussion to the isospin limit, especially, since there are isospin-conserving $3\pi$ contributions as well, see below.} The role of such $4\pi\simeq \pi\omega$ inelasticities is of phenomenological interest in the context of the electromagnetic form factor of the pion~\cite{Colangelo:2018mtw,Colangelo:2022prz,Stoffer:2023gba,Hanhart:2012wi,Chanturia:2022rcz,Heuser:2024biq}, while for the application to $B\to(P,V)\,\gamma^*$ form factors the $\omega\to\pi^0\gamma^*$ transition form factor and its analytic continuation would be required~\cite{Schneider:2012ez}. Possible left-hand singularities for the charged decay are summarized in Table~\ref{tab:triangle_singularities_Bdecays_piomega}, in analogy to Table~\ref{tab:triangle_singularities_Bdecays}. For the lower panel, with mass configuration $m_2=M_\omega$, $m_3=M_{\pi^0}$, the situation is qualitatively similar to the $\pi\pi$ intermediate states, as situations in which the external particle can decay into two particles in the loop still arise, while in the opposite case such decays are kinematically forbidden and the anomalous branch point always appears in the lower complex half-plane. We observe, however, that the imaginary part can again diverge in the chiral limit, for $p_3^2=M_{\rho}^2$ (and setting $M_\omega=M_\rho$), the anomalous branch points for $M_B\to \infty$, $\mpi\to 0$ become
\begin{equation}
t_\pm =M_B^2\bigg(\frac{1}{2}\mp i\frac{M_\rho}{\mpi}\bigg).
\end{equation}
The phenomenological consequences of such anomalous branch points deep in the complex plane remain to be explored. Most of the required couplings could again be estimated from $B$ decays, e.g., $\Br[B^+\to K^+\omega]=6.5(4)\times 10^{-6}$~\cite{CLEO:2000xjz,BaBar:2007nku,Belle:2013nby}, while in some cases only limits are available, e.g., $\Br[B^+\to K^{*+}\omega]<7.4\times 10^{-6}$~\cite{BaBar:2009mcf}. Moreover, for vector left-hand singularities now also amplitudes involving three vector particles arise, which take a more complicated structure than Eqs.~\eqref{P_amplitudes} and~\eqref{V_amplitudes}.

In the isoscalar channel, the $q^2$ spectrum is dominated by the narrow $\omega$ and $\phi$ resonances, and to go beyond a narrow-resonance description of the $B\to(P,V)\,\gamma^*$ form factors a more detailed understanding of the $\gamma^*\to 3\pi$ matrix element is required~\cite{Hoferichter:2012pm,Hoferichter:2014vra,Hoferichter:2018dmo,Hoferichter:2018kwz,Hoferichter:2019mqg,Hoferichter:2023bjm}. To apply a similar strategy as for the isovector channel, one could try to approximate the result in terms of a $\pi\rho$ effective two-body intermediate state~\cite{Stamen:2022uqh}, in which case the phenomenology should become similar to $\pi\omega$, while a full treatment of the three-particle intermediate state becomes significantly more complicated.

Finally, for $B_s$ decays the additional strangeness changes the kinematic configurations to some extent, see Table~\ref{tab:triangle_singularities_Bsdecays}. Compared to Table~\ref{tab:triangle_singularities_Bdecays}, the anomalous branch points in case B of the vector-meson final states move further into the complex plane, while case A still exhibits a branch point far on the negative real axis. The main change occurs for the pseudoscalar final states, since the vector mesons that describe the left-hand cut can no longer decay into the two pseudoscalars, the anomalous branch point moves from the unitarity cut into the lower complex half-plane.  

\begin{table}[t!]
	\centering
	\renewcommand{\arraystretch}{1.3}
	\begin{tabular}{ccccccc}
	\toprule 
		$p_3^2$ & \multicolumn{2}{c}{$M_{\phi}^2$} & $M_{\eta}^2$ & \multicolumn{2}{c}{$M_{K^{0*}}^2$} & $M_{K^0}^2$ \\
		$m_1^2$ & $M_{K^0}^2$ & $M_{K^{*0}}^2$ & $M_{K^{*0}}^2$ & $M_{\pi^0}^2$ & $M_{\rho^0}^2$ & $M_{\rho^0}^2$ \\\midrule
		Case & A & B & B & A & B & B \\
		$t_+$ $\big[\text{GeV}^2\big]$ & $-43.5$ & $0.7-15.4\iu$ & $13.5-8.3\iu$ & $-826$ & $1.6-17.9\iu$ & $14.5-11.2\iu$ \\
		$t_-$ $\big[\text{GeV}^2\big]$ & $-17.7$ & $0.7+15.4\iu$ & $13.5+8.3\iu$ & $-12.9$ & $1.6+17.9\iu$ & $14.5+11.2\iu$ \\
		\bottomrule
	\end{tabular}
	\renewcommand{\arraystretch}{1.0}
	\caption{Triangle singularities for $p_1^2 = M_{B^0_s}^2$ and $m_2^2=m_3^2=M_{K^0}^2$, using particle masses from Ref.~\cite{ParticleDataGroup:2022pth}. The results for the $K^+K^-$ channel, where applicable, are very similar.}
	\label{tab:triangle_singularities_Bsdecays}
\end{table}

\subsection{Charm loop}

The kinematic configurations for a representative case of the charm loop are summarized in Table~\ref{tab:triangle_singularities_Bdecays_DDbar}. The anomalous branch points are calculated for $\bar D^0 D^0$ intermediate states, because then for the $B^\pm$ channel all relevant left-hand singularities are allowed by charge conservation. One observes that in almost all cases the anomalous branch point lies in the lower complex half-plane, a consequence of the mass hierarchy in which the intermediate-state $D$-mesons are too heavy to decay. The only exception is defined by the case of an external $\pi^\pm$, since in this case the decay $D^*\to D\pi$ can occur, leading to an anomalous branch point on the unitarity cut. Barring this special case, however, the hadronization of the charm loop always leads to situations in which the contour is deformed into the lower complex half-plane, and only by a moderate amount. Our empirical findings for the $\pi\pi$ intermediate states suggest that in this situation the impact of the anomalous contributions should be smaller than in scenario A. On the other hand, we observed that also the strength of the left-hand singularities plays an important role, so that explicit calculations are necessary to draw robust conclusions.    

To perform such studies for the charm loop, one needs, as first ingredient, a good understanding of $D$-meson electromagnetic and transition form factors~\cite{Reinert:2020}, including their analytic continuation. Moreover, the analysis is complicated by the proximity of a number of $\bar D D$ states, whose phenomenology and interference patterns will therefore be important~\cite{Hanhart:2023fud,Husken:2024hmi}. Finally, to determine parameters, one can no longer rely on the dominance of a single resonance---the $\rho(770)$ in the example studied in this work---but a deeper understanding of the respective amplitudes such as $B\to \bar D D K^*$ will be required, for which experimental input would be extremely welcome. In this regard, an advantage compared to the $\pi\pi$ intermediate states arises from the mass of $D$-mesons, as the Dalitz plots are substantially smaller, and thus potentially less complicated than the ones for three light decay products.

\begin{table}[t!]
	\centering
	\renewcommand{\arraystretch}{1.3}
	\begin{tabular}{ccccccc}
	\toprule 
		$p_3^2$ & \multicolumn{2}{c}{$M_{K^{*\pm}}^2$} & $M_{K^\pm}^2$ & \multicolumn{2}{c}{$M_{\rho^\pm}^2$} & $M_{\pi^\pm}^2$ \\
		$m_1^2$ & $M_{D_s^\pm}^2$ & $M_{D_s^{*\pm}}^2$ & $M_{D_s^{*\pm}}^2$ & $M_{D^\pm}^2$ & $M_{D^{*\pm}}^2$ & $M_{D^{*\pm}}^2$ \\\midrule
		Case & B & B & B & B & B & C \\
		$t_+$ $\big[\text{GeV}^2\big]$ & $24.3-8.2\iu$ & $22.9-6.8\iu$ & $24.1-3.5\iu$ & $26.0-8.0\iu$ & $24.4-6.8\iu$ & $26.2$ \\
		$t_-$ $\big[\text{GeV}^2\big]$ & $24.3+8.2\iu$ & $22.9+6.8\iu$ & $24.1+3.5\iu$ & $26.0+8.0\iu$ & $24.4+6.8\iu$ & $25.5$ \\
		\bottomrule
	\end{tabular}
	\renewcommand{\arraystretch}{1.0}
	\caption{Triangle singularities for $p_1^2 = M_{B^\pm}^2$ and $m_2^2=m_3^2=M_{D^0}^2$, using particle masses from Ref.~\cite{ParticleDataGroup:2022pth}. In view of the small mass difference among the $D$-mesons, the kinematic situation for the neutral channel as well as other intermediate states, e.g., $D^+ D^-$, $\bar D^* D$, $\bar D^* D^*$, $\bar D_s D_s$, etc.\ is very similar (where applicable).}
	\label{tab:triangle_singularities_Bdecays_DDbar}
\end{table}

\section{Summary and outlook}
\label{sec:summary}

In this work we presented a first study of the possible impact of anomalous thresholds on $B\to (P,V)\,\gamma^*$ form factors. To this end, we started from a comprehensive discussion of anomalous thresholds in the triangle loop function, whose analytic properties coincide with the ones of the unitarity diagrams relevant for  the $B\to (P,V)\,\gamma^*$ form factors. In particular, we compiled a complete list of mass configurations---several of which are realized for different intermediate states and left-hand cuts---and derived the corresponding analytic continuations and anomalous discontinuities. As a first main result, this investigation leads to the prescriptions for the general analytic continuation given in App.~\ref{sec:triangle_position_general_case}, and the application to $B$-decay form factors in Table~\ref{tab:triangle_singularities_Bdecays}.  

Next, we studied the Bardeen--Tung--Tarrach decomposition of  the $B\to (P,V)\,\gamma^*$ matrix elements, to find the scalar functions for which dispersion relations can be written in the absence of kinematic singularities and zeros. Concentrating on $\pi\pi$ intermediate states as the dominant low-energy hadronization of the $u$-quark loop, we derived the unitarity relations and their solution using Muskhelishvili--Omn\`es techniques, in terms of the electromagnetic form factor of the pion and partial-wave amplitudes that describe the respective left-hand cuts in the $B\to (P,V)\,\pi\pi$ amplitude. We found that for a realistic phenomenology subtractions need to be introduced, reflecting the direct $B\to(P,V)\,\rho$ decay. The final dispersion relation for the form factors is given in Eq.~\eqref{eq:FF_dispersion_relation}, including the unitarized $P$-wave amplitudes  
from Eq.~\eqref{eq:pwaves_unitarized}.

To evaluate these general solutions phenomenologically, we determined parameters as follows: the modulus of most couplings can be extracted from measured branching fractions for  $B\to(P',V')\,\pi$, $B\to(P,V)\,\rho$, $V'\to (P,V)\,\pi$, and $P'\to V\pi$ decays (or crossed versions thereof), with $P=P'=K,\pi$, $V=K^*,\rho$, $V'=K^*,\rho,\omega$. For other couplings not directly accessible in this way $SU(3)$ relations apply, and the saturation of the amplitude by the $\rho$ resonance can be tested using $B\to(P,V)\,\pi\pi$ branching fractions. Moreover, helicity components can be disentangled from measured polarization fractions and QCD factorization predictions, leaving in most cases just the sign of the subtraction constant ambiguous. In principle, such relative phases can be determined from Dalitz plot analyses, available at present for $P=K^+,K^0$, while in the other cases we displayed results for both possible assignments. 

Our key findings for the relative size of anomalous contributions are illustrated in Figs.~\ref{fig:BtoPll_anomfrac}--\ref{fig:BtoKstarll_neutral_anomfrac}, compared to the respective normal contributions (the full non-local form factor being given by the sum of both). In all cases, we saw that the effect is suppressed at the $\rho(770)$ resonance, simply due to the resonance enhancement of the normal contribution in this case. Off-peak, however, the effects can become more sizable, depending on kinematic configuration and coupling strength. For the $P=K,\pi$ final states, the anomalous branch point lies on the unitarity cut, and while the relative size of normal and anomalous contributions grows to $100\%$ at the anomalous branch point, see Fig.~\ref{fig:BtoPll_anomfrac}, 
this separation 
can be avoided altogether by a suitable adjustment of the analytic continuation of the normal discontinuity. For the vector final states, we observed cases in which anomalous contributions become as large as $\Order(10\%)$, mainly for kinematic configurations in which the anomalous branch point lies on the negative real axis, see Figs.~\ref{fig:Btorholl_charged_anomfrac}--\ref{fig:BtoKstarll_neutral_anomfrac}. In cases of anomalous branch points that deform the contour into the lower complex half-plane, typically smaller effects were obtained, with the exact size depending on the strength and relative sign of the respective left-hand singularities.

Similar estimates should be possible in the future for higher intermediate states (such as $\pi\omega$), isoscalar intermediate states (including $\omega$ and $\phi$), and $B_s$ decays, to improve the phenomenology of the non-local contributions, by combining our dispersive representation with input from QCD factorization, light-cone sum rules, and lattice QCD. In particular, it would be important to delineate how the anomalous thresholds manifest themselves in a partonic calculation, e.g., to ensure that a matching in the space-like domain, where anomalous contributions can be enhanced compared to the resonance region, is not affected at the relevant level of precision. Moreover, our strategy to estimate the impact of anomalous thresholds  could be extended towards the hadronization of the charm  loop. While the qualitative features should be similar, we emphasize that the close proximity of different thresholds, $\bar D D$, $\bar D D^*$, $\bar D^*D^*$, $\bar D_s D_s$, etc., renders the phenomenological analysis more challenging. This includes both experimental and theoretical assignments, e.g., related to
\begin{enumerate}
 \item the  measurement of branching fractions such as $B\to\bar D D K^*$ required to determine coupling constants,
 \item the phenomenology and analytic continuation of $D$-meson electromagnetic and transition form factors,
 \item the interference patterns among the various $\bar D  D$ states. 
\end{enumerate}
On the other hand, as the discussion in Sec.~\ref{sec:generalizations} shows, the anomalous contributions for the charm loop (almost) all belong to the class in which the anomalous branch point lies in the lower complex half-plane, in which case we observed, empirically, the smallest effects for the $\pi\pi$ intermediate states in the hadronization of the $u$-quark loop. In conclusion, we believe that with the methods developed in this paper, it should be possible to improve estimates of the non-local form factors in the interpretation of $B\to (P,V)\,\ell^+\ell^-$ decays, and thereby consolidate potential hints for physics beyond the SM.

\acknowledgments   
We thank A.~Crivellin, N.~Gubernari, A.~Khodjamirian,  J.~Matias, M.~Reboud, D.~van Dyk, and J.~Virto for valuable discussions.
Financial support by the SNSF (Project No.\  PCEFP2\_181117), the Bonn--Cologne Graduate School of Physics and Astronomy (BCGS), the DFG through the funds provided to the Sino--German Collaborative
Research Center TRR110 ``Symmetries and the Emergence of Structure in QCD''
(DFG Project-ID 196253076 -- TRR 110), and 
the MKW NRW under the funding code NW21-024-A
 is gratefully acknowledged.  

\appendix

\section{Analytic continuation of the triangle function in the general case} \label{sec:triangle_position_general_case}

\begin{figure}[tb]
	\centering
	\begin{subfigure}{0.3\textwidth}
		\centering
		\includegraphics[width=\textwidth]{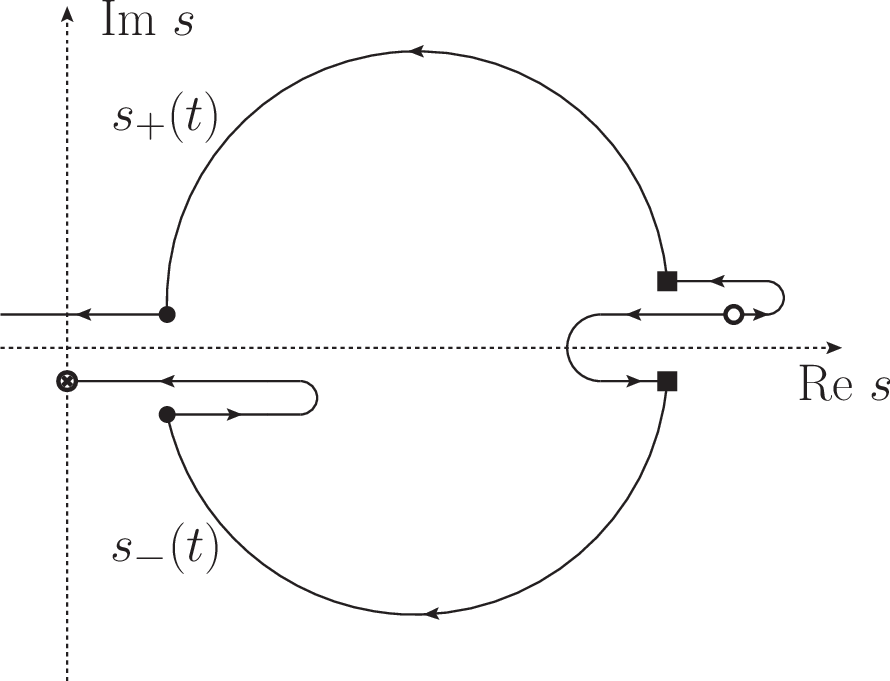}
		\caption*{I.a)}
	\end{subfigure}
	\hspace{10pt}
	\begin{subfigure}{0.3\textwidth}
		\centering
		\includegraphics[width=\textwidth]{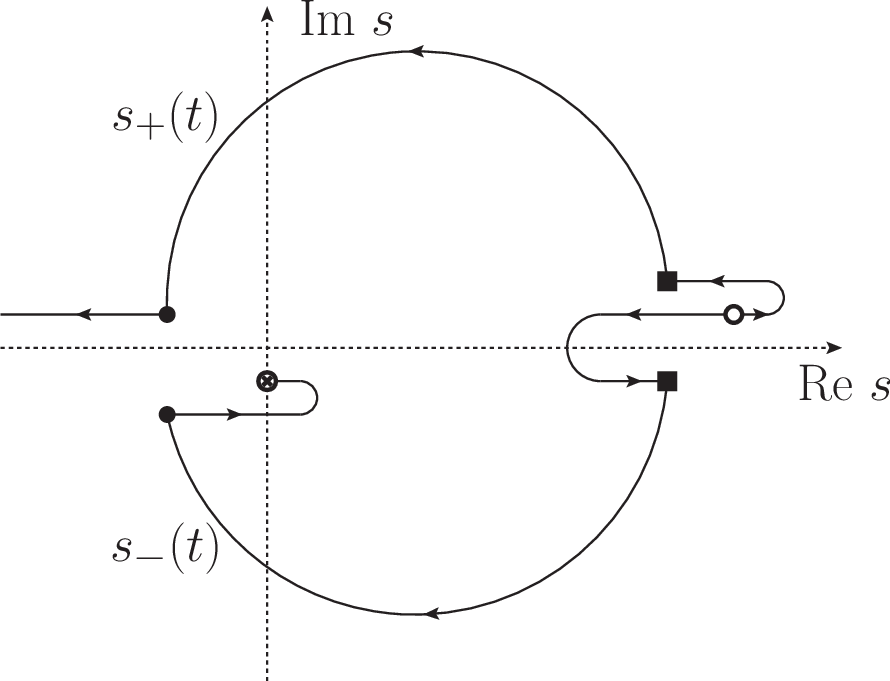}
		\caption*{I.b)}
	\end{subfigure}
	\hspace{10pt}
	\begin{subfigure}{0.3\textwidth}
		\centering
		\includegraphics[width=\textwidth]{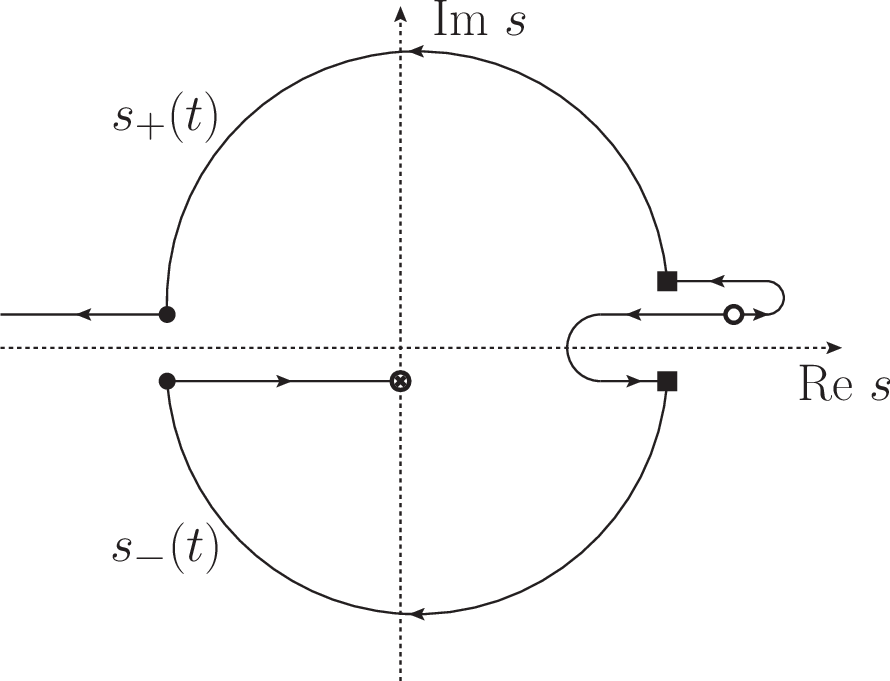}
		\caption*{I.c)}
	\end{subfigure}
	\begin{subfigure}{0.3\textwidth}
		\centering
		\includegraphics[width=\textwidth]{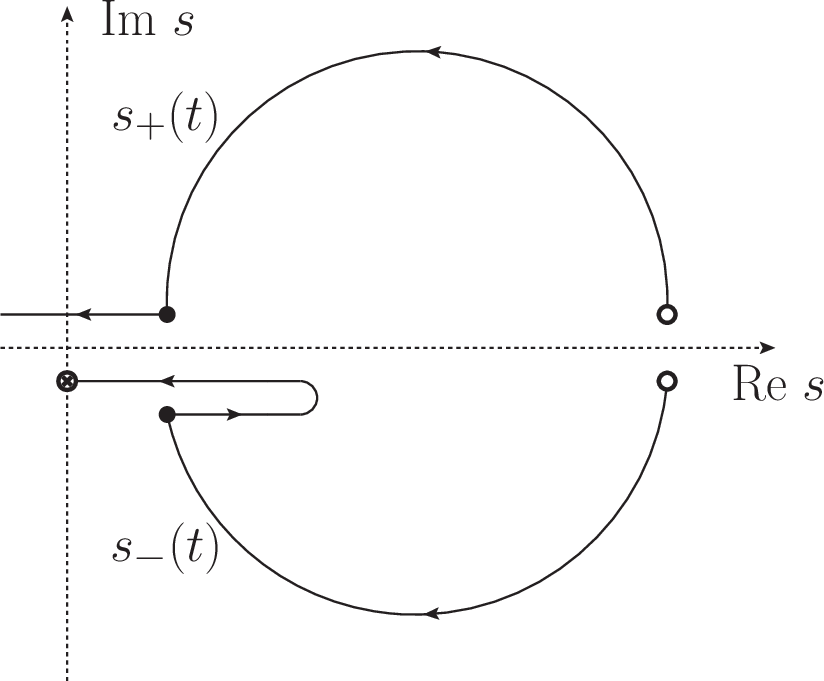}
		\caption*{II.a)}
	\end{subfigure}
	\hspace{10pt}
	\begin{subfigure}{0.3\textwidth}
		\centering
		\includegraphics[width=\textwidth]{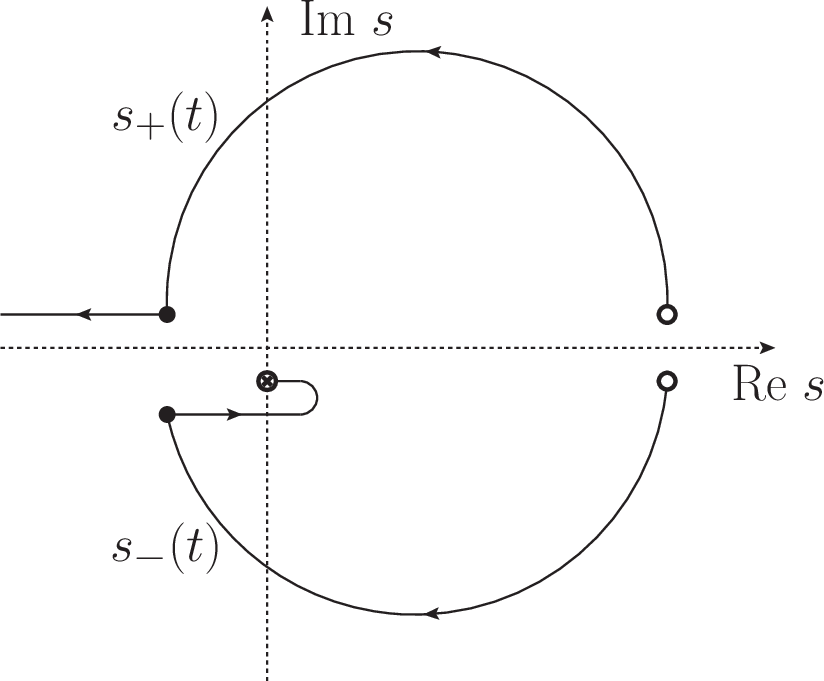}
		\caption*{II.b)}
	\end{subfigure}
	\hspace{10pt}
	\begin{subfigure}{0.3\textwidth}
		\centering
		\includegraphics[width=\textwidth]{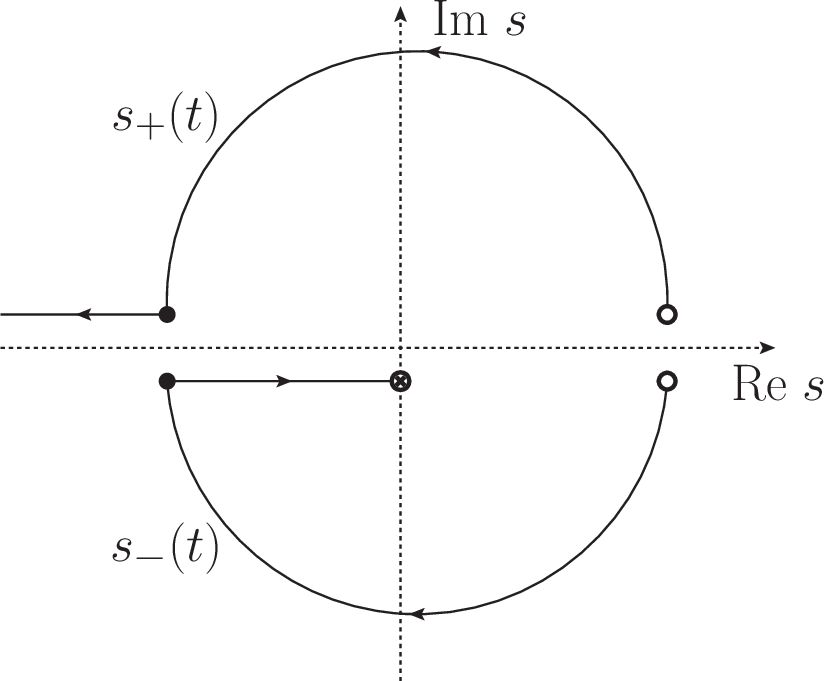}
		\caption*{II.c)}
	\end{subfigure}
	\begin{subfigure}{0.2\textwidth}
		\centering
		\includegraphics[width=\textwidth]{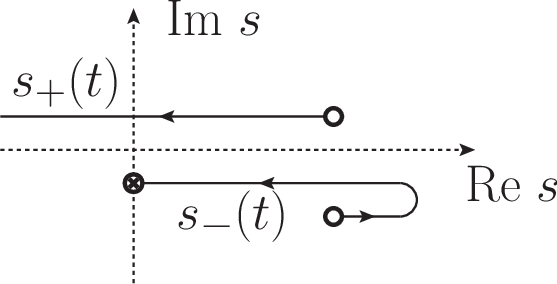}
		\caption*{III.a)}
	\end{subfigure}
	\hspace{10pt}
	\hspace{0.1\textwidth}
	\begin{subfigure}{0.2\textwidth}
		\centering
		\includegraphics[width=\textwidth]{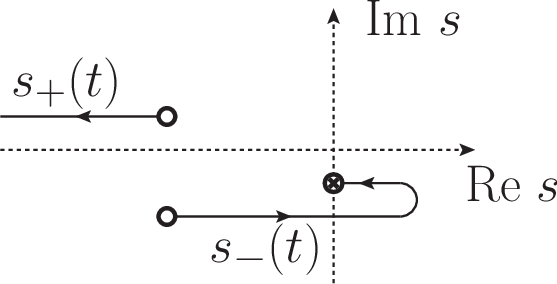}
		\caption*{III.b)}
	\end{subfigure}
	\hspace{10pt}
	\hspace{0.1\textwidth}
	\begin{subfigure}{0.2\textwidth}
		\centering
		\includegraphics[width=\textwidth]{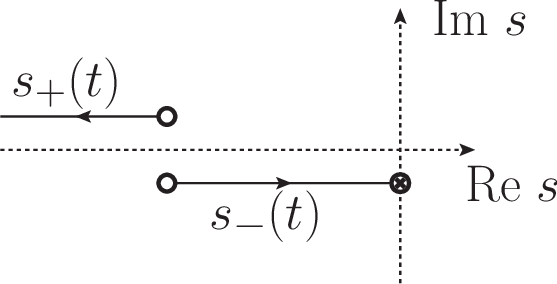}
		\caption*{III.c)}
	\end{subfigure}
	\caption{Paths of $s_\pm(t)$ in the complex $s$-plane where $t \in [t_1,\infty)$ for all nine kinematic configurations. The empty circles mark $s_\pm(t_1)$, the squares $s_\pm(t_2)$, the filled circles $s_\pm(t_3)$, and the crossed circles $s_-(\infty)$.}
	\label{fig:pinocchio_general}
\end{figure}

To systematically scan through all possible mass configurations and perform the analytic continuation of the scalar triangle function $C_0(t)$~\eqref{eq:triangle_function} for all these cases, we first need to identify all qualitatively different configurations of the physical regions in the Mandelstam plane.
Using the Mandelstam variables introduced in Eq.~\eqref{eq:triangle_mandelstam}, the boundaries of the physical region are the solutions to
\begin{equation}
	G(t,s,m_3^2,p_1^2,p_3^2,m_2^2) = 0,
\end{equation}
where $G$ is the four-particle function 
\begin{align}\label{eq:four_particle_function}
	G(x,y,z,u,v,w) &= x^2 y + y^2 x + z^2 u + u^2 z + v^2 w + w^2 v + xzw + xuv + yzv + yuw \notag\\
	&- xy(z+u+v+w) - zu(x+y+v+w) - vw(x+y+z+u).
\end{align}
Following the discussion in Ref.~\cite{Byckling:1971vca}, we find the following nine different configurations defined by
\begin{enumerate}[I)]
	\item $t_1<t_2$,
	\begin{enumerate}[{I}.a)]
		\item $\Delta < 0$ and $s_\pm(t_3) > 0$,
		\item $\Delta < 0$ and $s_\pm(t_3) \leq 0$,
		\item $\Delta \geq 0$,
	\end{enumerate}
	\item $t_2 \leq t_1 \leq t_3$,
	\begin{enumerate}[{II}.a)]
		\item $\Delta < 0$ and $s_\pm(t_3) > 0$,
		\item $\Delta < 0$ and $s_\pm(t_3) \leq 0$,
		\item $\Delta \geq 0$,
	\end{enumerate}
	\item $t_3 < t_1$,
	\begin{enumerate}[{III}.a)]
		\item $\Delta < 0$ and $s_\pm(t_1) > 0$,
		\item $\Delta < 0$ and $s_\pm(t_1) \leq 0$,
		\item $\Delta \geq 0$,
	\end{enumerate}
\end{enumerate}
where we used the short-hand notation 
\begin{align}
\Delta &= (p_1^2-m_2^2)(p_3^2-m_3^2),\notag\\
t_1 &= (m_2+m_3)^2,\qquad  
t_2 = \Big(\sqrt{p_1^2}-\sqrt{p_3^2}\Big)^2,\qquad  t_3 = \Big(\sqrt{p_1^2}+\sqrt{p_3^2}\Big)^2, 
\end{align}
and where $s_\pm(t)=s(t,z=\pm 1)$ with
\begin{align}
	s_\pm(t_1) &= \frac{m_3(p_1^2-m_2^2) + m_2(p_3^2-m_3^2)}{m_2 + m_3}, \notag\\
	s_\pm(t_2) &= \frac{\sqrt{p_1^2}(m_3^2-p_3^2) - \sqrt{p_3^2}(m_2^2-p_1^2)}{\sqrt{p_1^2} - \sqrt{p_3^2}},\notag \\
	s_\pm(t_3) &= \frac{\sqrt{p_1^2}(m_3^2-p_3^2) + \sqrt{p_3^2}(m_2^2-p_1^2)}{\sqrt{p_1^2} + \sqrt{p_3^2}}. 
\end{align}
For $t \in [t_1,\infty)$ these lead to the following paths of $s_\pm(t)$ in the complex $s$-plane shown in Fig.~\ref{fig:pinocchio_general}, where the $p_1^2 \to p_1^2 +\iu\delta$ and $p_3^2 \to p_3^2 +\iu\delta$ prescription was used, leading to an infinitesimal displacement to either side away from the real axis.

\begin{table}[t]
	\centering
	\renewcommand{\arraystretch}{1.3}
	\scalebox{0.785}{
	\begin{tabular}{lll}
	\toprule
		Configuration & $m_1^2 \in (\ldots)$ & Analytic continuation \\
		\midrule
		I.a.1 & $(s_2,\infty)$ & $D$ \\
		I.a.2 & $[s_\pm(t_1),s_2]$ & $C_2$ \\
		I.a.3 & $(s_\pm(t_2),s_\pm(t_1))$ & $C_2$ \\
		I.a.4 & $[s_1,s_\pm(t_2)]$ & $C_1$ \\
		I.a.5 & $(s_0,s_1)$ & $B$ \\
		I.a.6 & $[s_\pm(t_3),s_0]$ & $C_3$ \\
		I.a.7 & $(0,s_\pm(t_3))$ & $C_2$ \\
		I.b.1 & $(s_2,\infty)$ & $D$ \\
		I.b.2 & $[s_\pm(t_1),s_2]$ & $C_2$ \\
		I.b.3 & $(s_\pm(t_2),s_\pm(t_1))$ & $C_2$ \\
		I.b.4 & $[s_1,s_\pm(t_2)]$ & $C_1$ \\
		I.b.5 & $(s_0,s_1)$ & $B$ \\
		I.b.6 & $(0,s_0]$ & $C_3$ \\
		I.c.1 & $(s_2,\infty)$ & $D$ \\
		I.c.2 & $[s_\pm(t_1),s_2]$ & $C_2$ \\
		I.c.3 & $(s_\pm(t_2),s_\pm(t_1))$ & $C_2$ \\
		I.c.4 & $[s_1,s_\pm(t_2)]$ & $C_1$ \\
		I.c.5 & $(s_0,s_1)$ & $B$ \\
		I.c.6 & $(0,s_0]$ & $A$ \\
		\bottomrule
	\end{tabular}
	\begin{tabular}{lll}
	\toprule
		Configuration & $m_1^2 \in (\ldots)$ & Analytic continuation \\
		\midrule
		II.a.1 & $(s_\pm(t_1),\infty)$ & $D$ \\
		II.a.2 & $[s_2,s_\pm(t_1)]$ & $A$ \\
		II.a.3 & $(s_0,s_2)$ & $B$ \\
		II.a.4 & $(s_\pm(t_3),s_0]$ & $C_3$ \\
		II.a.5 & $(0,s_\pm(t_3)]$ & $C_2$ \\
		II.b.1 & $(s_\pm(t_1),\infty)$ & $D$ \\
		II.b.2 & $[s_2,s_\pm(t_1)]$ & $A$ \\
		II.b.3 & $(s_0,s_2)$ & $B$ \\
		II.b.4 & $(0,s_0]$ & $C_3$ \\
		II.c.1 & $(s_\pm(t_1),\infty)$ & $D$ \\
		II.c.2 & $[s_2,s_\pm(t_1)]$ & $A$ \\
		II.c.3 & $(s_0,s_2)$ & $B$ \\
		II.c.4 & $(0,s_0]$ & $A$ \\
		III.a.1 & $(s_0,\infty)$ & $D$ \\
		III.a.2 & $[s_\pm(t_1),s_0]$ & $C_2$ \\
		III.a.3 & $(s_0,s_\pm(t_1))$ & $C_1$ \\
		III.b.1 & $(s_0,\infty)$ & $D$ \\
		III.b.2 & $(0,s_0]$ & $C_2$ \\
		III.c.1 & $(0,\infty)$ & $D$\\\bottomrule
	\end{tabular}
	}
	\renewcommand{\arraystretch}{1.0}
	\caption{Mass configurations and their corresponding analytic continuations.}
	\label{tab:mass_configurations_c0}
\end{table}

Having identified the possible kinematic configurations of $s_\pm(t)$, we can perform the analytic continuation of
\begin{equation}
	\disc C_0(t) = 2\pi\iu \, \theta(t-t_1) \, \frac{1}{\lambda^{\frac12}(t,p_1^2,p_3^2)} \int_{s_-(t)}^{s_+(t)}\dd{s} \frac{1}{s-m_1^2},
\end{equation}
similarly to Sec.~\ref{sec:triangle_function_dispersion} by starting with a sufficiently large value for $m_1^2$ and then decreasing it to its physical value.\footnote{The condition $m_1^2 < s_\pm(t_1)$ is equivalent to the condition from Eq.~\eqref{eq:triangle_singularity_condition}.} In this process we encounter the 38 different configurations in Table~\ref{tab:mass_configurations_c0}, where we set 
\begin{align}
	s_0 &= \min \Big\{ \Big(\sqrt{p_1^2}-m_2\Big)^2, \Big(\sqrt{p_3^2}-m_3\Big)^2 \Big\}, \notag\\
	s_1 &= \min \Big\{ \Big(\sqrt{p_1^2}+m_2\Big)^2, \Big(\sqrt{p_3^2}+m_3\Big)^2 \Big\}, \notag\\
	s_2 &= \max \Big\{ \Big(\sqrt{p_1^2}-m_2\Big)^2, \Big(\sqrt{p_3^2}-m_3\Big)^2 \Big\}, \notag\\
	s_3 &= \max \Big\{ \Big(\sqrt{p_1^2}+m_2\Big)^2, \Big(\sqrt{p_3^2}+m_3\Big)^2 \Big\}.
\end{align}
These lead to the analytic continuations $A$, $B$, $C_1$, $C_2$, $C_3$, and $D$, to be defined in the following. For $A$ and $B$ the anomalous threshold $t_+$ lies on the physical sheet, so that we need to include an anomalous integral. For $C_1$, $C_2$, and $C_3$ both $t_+$ and $t_-$ lie along the unitarity cut, in such a way that the anomalous contribution can be absorbed into the normal dispersion integral. In the special cases of I.a.2, I.b.2, I.c.2, III.a.2, and III.b.2 (see Table~\ref{tab:mass_configurations_c0}), all of which have the analytic continuation $C_2$ while still fulfilling $m_1^2 \geq s_\pm(t_1)$, the anomalous threshold $t_-$ lies on the physical boundary, leading to a resonance-like peak. For $D$ both anomalous thresholds lie on the second sheet, so no anomalous contribution is needed. In all cases the position of the anomalous threshold is given by Eq.~\eqref{eq:triangle_singularities_analytic_continuation} and for both $A$ and $B$ the anomalous discontinuity is
\begin{equation}
	\disca C_0(t) = -\frac{4\pi^2}{\lambda^{\frac12}(t,p_1^2,p_3^2)}.
\end{equation}
The analytic continuation of $\disc C_0(t)$ is provided by
\begin{equation}\label{eq:disc_C0_case_AB}
	\disc C_0(t) = \theta(t-t_1) \begin{cases}
		\frac{-2\pi\iu}{\sqrt{\lambda(t,p_1^2,p_3^2)}} \left[ \log\frac{Y(t)+\kappa_+(t)}{Y(t)-\kappa_+(t)}  + 2\pi\iu \theta(t_2 - t) \right], & \lambda(t,p_1^2,p_3^2)>0, \\
		\frac{-4\pi\iu}{\sqrt{-\lambda(t,p_1^2,p_3^2)}} \left[ \arctan\frac{\kappa_-(t)}{Y(t)} + \pi \theta(t_0-t) \right], & \lambda(t,p_1^2,p_3^2)<0,
	\end{cases}
\end{equation}
in case $A$ and $B$, by
\begin{equation}\label{eq:disc_C0_case_C1}
	\disc C_0(t) = \theta(t-t_1) \begin{cases}
		\frac{-2\pi\iu}{\sqrt{\lambda(t,p_1^2,p_3^2)}} \Big[ \log\frac{Y(t)+\kappa_+(t)}{Y(t)-\kappa_+(t)}  + 2\pi\iu \theta(t_2 - t) \\ \hspace{55pt}- \pi\iu \theta(t_+ - t) - \pi\iu \theta(t_- - t) \Big], & \lambda(t,p_1^2,p_3^2)>0, \\
		\frac{-4\pi\iu}{\sqrt{-\lambda(t,p_1^2,p_3^2)}} \Big[ \arctan\frac{\kappa_-(t)}{Y(t)} + \pi \theta(t_0-t) \Big], & \lambda(t,p_1^2,p_3^2)<0,
	\end{cases}
\end{equation}
in case $C_1$, by
\begin{equation}\label{eq:disc_C0_case_C2}
	\disc C_0(t) = \theta(t-t_1) \begin{cases}
		\frac{-2\pi\iu}{\sqrt{\lambda(t,p_1^2,p_3^2)}} \Big[ \log\frac{Y(t)+\kappa_+(t)}{Y(t)-\kappa_+(t)} \\ \hspace{55pt}+ \pi\iu \theta(t_+ - t) - \pi\iu \theta(t_- - t) \Big], & \lambda(t,p_1^2,p_3^2)>0, \\
		\frac{-4\pi\iu}{\sqrt{-\lambda(t,p_1^2,p_3^2)}} \Big[ \arctan\frac{\kappa_-(t)}{Y(t)} \Big], & \lambda(t,p_1^2,p_3^2)<0,
	\end{cases}
\end{equation}
in case $C_2$, by
\begin{equation}\label{eq:disc_C0_case_C3}
	\disc C_0(t) = \theta(t-t_1) \begin{cases}
		\frac{-2\pi\iu}{\sqrt{\lambda(t,p_1^2,p_3^2)}} \Big[ \log\frac{Y(t)+\kappa_+(t)}{Y(t)-\kappa_+(t)}  + 2\pi\iu \theta(t - t_3) \\ \hspace{55pt}- \pi\iu \theta(t - t_-) - \pi\iu \theta(t - t_+) \Big], & \lambda(t,p_1^2,p_3^2)>0, \\
		\frac{-4\pi\iu}{\sqrt{-\lambda(t,p_1^2,p_3^2)}} \Big[ \arctan\frac{\kappa_-(t)}{Y(t)} - \pi \theta(t-t_0) \Big], & \lambda(t,p_1^2,p_3^2)<0,
	\end{cases}
\end{equation}
in case $C_3$, and by
\begin{equation}\label{eq:disc_C0_case_D}
	\disc C_0(t) = \theta(t-t_1) \begin{cases}
		\frac{-2\pi\iu}{\sqrt{\lambda(t,p_1^2,p_3^2)}} \log\frac{Y(t)+\kappa_+(t)}{Y(t)-\kappa_+(t)}, & \lambda(t,p_1^2,p_3^2)>0, \\
		\frac{-4\pi\iu}{\sqrt{-\lambda(t,p_1^2,p_3^2)}} \arctan\frac{\kappa_-(t)}{Y(t)}, & \lambda(t,p_1^2,p_3^2)<0,
	\end{cases}
\end{equation}
in case $D$, where $Y(t)$, $t_0$, and $\kappa_\pm(t)$ were introduced in Eqs.~\eqref{Y_def}, \eqref{eq:disc_C0_t0}, and~\eqref{kappapm_def}, respectively. In the case $C_1$ the discontinuity has a square-root singularity at $t_2$, which needs to be treated as described in Sec.~\ref{sec:numerical_treatment_pseudothreshold}, and likewise for $t_3$ in case $C_3$.

\section{Numerical treatment of the pseudothreshold singularity} \label{sec:numerical_treatment_pseudothreshold}

We consider the following integral
\begin{equation}
	I(t) = \int_{t_\text{thr}}^{\infty} \dd{t^\prime} \frac{S(t^\prime)}{t^\prime-t\mp\iu\epsilon},
\end{equation}
to be integrated numerically, where $S(t)$ has a singularity at the pseudothreshold $t_\text{ps} = \big(\sqrt{p_1^2+\iu\delta}-\sqrt{p_3^2}\big)^2$ of the type $S(t) \sim (t_\text{ps}-t^\prime)^{-(2l+1)/2}$ (non-vanishing values of $l$ arise for higher partial waves). In order to have a stable numerical integration both this singularity and the Cauchy kernel $(t^\prime-t\mp\iu\epsilon)^{-1}$ must be tamed. In order to achieve this, we follow the procedure presented in~App.~C of Ref.~\cite{Stamen:2022eda}.\footnote{These singularities are related to non-Landau singularities, also called singularities of the second type~\cite{Fairlie:1962}, which can arise on the non-principal Riemann sheets due to a pinching of the integration contour at infinity, see Refs.~\cite{Itzykson:1980rh,Mutke:2024} for more details.}

First, let us define $T(t) \equiv (t_\text{ps}-t^\prime)^{(2l+1)/2} S(t)$ which removes this singularity to rewrite the integral as
\begin{equation}
	I(t) = \int_{t_\text{thr}}^{\infty} \dd{t^\prime} \frac{T(t^\prime)}{(t_\text{ps}-t^\prime)^{(2l+1)/2}(t^\prime-t\mp\iu\epsilon)}.
\end{equation}
Next, we have to distinguish between the cases for which $t$ does and does not lie on the integration path. We begin with the former where $\Im{t} \neq 0$ or $t<t_\text{thr}$. In this case the Cauchy kernel cannot become singular, so we only need to treat the pseudothreshold singularity. We do so by adding and subtracting the first $(l+1)$ terms of the Taylor expansion
\begin{equation}
	T(t) = T(t_\text{ps}) + (t-t_\text{ps}) T^\prime(t_\text{ps}) + \frac{(t-t_\text{ps})^2}{2} T^{\prime\prime}(t_\text{ps}) + \ldots,
\end{equation}
around $t_\text{ps}$ in the integrand to obtain
\begin{equation}
	I(t) = \int_{t_\text{thr}}^{\Lambda^2} \dd{t^\prime} \frac{\tilde{T}(t^\prime)}{(t_\text{ps}-t^\prime)^{(2l+1)/2}(t^\prime-t\mp\iu\epsilon)} + \sum_{k=0}^{l} \frac{(-1)^k}{k!} T^{(k)}(t_\text{ps})\, \mathcal{Q}_{\frac{2k+1}{2}}(t,t_\text{thr},\Lambda^2),
\end{equation}
where $T^{(k)}(t)\equiv\dv[k]{t}T(t)$ and we introduced a high-energy cutoff $\Lambda^2$ as well as the functions
\begin{equation}
	\tilde{T}(t) \equiv T(t) - \sum_{k=0}^{l} \frac{(-1)^k}{k!} T^{(k)}(t_\text{ps})\, \mathcal{Q}_{\frac{2k+1}{2}}(t,t_\text{thr},\Lambda^2),
\end{equation}
and
\begin{equation}
	\mathcal{Q}_{\frac{2k+1}{2}}(t,x,y) = \int_{x}^{y} \dd{t^\prime} \frac{1}{(t_\text{ps}-t^\prime)^{\frac{2k+1}{2}}(t^\prime-t)}.
\end{equation}
The latter can be integrated analytically to yield~\cite{Stamen:2022eda}
\begin{equation}
	\mathcal{Q}_\frac12(t,x,y) = \frac{1}{\sqrt{t_\text{ps}-t}}  \left( \log\frac{\sqrt{t_\text{ps}-t}+\sqrt{t_\text{ps}-x}}{\sqrt{t_\text{ps}-t}-\sqrt{t_\text{ps}-x}} -2\iu \arctan\frac{\sqrt{y-t_\text{ps}}}{\sqrt{t_\text{ps}-t}} \right),
\end{equation}
and
\begin{equation}
	\mathcal{Q}_{\frac{2k+1}{2}}(t,x,y) = \frac{1}{t_\text{ps}-t}\left( \frac{2}{(2k-1)\sqrt{y-t_\text{ps}}^{2k-1}} - \frac{-2\iu}{(2k-1)\sqrt{t_\text{ps}-x}^{2k-1}} + \mathcal{Q}_{\frac{2k-1}{2}}(t,x,y) \right).
\end{equation}

In the other case where $t$ lies on the integration path we introduce an artificial cutoff point $p = (t+t_\text{ps})/2$ to split the integral into two parts that each contain one of the two singularities. For $t<t_\text{ps}$ this leads to
\begin{align}
	I(t) &= \int_{t_\text{thr}}^{p} \dd{t^\prime} \frac{T(t^\prime)-T(t)}{(t_\text{ps}-t^\prime)^{(2l+1)/2}(t^\prime-t)} + T(t)\, \mathcal{R}_{\frac{2l+1}{2}}(t,t_\text{thr},p)\notag \\
	&+\int_{p}^{\Lambda^2} \dd{t^\prime} \frac{\tilde{T}(t^\prime)}{(t_\text{ps}-t^\prime)^{(2l+1)/2}(t^\prime-t)} + \sum_{k=0}^{l} \frac{(-1)^k}{k!} T^{(k)}(t_\text{ps})\, \mathcal{Q}_{\frac{2k+1}{2}}(t,p,\Lambda^2), 
\end{align}
where
\begin{equation}
	\mathcal{R}_{\frac{2k+1}{2}}(t,x,y) = \int_{x}^{y} \dd{t^\prime} \frac{1}{(t_\text{ps}-t^\prime\mp\iu\epsilon)^{\frac{2k+1}{2}}(t^\prime-t)}
\end{equation}
can also be evaluated analytically~\cite{Stamen:2022eda}
\begin{equation}
	\mathcal{R}_\frac12(t,x,y) = \frac{1}{\sqrt{t_\text{ps}-t}} \left( \log\frac{\sqrt{_\text{ps}-x}+\sqrt{t_\text{ps}-t}}{\sqrt{t_\text{ps}-x}-\sqrt{t_\text{ps}-t}} - \log\frac{\sqrt{_\text{ps}-t}+\sqrt{t_\text{ps}-y}}{\sqrt{t_\text{ps}-t}-\sqrt{t_\text{ps}-y}} \pm \iu\pi \right),
\end{equation}
and
\begin{equation}
	\mathcal{R}_{\frac{2k+1}{2}}(t,x,y) = \frac{1}{t_\text{ps}-t} \left( \frac{2}{(2k-1)\sqrt{t_\text{ps}-y}^{2k-1}} - \frac{2}{(2k-1)\sqrt{t_\text{ps}-x}^{2k-1}} + \mathcal{R}_{\frac{2k-1}{2}}(t,x,y) \right).
\end{equation}
For $t>t_\text{ps}$ on the other hand we find
\begin{align}
	I(t) &= \int_{p}^{\Lambda^2} \dd{t^\prime} \frac{T(t^\prime)-T(t)}{(t_\text{ps}-t^\prime)^{(2l+1)/2}(t^\prime-t)} + T(t)\, \mathcal{R}_{\frac{2l+1}{2}}(t,p,\Lambda^2) \notag\\
	&+\int_{t_\text{thr}}^{p} \dd{t^\prime} \frac{\tilde{T}(t^\prime)}{(t_\text{ps}-t^\prime)^{(2l+1)/2}(t^\prime-t)} + \sum_{k=0}^{l} \frac{(-1)^k}{k!} T^{(k)}(t_\text{ps})\, \mathcal{Q}_{\frac{2k+1}{2}}(t,t_\text{thr},p).
\end{align}
In this form, the numerical integration is stable and can be performed, e.g., via a Gauss--Legendre quadrature routine. To significantly increase the computation speed both $T(t^\prime)$ and ${\tilde{T}(t^\prime)/(t_\text{ps}-t^\prime)^{(2l+1)/2}}$ are sampled and interpolated via cubic splines beforehand.

\begin{table}[tb]
	\centering
	\renewcommand{\arraystretch}{1.3}
 \scalebox{0.92}{
	\begin{tabular}{ll}\toprule
		Particle & Mass $[\text{GeV}]$\\
		\midrule
		$\pi^\pm$ & \SI{0.13957039(18)}{} \\
		$\pi^0$ & \SI{0.1349768(5)}{} \\
		$\rho^\pm$ & \SI{0.77511(34)}{} \\
		$\rho^0$ & \SI{0.77526(23)}{} \\
		$\omega$ & \SI{0.78266(13)}{} \\
		$K^\pm$ & \SI{0.493677(16)}{} \\
		$K^0$ & \SI{0.497611(13)}{} \\
		$K^{*\pm}$ & \SI{0.89167(26)}{} \\
		$K^{*0}$ & \SI{0.89555(20)}{} \\
		$B^\pm$ & \SI{5.27934(12)}{} \\
		$B^0$ & \SI{5.27966(12)}{} \\\bottomrule
	\end{tabular}
 \quad
	\begin{tabular}{lll}\toprule
		Decay & Branching fraction & References\\
		\midrule
		$B^0 \to K^+ \pi^-$& $\SI{2.00(4)e-5}{}$& \cite{CLEO:2003oxc,BaBar:2006pvm,Belle:2012dmz,Belle-II:2023ksq}\\
		$B^0 \to K^{*+} \pi^-$ & $ \SI{7.5(4)e-6}{}$& \cite{CLEO:2002jwu,Belle:2006ljg,BaBar:2009jov,BaBar:2011vfx,LHCb:2017pkd}\\
		$B^0 \to \pi^+ \pi^-$& $ \SI{5.37(20)e-6}{}$& \cite{CLEO:2003oxc,BaBar:2006pvm,CDF:2011ubb,LHCb:2012ihl,Belle:2012dmz,Belle-II:2023ksq}\\
		$B^0 \to \rho^+ \pi^-$& $\frac12 \times \SI{2.30(23)e-5}{}$&\cite{CLEO:2000xjz,BaBar:2003hgn,Belle:2007jkw}\\
		$B^+ \to K^0 \pi^+$& $\SI{2.39(6)e-5}{}$&\cite{CLEO:2003oxc,BaBar:2006enb,Belle:2012dmz,Belle-II:2023ksq}\\
		$B^+ \to K^{*0} \pi^+$ &  $\SI{1.01(8)e-5}{}$&\cite{Belle:2005rpz,BaBar:2008lpx,BaBar:2015pwa}\\
		$B^+ \to \pi^0 \pi^+$&$\SI{5.31(26)e-6}{}$&\cite{CLEO:2003oxc,Belle:2006uab,BaBar:2007uoe,Belle:2012dmz,Belle-II:2023ksq}\\
		$B^+ \to \rho^0 \pi^+$& $\SI{0.83(12)e-5}{}$& \cite{CLEO:2000xjz,Belle:2002ezq,BaBar:2009vfr}\\
		$B^+ \to \omega \pi^+$&$\SI{6.9(5)e-6}{}$& \cite{CLEO:2000xjz,Belle:2006tin,BaBar:2007nku}\\
		\bottomrule
	\end{tabular}}
	\renewcommand{\arraystretch}{1.0}
	\caption{Meson masses (left) and branching fractions (right) from Ref.~\cite{ParticleDataGroup:2022pth}. The $B$-meson lifetimes are $\tau_{B^0} =  \SI{1519(4)e-15}{\second}$ and $\tau_{B^+} = \SI{1638(4)e-15}{\second}$~\cite{HFLAV:2022esi}. The factor $1/2$ for $B^0\to\rho^+\pi^-$ accounts for the sum over charges in the listed branching fraction.}
	\label{tab:meson_masses}
\end{table} 

\section{Decay rates and coupling constants}
\label{app:coupling}

In this appendix, we collect the relations between the couplings describing
 $P_1\to P_2 P_3$, $P_1 \to V_2 P_3$, and $V_1 \to P_2 P_3$ decays and the corresponding branching ratios. We give all relations for distinguishable particles, and symmetry factors need to be added otherwise. 
 \begin{enumerate}
  \item For the $P_1\to P_2 P_3$ decay we use the amplitude
\begin{equation}
	\mathcal{M}\big[P_1(p) \to P_2(q_1) P_3(q_2)\big] = g_{P_1P_2P_3}
\end{equation}
to compute the branching fraction
\begin{equation} \label{eq:coupling_from_br_PPP}
	\Br [P_1 \to P_2 P_3] = \tau_{P_1} \frac{\lambda^{1/2}(M_{P_1}^2,M_{P_2}^2,M_{P_3}^2)}{16 \pi M_{P_1}^3} |g_{P_1P_2P_3}|^2,
\end{equation}
which can be solved to obtain the coupling constant $|g_{P_1P_2P_3}|$.
\item For the $P_1\to V_2 P_3$ decay we have
\begin{equation}
	\mathcal{M}\big[P_1(p) \to V_2(q_1) P_3(q_2)\big] = g_{P_1V_2P_3} \eta_{V_2,\lambda}^{*\alpha} (p + q_2)_\alpha,
\end{equation}
leading to
\begin{equation} \label{eq:coupling_from_br_PVP}
	\Br [P_1 \to V_2P_3] = \frac{\Gamma[P_1 \to V_2P_3]}{\Gamma_{P_1}^{\mathrm{tot}}}  = \tau_{P_1} \frac{\lambda^{3/2}(M_{P_1}^2,M_{V_2}^2,M_{P_3}^2)}{16 \pi M_{P_1}^3 M_{V_2}^2} |g_{P_1V_2P_3}|^2,
\end{equation}
which can be solved for the coupling constant $|g_{P_1V_2P_3}|$.
\item The third case of interest is the decay $V_1 \to P_2 P_3$ with amplitude
\begin{equation}
	\mathcal{M}\big[V_1(k) \to P_2(q_1) P_3(q_2)\big] = g_{V_1P_2P_3} \eta_{V_1,\lambda}^{*\alpha} (q_1 - q_2)_\alpha
\end{equation}
and decay width
\begin{equation} \label{eq:coupling_from_br_VPP}
	\Gamma [V_1 \to P_2P_3]  = \frac{\lambda^{3/2}(M_{V_1}^2,M_{P_2}^2,M_{P_3}^2)}{48 \pi M_{V_1}^5} |g_{V_1P_2P_3}|^2,
\end{equation}
which we can solve for $|g_{V_1P_2P_3}|$.
 \end{enumerate}
 
\begin{table}[tb]
	\centering
	\renewcommand{\arraystretch}{1.3}
	\begin{tabular}{llll}\toprule
	$g_{B^0 K^+ \pi^-}$ & $\SI{4.82(5)e-8}{\giga\electronvolt}$ & $g_{B^+ K^0 \pi^+}$ & $\SI{5.07(6)e-8}{\giga\electronvolt}$ \\
		$g_{B^0 K^{*+} \pi^-} $&$ \SI{9.82(26)e-10}{}$ & $g_{B^+ K^{*0} \pi^+} $&$ \SI{1.10(4)e-9}{}$\\
		$g_{B^0 \pi^+ \pi^-} $&$ \SI{2.49(5)e-8}{\giga\electronvolt}$ & $g_{B^+ \pi^0 \pi^+} $&$ \SI{2.38(6)e-8}{\giga\electronvolt}$ \\
		$g_{B^0 \rho^{+} \pi^-} $&$ \SI{1.05(5)e-9}{}$ & $g_{B^+ \rho^{0} \pi^+} $&$ \SI{8.6(6)e-10}{}$ \\
		$g_{K^{*+}K^{*0}\pi^+} $&$ \SI{10.9}{\per\GeV}$ & $g_{B^+ \omega \pi^+} $&$ \SI{7.88(29)e-10}{}$ \\
		$g_{K^{*0} K^{+} \pi^-} $&$ \SI{4.398(23)}{}$ & $g_{K^{*+} K^{0} \pi^+} $&$ \SI{4.69(4)}{}$ \\
		$g_{\rho^0 \pi^{+} \pi^-} $&$ \SI{5.94(2)}{}$ & $g_{\rho^+ \pi^{0} \pi^+} $&$ \SI{5.96(2)}{}$ \\
		$g_{\rho^0\omega\pi^0} $&$ \SI{15.4}{\per\GeV}$ & $g_{\rho^+\omega\pi^+} $&$ \SI{15.4}{\per\GeV}$\\\bottomrule
	\end{tabular}
	\renewcommand{\arraystretch}{1.0}
	\caption{Final values for the various couplings discussed in the main text. $g_{\rho^0\omega\pi^0}$ and $SU(3)$-related couplings are only estimates and we do not try to assign an uncertainty. Phases cannot be extracted from the branching fractions, so that all numbers refer to the modulus of the respective coupling (the exception being cases in which $SU(3)$ symmetry determines relative signs).}
	\label{tab:couplings}
\end{table}

To infer numerical values for the couplings, we use the particle masses and branching ratios given in Table~\ref{tab:meson_masses}. Moreover, for the $K^*$-meson the total decay widths $\Gamma[K^{*0} \to (K \pi)^0] = \SI{47.2(5)}{\MeV}$~\cite{Aston:1987ir,FOCUS:2005iqy,CLEO:2008jus,BaBar:2010vmf,BESIII:2015hty,ParticleDataGroup:2022pth}, $\Gamma[K^{*+} \to (K \pi)^+] = \SI{51.4(8)}{\MeV}$~\cite{Paler:1975qf,Bombay-CERN-CollegedeFrance-Madrid:1977glg,Toaff:1981yk,Birmingham-CERN-Glasgow-MichiganState-Paris:1984ppi,CrystalBarrel:2019zqh,ParticleDataGroup:2022pth},
in combination with isospin symmetry, imply
\begin{align}
	\Gamma[K^{*0} \to K^+ \pi^-] &= \frac23 \Gamma[K^{*0} \to (K \pi)^0] = \SI{31.5(3)}{\MeV},\notag\\
	\Gamma[K^{*0} \to K^0 \pi^0] &= \frac13 \Gamma[K^{*0} \to (K \pi)^0] = \SI{15.7(2)}{\MeV},\notag\\
	\Gamma[K^{*+} \to K^0 \pi^+] &= \frac23 \Gamma[K^{*+} \to (K \pi)^+] = \SI{34.3(5)}{\MeV},\notag\\
	\Gamma[K^{*+} \to K^+ \pi^0] &= \frac13 \Gamma[K^{*+} \to (K \pi)^+] = \SI{17.1(3)}{\MeV}.
\end{align}
For the $\rho$-meson a narrow-width approximation already becomes less well justified, with systematic uncertainties depending on the assumed spectral function. For definiteness, we use the parameters given in Ref.~\cite{ParticleDataGroup:2022pth}, $\Gamma[\rho^0 \to \pi^+ \pi^-] = \SI{147.4(8)}{\MeV}$~\cite{CMD-2:2006gxt,BaBar:2012bdw,SND:2020nwa} and $\Gamma[\rho^+ \to \pi^+ \pi^0] = \SI{149.1(8)}{\MeV}$~\cite{CLEO:1999dln,Achasov:2001hb,ALEPH:2005qgp,Belle:2008xpe}. For the last set of couplings we use the estimate $g_{\rho^0\omega\pi^0} = \SI{15.4}{\per\GeV}$~\cite{Zanke:2021wiq}, and from that infer the couplings $g_{\rho^+\omega\pi^+}$ and $g_{K^{*+}K^{*0}\pi^+}$ by $SU(3)$ symmetry~\cite{Leupold:2008bp,Lutz:2008km}, leading to $g_{\rho^+\omega\pi^+} = g_{\rho^0\omega\pi^0} =\sqrt{2}g_{K^{*+}K^{*0}\pi^+}$.
Combining all these constraints with the parameters from Table~\ref{tab:meson_masses}, we find the values for the couplings in Table~\ref{tab:couplings}.

\begin{table}[tb]
	\centering
	\renewcommand{\arraystretch}{1.3}
	\begin{tabular}{lllll}\toprule
	Decay & Branching fraction & $f_L$& $f_\perp$ & References\\\midrule
	$B^0 \to \rho^0 K^0$ & $\SI{0.34(11)e-5}{}$ & & & \cite{Belle:2006ljg,BaBar:2009jov,LHCb:2017pkd}\\
	$B^0 \to \rho^0 K^{*0}$ & $\SI{0.39(13)e-5}{}$ & $\SI{0.173(26)}{}$ &$\SI{0.40(4)}{}$ & \cite{Belle:2009roe,BaBar:2011ryf,LHCb:2018hsm}\\
	$B^0 \to \rho^0 \pi^0$ &  $\SI{0.20(5)e-5}{}$ &&& \cite{CLEO:2000xjz,BaBar:2003gxx,Belle:2007jkw}\\
	$B^0 \to \rho^0 \rho^0$ &  $\SI{0.096(15)e-5}{}$ & $0.71^{+0.08}_{-0.09}$&& \cite{BaBar:2008xku,Belle:2012ayg,LHCb:2015zxm}\\
	$B^+ \to \rho^0 K^+$ & $\SI{0.37(5)e-5}{}$ &&& \cite{Belle:2005rpz,BaBar:2008lpx}\\
	$B^+ \to \rho^0 K^{*+}$ & $\SI{0.46(11)e-5}{}$ & $\SI{0.78(12)}{}$&&\cite{BaBar:2010evf}\\
	$B^+ \to \rho^0 \pi^+$ & $\SI{0.83(12)e-5}{}$ &&& \cite{CLEO:2000xjz,Belle:2002ezq,BaBar:2009vfr}\\
	$B^+ \to \rho^0 \rho^+$ &  $\SI{2.40(19)e-5}{}$ & $\SI{0.950(16)}{}$&&
	\cite{Belle:2003lsm,BaBar:2009rmk}\\\bottomrule
	\end{tabular}
	\renewcommand{\arraystretch}{1.0}
	\caption{Branching and polarization fractions needed to determine the subtraction constants, taken from Ref.~\cite{ParticleDataGroup:2022pth}. $f_\perp$ has not been measured for the other vector channels.}
	\label{tab:BR_fL}
\end{table} 

Finally, to determine the subtraction constants we need experimental input for 
$B \to \rho (P,V)$ branching and polarization fractions. These are collected in Table~\ref{tab:BR_fL}.

\bibliographystyle{apsrev4-1_mod_2}
\bibliography{ref}
	
\end{document}